\documentclass[reprint,prb,footinbib,superscriptaddress]{revtex4-1}
\usepackage{stmaryrd}
\usepackage[dvipsnames]{color,xcolor}
\usepackage{graphicx,overpic}
\usepackage{amsmath,amssymb}
\usepackage[normalem]{ulem}
\usepackage{bbold,units}

\usepackage{mathtools}
\DeclarePairedDelimiter{\floor}{\lfloor}{\rfloor}
\DeclareFontFamily{U}{cry}{\hyphenchar\font=-1}
\DeclareFontShape{U}{cry}{m}{n}{ <-> cryst}{}
\newcommand{\cry}[1]{{\usefont{U}{cry}{m}{n} \symbol{#1}}}
\renewcommand{\vec}[1]{\underline{#1}}
\newcommand{\cc}[1]{\overline{#1}}
\newcommand{\mat}[1]{\underline{\underline{#1}}\,}
\newcommand{\matns}[1]{\underline{\underline{#1}}}

\newcommand{\uda}{\uparrow\!\downarrow}
\newcommand{\dua}{\downarrow\!\uparrow}
\newcommand{\da}{\downarrow}
\newcommand{\ua}{\uparrow}

\begin{document} 

%\title{Circular Polarization Effects in Eight-Fold Intergrown Dielectric Gyroid Nets:\\ From Group Theory to Designs for Nano-Fabricated Photonic Materials}

%\title{Eight-Fold Intergrowth of Dielectric Gyroid Nets:\\ Huge Optical Activity with Polarisation Conversion and Circular Dichroism Suppressed by 4-fold Rotational Symmetry}

\title{Group Theory of Circular-Polarisation Effects in Chiral Photonic Crystals with Four-Fold Rotation Axes, Applied to the Eight-Fold Intergrowth of Gyroid Nets} 
\date{\today}

\author{Matthias Saba}
\email[E-mail:]{Matthias.Saba@fau.de}
\affiliation{Theoretische~Physik, Friedrich-Alexander~Universit\"at~Erlangen-N\"urnberg, 91058~Erlangen, Germany}
\author{Mark D.~Turner}
\affiliation{CUDOS~\&~Centre~for~Micro-Photonics, Swinburne~University~of~Technology, Victoria~3122, Australia}
\author{Klaus Mecke}
\affiliation{Theoretische~Physik, Friedrich-Alexander~Universit\"at~Erlangen-N\"urnberg, 91058~Erlangen, Germany}
\author{Min Gu}
\affiliation{CUDOS~\&~Centre~for~Micro-Photonics, Swinburne~University~of~Technology, Victoria~3122, Australia}
\author{Gerd E.~Schr\"oder-Turk}
\email[E-mail:]{Gerd.Schroeder-Turk@fau.de}
\affiliation{Theoretische~Physik, Friedrich-Alexander~Universit\"at~Erlangen-N\"urnberg, 91058~Erlangen, Germany}


%%%ABSTRACT
\begin{abstract}
We use group or representation theory and scattering matrix calculations to derive analytical results for the band structure topology and the scattering parameters, applicable to any chiral photonic crystal with body-centered cubic symmetry $I432$ for circularly-polarised incident light. We demonstrate in particular that all bands along the cubic $[100]$ direction can be identified with the irreducible representations $E_\pm$, $A$ and $B$ of the $C_4$ point group. $E_+$ and $E_-$ modes represent the only transmission channels for plane waves with wave vector along the $\Delta$ line, and can be identified as non-interacting transmission channels for right- ($E_-$) and left-circularly polarised light ($E_+$), respectively. Scattering matrix calculations provide explicit relationships for the transmission and reflectance amplitudes through a finite slab which guarantee equal transmission rates for both polarisations and vanishing ellipticity below a critical frequency, yet allowing for finite rotation of the polarisation plane. All results are verified numerically for the so-called {\bf 8-srs} geometry, consisting of eight interwoven equal-handed dielectric Gyroid networks embedded in air. The combination of vanishing losses, vanishing ellipticity, near-perfect transmission and optical activity comparable to that of metallic meta-materials makes this geometry an attractive design for nanofabricated photonic materials.
\end{abstract}

\maketitle

%%%SECTION INTRODUCTION
Optical properties, such as optical rotation or circular dichroism, that are caused by a chiral structure of a light-transmitting medium or of its constituent molecules, remain of great interest in many different contexts. Circular dichroism spectroscopy of optically active molecules in solution is used in bio-chemistry where left-handed (LH) and right-handed (RH) molecular architectures cause different absorption properties for left-circularly polarised (LCP) and right-circularly polarised (RCP) light \cite{Johnson:1990}. Optical activity of natural crystals such as quartz (see e.g.~\cite{Barron:2009}) and of liquid crystals, both in the twisted nematic \cite{SchadtHelfrich:1971} and the blue phases \cite{GreletCollingsLiNguyen:2001,BensimonDomanyShtrikman:1983}, is well-known. Circularly polarised (CP) reflections of insect cuticles were observed by Michelson a century ago \cite{Michelson:1911}, with circular-polarisation effects an active topic in bio-photonics of beetles \cite{JewellVukusicRoberts:2007,SharmaCrneParkSrinivasarao:2009}, crustaceans \cite{KleinlogelWhite:2008,ChiouKleinlogelCroninCaldwellLoefflerSiddiqiGoldizenMarshal:2008} and butterflies \cite{SabaThielTurnerHydeGuBrauckmannNeshevMeckeSchroederTurk:2011}, and also the plant kingdom \cite{VignoliniRudallRowlandReedMoyroudFadenBaumbergGloverSteiner:2012}. Nano-fabrication technology nowadays allows for the fabrication of custom-designed chiral materials, both dielectric photonic crystals \cite{ThielRillVonFreymannWegener:2009,TurnerSchroederTurkGu:2011,TurnerSabaZhangCummingSchroederTurkGu:2013} and metallic meta-materials \cite{DeckerRutherKrieglerZhouSoukoulisLindenWegener:2009,GanselThielRillDeckerBadeSaileVonFreymannLindenWegener:2009,KuwataGonokamiSaitoInoKauranenJefimovsValliusTurunenSvirko:2005}, which the potential for technological photonic devices. This ability to fabricate custom-designed structures has led to noteworthy chiral-optical behaviour, including strong circular dichroism \cite{DeckerKleinWegenerLinden:2007}, negative refractive index \cite{ZhangParkLiLuZhangZhang:2009,*PlumZhouDongFedotovKoschnySoukoulisZheludev:2009} based on Pendry's prediction \cite{Pendry:2004}, optically-induced torque \cite{LiuZentgrafLiuBartalZhang:2010}, handedness switching in meta-molecules \cite{ZhangZhouParkRhoSinghNamAzadChenYinTaylorZhang:2012} and circular-polarised beam splitting \cite{TurnerSabaZhangCummingSchroederTurkGu:2013}. Metallic or plasmonic meta-materials have been designed to give strong optical activity \cite{KuwataGonokamiSaitoInoKauranenJefimovsValliusTurunenSvirko:2005,DeckerRutherKrieglerZhouSoukoulisLindenWegener:2009,RenPlumXuZheludev:2012,SchaeferlingDregelyHentschelGiessen:2012,RenPlumXuZheludev:2012} that is orders of magnitudes stronger than in the natural materials.
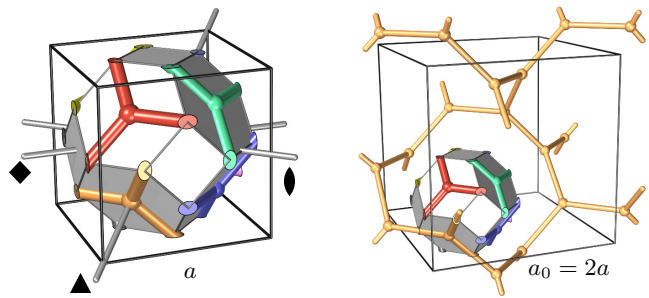
\begin{figure}[t]
\begin{minipage}{.49\columnwidth}
    \begin{overpic}[width=\textwidth]{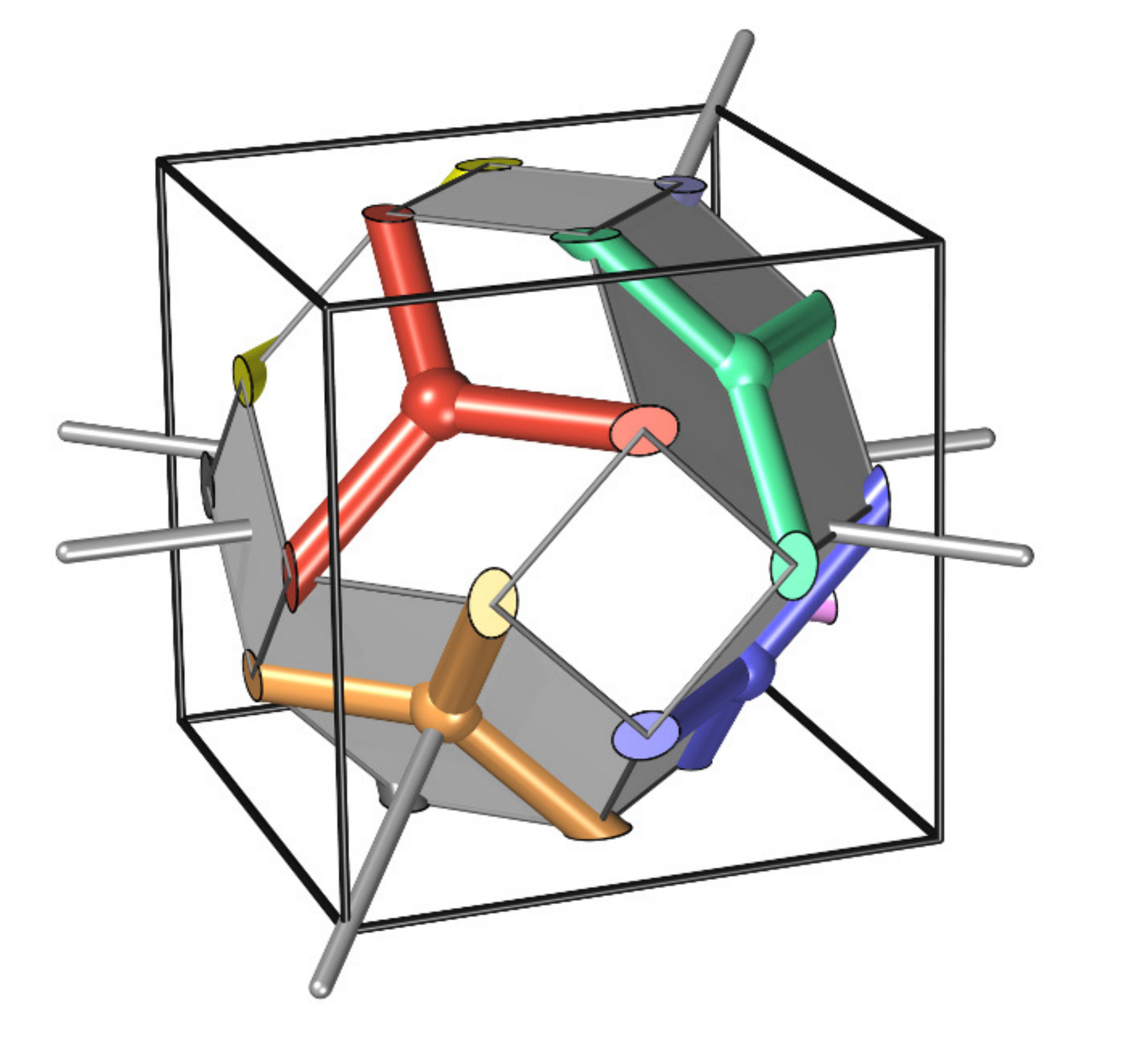}
    	\put(1,37){\color{black} \Large \cry{4}}
    	\put(87,34){\color{black} \Large \cry{2}}
    	\put(20,1){\color{black} \Large \cry{3}}
    	\put(56,7){\color{black} $a$}
    \end{overpic}
\end{minipage} \hfill
\begin{minipage}{.49\columnwidth}
    \begin{overpic}[width=\textwidth]{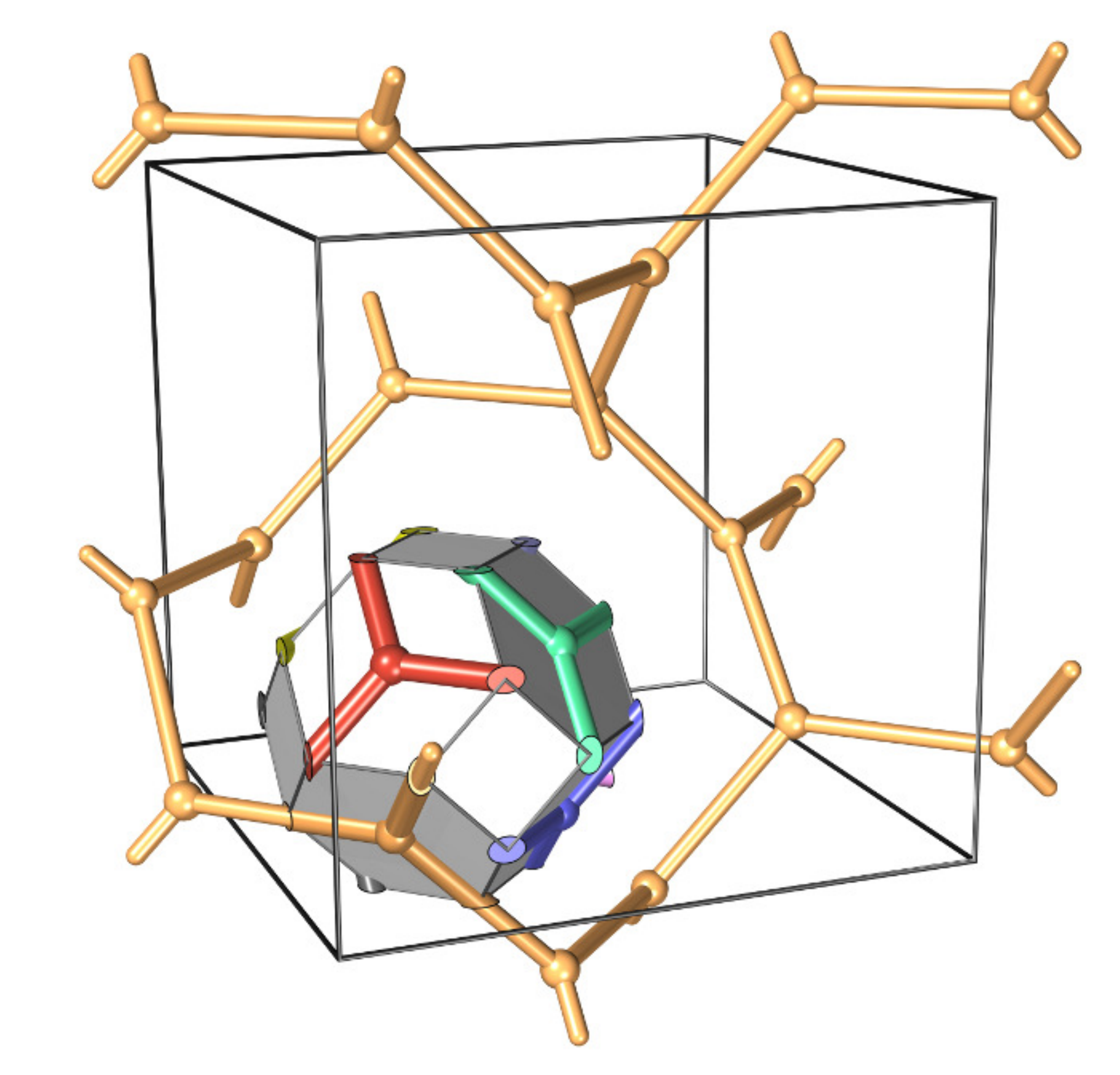}
    	\put(60,9){\color{black} $a_0=2a$}
    \end{overpic}
\end{minipage}
\caption{(Color Online) Construction of the {\bf 8-srs} structure: (left) The BCC translational unit cell is obtained by placing degree-three vertices at the mid-points of the hexagonal facets of a Kelvin body. Edges connect the hexagon's center point to every second vertex, such that rotational symmetries of the Kelvin body are maintained.  (right) When repeated periodically, the {\bf 8-srs} consists of 8 equal-handed interwoven {\bf srs} nets; only one of the eight nets is shown for clarity. }
\label{fig:1srs-8srs-prb}
\end{figure}

This article makes a two-fold contribution to a deeper understanding of circular polarisation effects in chiral materials. First, we combine group theory and scattering matrix treatment of chiral photonic crystals (PCs) to predict those properties of the band structure that are relevant for coupling to circularly-polarised light. This analysis applies to {\em all} structures with cubic symmetry with at least one point that has 2-, 3- and 4-fold point symmetry. Many of the ideas underlying this formalism are not specific to this symmetry group and apply, upon suitable adjustments, to structures with other symmetries. Second, we analyse in detail a specific chiral geometry that fulfils the symmetry requirements, namely the so-called {\bf 8-srs} structure consisting of eight interwoven non-overlapping Gyroid (or {\bf srs}) nets. Numeric data for its band structure and transmission coefficient, obtained by Finite element simulations and finite-difference time-domain methods, is in perfect agreement with our theoretical predictions. Further, these simulations demonstrate that the {\bf 8-srs} geometry exhibits a particularly strong chiral-optical response. Specifically, in parts of the frequency spectrum, the lossless dielectric {\bf 8-srs} material with $\epsilon=5.76$ is fully transparent for left-~and right-circularly polarised (LCP and RCP) light, yet exhibits an optical activity that is comparable to metallic meta-materials. This suggests that the use of multiply intertwined Gyroid nets, that can be realised with current nano-fabrication methods, is promising for custom-designed photonic materials.

{\em Polarisation conversion} measures the relative ratio of LCP to RCP light transmitted through a photonic material if the incident light was purely LCP. We will here use the term {\em optical activity} (OA) to denote the rotation of the polarisation plane as a linearly polarised plane wave transmits through a chiral medium and {\em Circular dichroism} (CD) is utilized to describe the difference in transmission rates for LCP and RCP light. The definition of OA and CD represent a slight deviation from the conventional nomenclature where both refer to a respective phenomenological origin, i.e.~a respective difference of the refractive indices and the absorption coefficients between LCP and RCP light.

Our use of these terms to describe differences in scattering parameters for CP light represents a natural adaption for a slab made of lossless but inhomogeneous material. With $t$ and $r$ denoting the scalar complex transmission and the reflection amplitudes, respectively, OA is the phase difference and CD as the relative difference in absolute values between the complex scattering amplitudes $s_\pm \in \{t_\pm,r_\pm\}$ for an incoming left (LCP, $+$) or right (RCP, $-$) circularly polarized plane wave:
$$
	\text{CD}_s = \frac{\left|s_+\right| - \left|s_-\right|}{\left|s_+\right| + \left|s_-\right|}\text{;} \quad
	\text{OA}_s = \frac{\varphi^{(s)}_+ - \varphi^{(s)}_-}{2}\text{;} \quad e^{\imath\varphi^{(s)}_\pm} = \frac{s_\pm}{\left|s_\pm\right|}
$$

\begin{figure}[t]
\begin{minipage}{.49\columnwidth}
    \centering
    {\large $t=0$ ("eyes")}
    \vspace{0.2cm}\\
    \begin{overpic}[width=\textwidth]{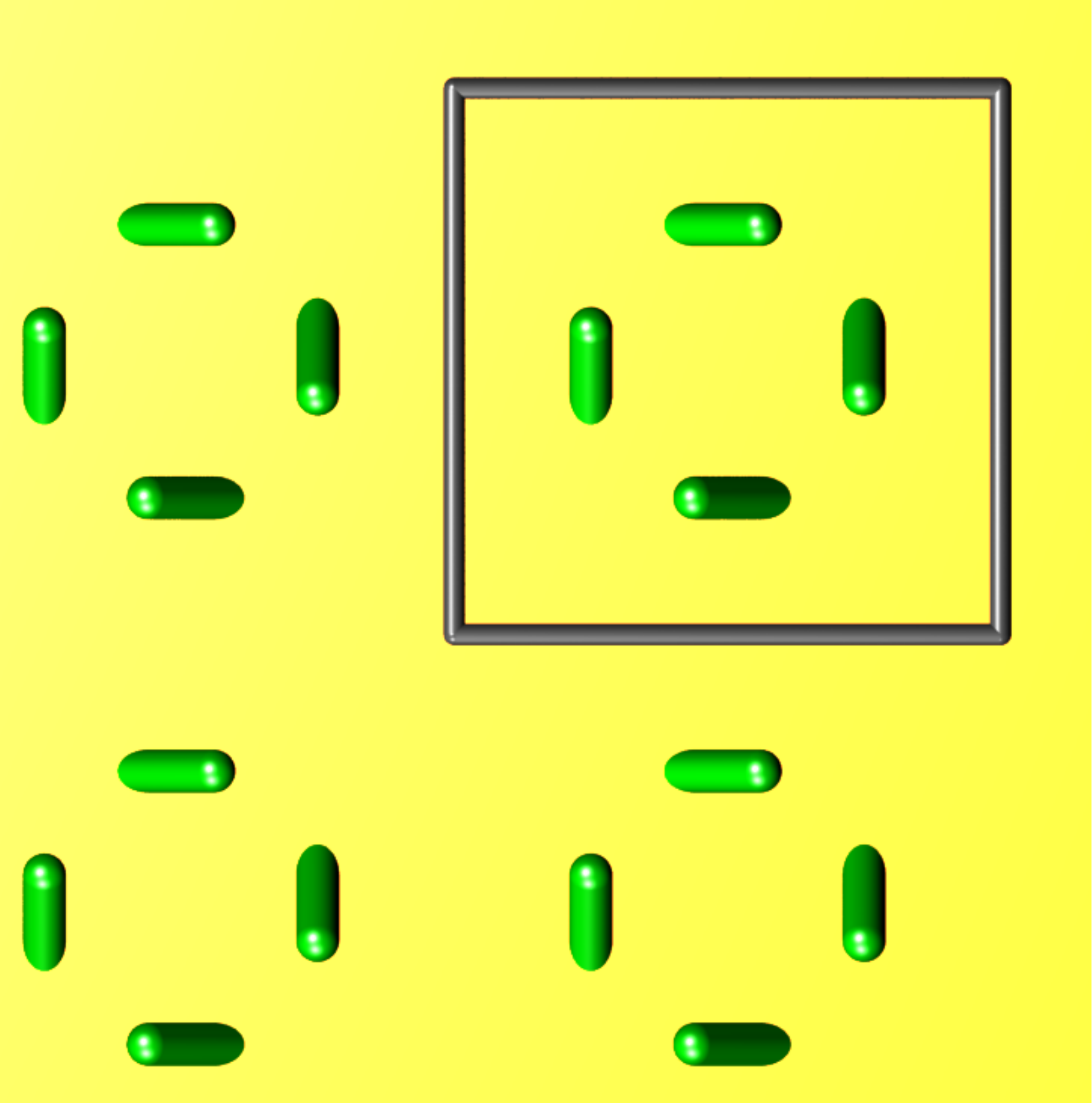}
    	\put(15,37.8){\color{blue} \Large $432$}
	\put(35.7,37.8){\color{blue} \huge \cry{4}}
	\put(85.5,37.8){\color{blue} \huge \cry{4}}
	\put(35.7,87){\color{blue} \huge \cry{4}}
	\put(85.5,87){\color{blue} \huge \cry{4}}
    	\put(60.6,62.3){\huge \cry{4}}
    	\put(60.6,37.8){\huge \cry{42}}
    	\put(60.6,87){\huge \cry{42}}
    	\put(35.7,62.3){\huge \cry{42}}
    	\put(85.5,62.3){\huge \cry{42}}
    \end{overpic}\\
\end{minipage} \hfill
\begin{minipage}{.49\columnwidth}
    \centering
    {\large $t=a/4$ ({\color{Brown} "dog bones"})}
    \vspace{0.2cm}\\
    \begin{overpic}[width=\textwidth]{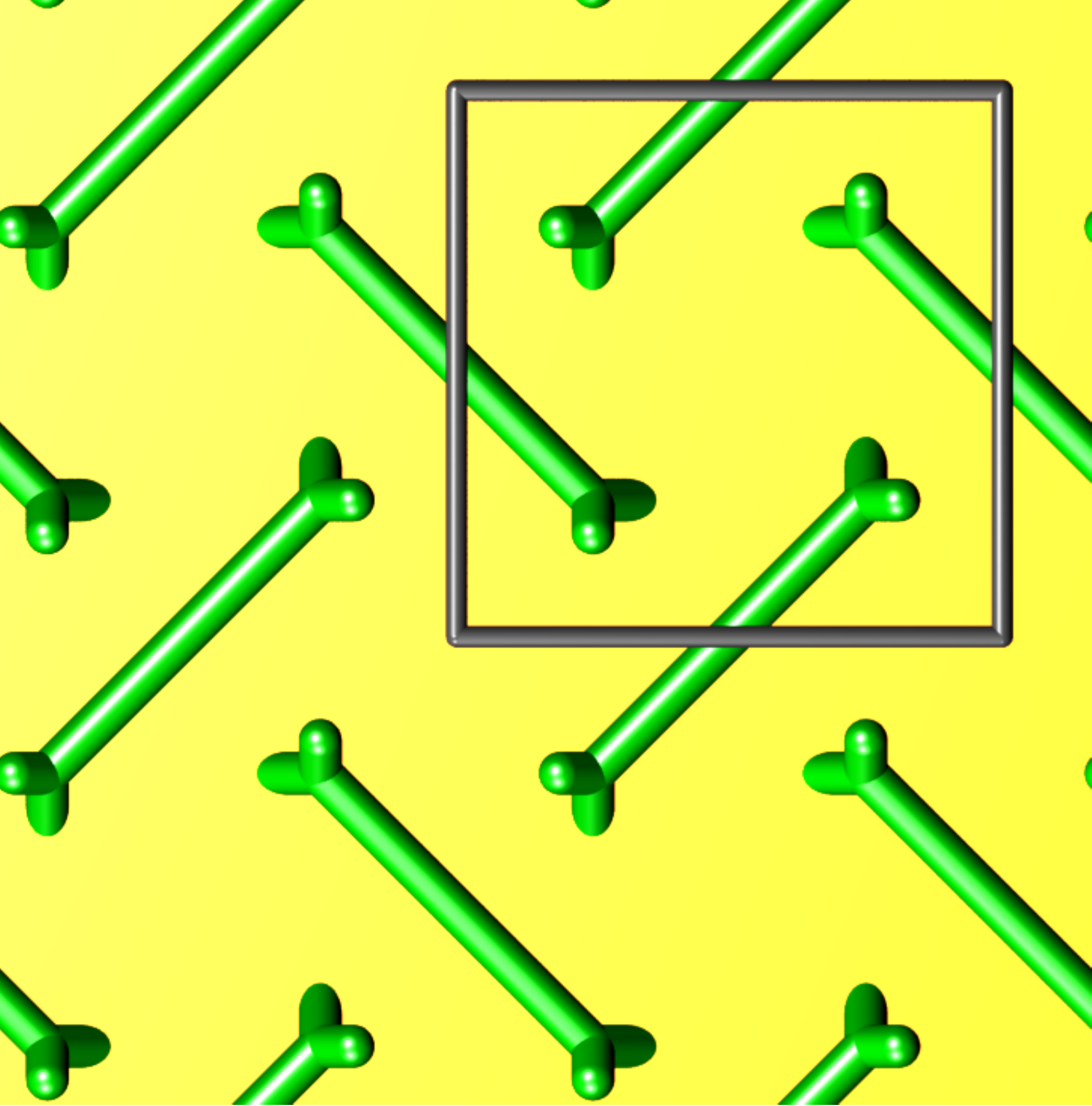}
    	\put(35.7,37.8){\huge \cry{4}}
    	\put(85.5,37.8){\huge \cry{4}}
    	\put(35.7,87){\huge \cry{4}}
    	\put(85.5,87){\huge \cry{4}}
    	\put(60.6,62.3){\huge \cry{4}}
    	\put(60.6,37.8){\huge \cry{42}}
    	\put(60.6,87){\huge \cry{42}}
    	\put(35.7,62.3){\huge \cry{42}}
    	\put(85.5,62.3){\huge \cry{42}}
    \end{overpic}\\
\end{minipage}
\caption{(Color Online) Different cross-sections with $[100]$ inclination through the {\bf 8-srs} reveal its four-fold symmetry. 
The parameter $t$ denotes the position of the termination plane within the cubic unit cell.
$4$-fold rotation and $4_2$ screw axes are marked by the symbols \cry{4} and \cry{42}, respectively. The grey square represents the cross section of the cubic unit cell whose vertices are located at the $432$ (Schoenfliess $O$) symmetry point marked by a blue \cry{4} symbol (same choice as in \cite{InternationalTablesForCrystallography:1992}).}
\label{fig:8srs-cross-section}
\end{figure}
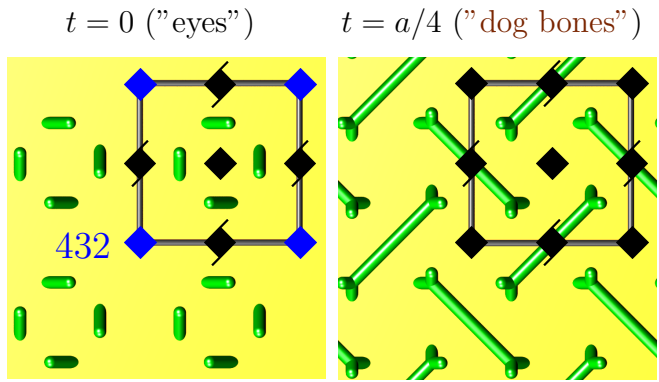

Group theoretic approaches to the optical properties of photonic crystals are motivated by the obvious importance of spatial symmetries in these systems. The common textbook example for this is the strict classification of discrete dispersion bands $\omega_n(\vec{k})$  of a two dimensional PC into transverse-electric and transverse-magnetic modes if the wave vector $\vec{k}$ is restricted to the plane of periodicity \cite{JoannopoulosJohnsonWinnMeade:2008}. In terms of circular polarisation, the role of four-fold rotational symmetries, and more generally of $(m\ge 3)$-fold rotations, for the transmission and reflectance coefficients has been recognised \cite{KaschkeGanselWegener:2012,BaiSvirkoTurunenVallius:2007,BaiVentolaTervoZhang:2012,MenzelRockstuhlLederer:2010,RockstuhlMenzelPaulLederer:2009,MaslovskiMoritsTretyakov:2009,Dmitriev:2011,Dmitriev:2013,KwonWernerWerner:2008}. In particular, it has recently been shown that any lossless structure with a $4$-fold symmetry axis inclined perpendicular to this axis shows no circular dichroism and polarization conversion for normal incidence \cite{KaschkeGanselWegener:2012}. Our results on the scattering parameters for chiral PCs presented in this article are generalizations of previous works that derived related restrictions on reflection and transmission amplitudes for two-dimensional diffraction gratings made of quasi-planar particles \cite{BaiSvirkoTurunenVallius:2007,BaiVentolaTervoZhang:2012,MenzelRockstuhlLederer:2010,MaslovskiMoritsTretyakov:2009} and also particles with 3D symmetry \cite{Dmitriev:2011,Dmitriev:2013}. Group theory, or more precisely representation theory, are the natural language to deal with the influence of spatial symmetry on physical properties in general \cite{Dresselhaus:2008,Sakoda:2005} and with photonic properties in particular \cite{Sakoda:2005}. 

A particularly intricate design for a chiral photonic material is the so-called {\em single Gyroid} (SG) or {\bf srs} net \cite{HydeOKeeffeProserpio:2008}, with edges inflated to solid tubes with a given volume fraction $\phi$. This periodic network is composed of identical three-coordinated nodes and has cubic chiral symmetry $I4_132$, without pure four-fold rotation axes \footnote{The network’s single type of 3-coordinated vertices are at Wyckhoff sites 8a (.32) in $I4_132$ (no. 214 in \cite{InternationalTablesForCrystallography:1992})}; the three-fold and four-fold screw axes correspond to three-fold and four-fold helical structures along the [111] and [100] directions, respectively. The theoretical prediction of circular dichroism for a dielectric photonic {\bf srs} crystal \cite{SabaThielTurnerHydeGuBrauckmannNeshevMeckeSchroederTurk:2011} has been verified by nano-fabrication experiments \cite{TurnerSchroederTurkGu:2011,TurnerSabaZhangCummingSchroederTurkGu:2013}. Recently, the first prediction of fully three-dimensional Weyl points has been published using photonic designs based on the SG \cite{LuFuJoannopoulosSoljacic:2013}. For metallic {\bf srs} nets, discrimination of LCP and RCP modes is observed \cite{OhDemetriadouWuestnerHess:2013,HurFrancescatoGianniniMaierHennigWiesner:2011}, but its magnitude is lower than what might be expected from its helical nature \cite{OhDemetriadouWuestnerHess:2013}. The relevance of the SG geometry, which is inspired by the biological PCs in wing scales of several butterflies \cite{MichielsenStavenga:2008,*SaranathanOsujiMochrieNohNarayananSandyDufresnePrum:2010,*SchroederTurkWickhamAverdunkBrinkFitzGeraldPoladianLargeHyde:2011}, as a chiral photonic material is increased by the ability to generate this structure at different length scales by molding from the self-assembly copolymer structure with unit cell size $a=50\mathrm{nm}$ \cite{VignoliniYufaCunhaGuldinRushkinStefikHurWiesnerBaumbergSteiner:2011}, replication of the butterfly structures with $a\approx 300 \mathrm{nm}$ \cite{MilleTyrodeCorkery:2013,*MilleTyrodeCorkery:2011}, direct laser writing with $a \gtrsim 1 \mu\mathrm{m}$ \cite{TurnerSabaZhangCummingSchroederTurkGu:2013,TurnerSchroederTurkGu:2011} and cm-scale replica for micro-wave experiments \cite{PouyaVukusic:2012}.

This article investigates a related geometry called {\bf 8-srs} consisting of eight identical equal-handed intertwined copies of the {\bf srs} net \cite{HydeRamsden:2000,HydeOguey:2000,EvansRobinsHyde:2013a}. As Fig.~1 of ref.~\onlinecite{SabaTurnerGuSchroeder_prl:2013} demonstrates, the {\bf 8-srs} can be obtained by arranging translated copies of the {\bf srs} net, such that all 8 networks remain disjoint and to yield body-centered cubic (BCC) symmetry $I432$; this is achieved by three orthogonal copy-translations along the perpendicular [100] directions of the original {\bf 1-srs} net by $a=a_0/2$ where $a_0$ is the crystallographic lattice parameter of the {\bf 1-srs} in its space group $I4_132$. Figure \ref{fig:1srs-8srs-prb} illustrates an alternative construction where the {\bf 8-srs} is obtained by the decoration of the hexagonal facets of a {\em Kelvin body}\footnote{The BCC Wigner-Seitz cell forms a Kelvin body that is a cube truncated along its [100] directions to give a polytope with 14 facets, eight of which are regular hexagons and six of which are squares, also known as truncated octahedron. The use of the term {\em Kelvin body} for this cell is motivated by it being Kelvin's proposition for the cell with least surface area (given fixed volume) that can tessellate space \cite{Thomson:1888}.} by a degree-three network, such that $4$ (\cry{4}), $3$ (\cry{3}) and $2$ (\cry{2}) fold rotations of the Kelvin body are maintained. This provides the BCC unit cell of the uncolored {\bf 8-srs} structures, with all components undistinguishable, whose lattice parameter $a$ is half the size of the parameter $a_0$ of the individual {\bf 1-srs} nets. Note also the similar construction of the {\bf 8-srs} as an embedding on a Schwarz periodic Primitive minimal surface in Fig.~12 of ref.~\cite{EvansRobinsHyde:2013a}. In crystallographic notation, the {\bf 8-srs} has a single vertex at $(1/4,1/4,1/4)$ at Wyckhoff site 8c (symmetry $32$) and a single edge with mid-point $(1/4,0,1/2)$ at Wyckhoff position 12d (symmetry $222$), in the body-centered cubic (BCC) space group $I432$ (n$^o$ 211 in \cite{InternationalTablesForCrystallography:1992}) with lattice parameter $a=a_0/2$ \footnote{See also the reticular chemistry structure resource {\tt www.rscr.anu.edu.au} \cite{OKeeffePeskovRamsdenYaghi:2008} for details, where the {\bf 2-srs}, the {\bf 4-srs} and the {\bf 8-srs} are denoted srs-c2*, srs-c4 and srs-c8, respectively.}. The {\bf 8-srs} has four-fold rotation axes {\em and} four-fold screw axes along the [100] lattice directions (see Fig.~\ref{fig:8srs-cross-section}).

\begin{figure}[t]
	\includegraphics[width=\columnwidth]{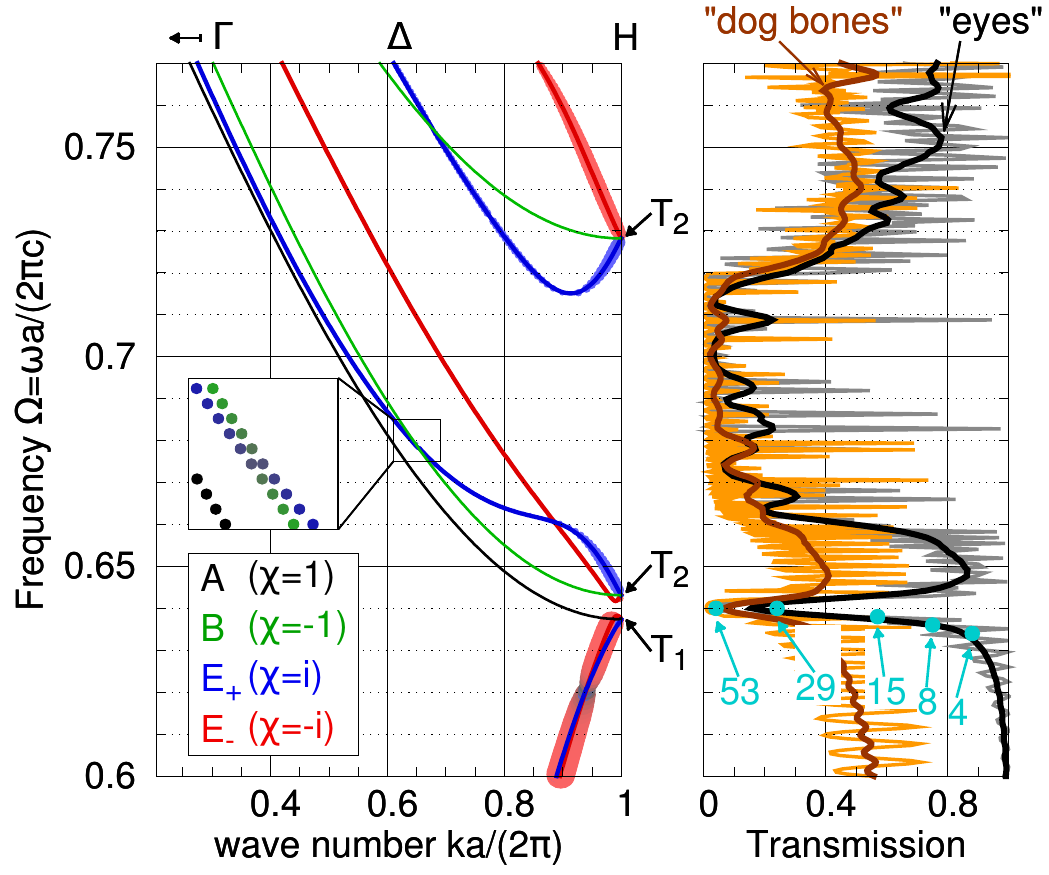}
	\caption{(Color online) Photonic band structure and transmission spectrum of the {\bf 8-srs} PC: (Left) Band structure for $k$ along $\Delta$ (see Fig.~\ref{fig:8srs-cross-section}). The bands are colored according to their symmetry behaviour corresponding to the irreducible representations $i\in\{A,B,E_+,E_-\}$ of the $C_4$ point symmetry (cf.~Tab.~\ref{tab:O-C_4}). The $E_\pm$ bands that are able to couple to plane waves at normal incidence are further underlaid by dots of size proportional to the coupling constant $\beta$. Dots are smaller than the linewidth for $\beta\lessapprox 0.1$ and hence invisible. See discussion in Sec.~\ref{sssec:Delta} on page \pageref{sssec:Delta} for the meaning of the inset. (Right) Transmission of light at normal incidence through a quasi-infinite slab of thickness $N_z=53$, $[100]$ inclination and the two terminations shown in Fig.~\ref{fig:8srs-cross-section}: $t=0$ or the {\it eyes} cut shown in black and $T=0.25a$ or the {\it dog bones} cut shown in brown. The spectrum is the same for any polarization state and is illustrated by the thin lines. The thick and more saturated lines represent a convolution with a Gaussian of width $\delta\Omega = 0.002$ eliminating the sharp Fano and Fabry-P\'erot resonances. The teal points in the transmission spectrum mark the transmission minima at the \emph{pseudo}-bandgap at roughly $\Omega_g=0.64$ for a slab of thickness $N_Z=4, \dots, 53$, respectively.}
        \label{fig:transmission-bandstructure}
\end{figure}

A photonic crystal geometry with finite volume fraction is obtained by inflating all edges of the {\bf 8-srs} structure to solid cylindrical struts with permittivity $\epsilon$, embedded in air or vacuum. For the quantitative results in figures \ref{fig:transmission-bandstructure}\footnote{Transmission simulations are performed with the open source finite difference time domain package MEEP \cite{OskooiRoundyIbanescuBermelJoannopoulosJohnson:2010}. Simulations are for a slab of $53$ unit cells of {\bf 8-srs} structure. Periodic boundary conditions are assumed in all three directions of the simulation box with a Yee grid of $64$ points per unit cell and a size of $1x1x61$ unit cells. A combination of respectively $2$ unit cells vacuo and perfectly matched layers on each side are used with a Gaussian source in vacuum on one side and energy flux measured on the other. Band structure frequencies and eigenmodes are calculated with the open source plane wave frequency domain eigensolver MPB \cite{JohnsonJoannopoulos:2001} with a $128^3$ structural resolution and a $32^3$ Fourier grid in the primitive body centered cubic basis.} and \ref{fig:optical-activity}\footnote{Simulations are performed with the commercial software package CST Microwave Studio. We used the finite element frequency domain solver with periodic boundary conditions in the transverse plane and a $3$ [$4$] unit cells wide slab of the {\bf 8-srs} structure. Perfectly matched layers are used for the $z$ boundaries of the simulation box.}  we assume a value $\epsilon=5.76$, that closely resembles high-refractive index Chalcogenide glass at telecommunication wavelengths \cite{NicolettiZhouJiaVenturaDouglasLutherDaviesGu:2008} or titanium-dioxide at optical wave lengths \cite{MilleTyrodeCorkery:2013}; the solid volume fraction is set to $\phi\approx31\%$, corresponding to a rod diameter of $d\approx0.115$; this value is well below the threshold where the distinct nets overlap to form a single connected component. We consider in particular an infinitely wide slab of size $\infty \times \infty \times n_z a$ ($n_z=4$ in Fig.~\ref{fig:optical-activity} and $n_z=53$ in Fig.~\ref{fig:transmission-bandstructure}) \footnote{The infinite size in $x$ and $y$ direction is achieved by use of periodic boundary conditions with a single unit cell of the {\bf 8-srs} structure}, with the crystallographic lattice vector [100] aligned with the $z$-axis; all analyses are for wave vectors $\vec{k}$ along that axis. The two different terminations \footnote{The termination determines the planes at which the infinite periodic crystal is clipped to give the slab of height $n_z\,a$. The termination is indicated by the vertical clipping position $t$ in crystallographic coordinates, with $t=0$ corresponding to a clipping plane through the point with symmetry $432$ (the origin in the notation of ref.~\cite{InternationalTablesForCrystallography:1992}).} illustrated in figure \ref{fig:8srs-cross-section} for the {\bf 8-srs} crystal are chosen for the transmission simulation in Fig.~\ref{fig:transmission-bandstructure}.

The spontaneous formation of the {\bf 8-srs} geometry by self-assembly appears as an uncertain long-term goal, despite progress in self-assembly of simpler poly-network geometries \cite{HanZhangChngSunZhaoZouYing:2009,HydeDiCampOguey:2009,Kirkensgaard:2012,SchroederTurk:2013}. However, the {\bf 8-srs} structure provides a design pattern suitable for nano-fabrication by direct laser writing technologies. Considering the realisation of {\bf 1-srs} structures in low-dielectric polymer \cite{TurnerSchroederTurkGu:2011,TurnerSabaZhangCummingSchroederTurkGu:2013} and in high-dielectric Chalcogenide glasses \cite{Cumming:2012}, the fabrication of the {\bf 8-srs} material appears feasible.

This article is organised as follows: Section \ref{sec:representation-theory} develops a group theoretical treatment valid for any photonic crystal with symmetry group $I432$ (we henceforth refer to any such crystal as a {\em I432 PC}), including the {\bf 8-srs} geometry as a specific case, that can be readily extended to symmetries $P432$ or $F432$. This theory predicts the topology of the band structure at the $\Gamma$ and $H$ point and identify LCP incident light with modes with irreducible representation $E_-$, and RCP with $E_+$. Numerical data for the band structure of the {\bf 8-srs} (Figs.~\ref{fig:transmission-bandstructure} and \ref{fig:SCvsBCC}) is in perfect agreement with these predictions, provided that the symmetry of the Fourier grid is equally high as the structural symmetry (Fig.~\ref{fig:symm-mismatch}). A gap map providing the width and position of the band gap in [100] direction as a function of dielectric contrast and volume fraction is provided (Fig.~\ref{fig:gapwidth}). Section \ref{sec:scattering-parameter-predictions} uses scattering matrix methods to provide analytical results for the scattering parameters, both transmission and reflectance, that are valid for any structure with $4$-fold symmetry that is inclined normal to its symmetry axis. The predictions that OA and CD are always zero in reflection, and that CD is zero below a threshold frequency in transmission are shown to be correctly reproduced by simulations of the {\bf 8-srs} (Fig.~\ref{fig:optical-activity}). The potential of the {\bf 8-srs} as a design for photonic materials and in particular the magnitude of its optical activity relative to metallic meta-materials, is discussed in the conclusion section.

The paragraphs (a)-(f) that extend over chapter \ref{sec:representation-theory} and \ref{sec:scattering-parameter-predictions} correspond to the statements with the same numbering in \cite{SabaTurnerGuSchroeder_prl:2013}.

%%%%SECTION REPRESENTATION THEORY
\section{\label{sec:representation-theory}Representation Theory}

A photonic band structure (PBS) can be seen as a classification of the eigenmodes of an infinite PC by their transformation behaviour under its translational symmetry operations characterized by the Bloch wave vector $\vec{k}$. This classification yields a deeper understanding of the underlying physics and is also if practical use, the transverse dispersion for example yields a matching condition at interfaces. In this context, $\vec{k}$ acts as a continuous {\em quantum number}.\\
Here, we additionally classify the band structure modes in the crystallographic $[100]$ direction by their transformation behaviour under the PCs point symmetries (rotations, mirrors etc.) and introduce a corresponding second discrete quantum number $i$.

The PBS eigenmodes are shown to belong to several orthogonal classes characterized by their symmetry transformation behaviour. Group theory (or more specifically representation theory \cite{Dresselhaus:2008,Sakoda:2005}) is used to show that there are four such classes with respect to the $4$-fold rotation axis along the $[100]$ direction that can be seen as non-interacting transmission channels. A circularly polarized and normally incident plane wave decomposes into one of the $4$ symmetry classes alone and therefore couples light into the respective transmission channel only.\\
In the following, we use Dirac notation where the basis functions are denoted $|\alpha i\rangle$ and any operator acting on elements of the corresponding Hilbert space is marked with a hat. The basis functions are ortho-normal, see theorem (iv) below. \\
In this notation, a point group is a mathematical (in general non-Abelian) group with (unitary or length preserving) point symmetry operations $\hat{R}$ as its elements and the operator $\cdot$ that is defined by the action onto an arbitrary function $|f\rangle$ via $\left(\hat{R}_1\cdot \hat{R}_2\right)|f\rangle \coloneqq \hat{R}_2\left(\hat{R}_1|f\rangle\right)$.

The operator $\hat{R}$ defines the operation of a point symmetry $R\in S$ where the point group $S$ is a finite subgroup of the orthogonal group $O(3,\mathbb{R})$ for $I432$ and any other symmorphic space group. It can be represented by complex square matrices $\mat{D}(R)$ so that the corresponding map is linear, i.e. $\mat{D}(R_1\cdot R_2) = \mat{D}(R_2)\mat{D}(R_1)$. A matrix representation $\mat{D}(R)$ is called irreducible if no similarity transformation $\mat{D}^{'}(R)=\mat{D}(R_0)\mat{D}(R)\mat{D}^{-1}(R_0)$ with $R_0\in S$ exists that simultaneously transforms all $\mat{D}(R)$ into the same block form, i.e.~the representation cannot be split into representations of lower matrix dimension. Each representation has a set of basis functions $|\alpha\rangle$ for which the symmetry operator can be replaced by the representation matrix $\hat{R}\,|\alpha\rangle = \sum_{\beta} D_{\alpha\beta}(R)|\beta\rangle$; with $\alpha$ and $\beta$ being {\em partners} of $i$ and the corresponding indices denoting the rows and columns of $\mat{D}$. In general, the matrices of an irreducible representation of dimension $> 1$ and the corresponding basis functions are not unique due to similarity transformation gauge freedom. An irreducible representation $i$ is uniquely characterized by the similarity transformation invariant trace of the respective matrix $\chi^{(i)}(R) = \sum_\alpha D_{\alpha\alpha}^{(i)}(R)$ that is also known as the character \footnote{Note, that for compactness of notation we suppress that $\alpha$ itself depends on $i$, i.e.~$\alpha=\alpha_i$.}:
	$$\left(\forall R \in S: \chi^{(i)}(R)=\chi^{(j)}(R) \right)\quad\Leftrightarrow\quad i=j\mbox{.}$$

\subsection{General Theorems}
For our photonic mode analysis below we use four representation theorems that are all implications of the {\it Wonderful Orthogonality Theorem} \cite{Dresselhaus:2008} and the length preserving nature of all point symmetry operations of $I432$:
\begin{enumerate}
\item The eigenfunctions $|n\rangle$ of any operator $\hat{\vartheta}$ that commutes with all operations $\hat{R}$ of a point group $S$ are generally given by the superposition of basis functions $|i\alpha\rangle$ of one irreducible representation $i$ of $S$ only:
	$$\left( \forall \hat{R}\in S: [\hat{\vartheta},\hat{R}]=0\right)$$
	$$\Rightarrow\quad |n\rangle = \sum_\alpha c_\alpha^{(i)}|i\alpha\rangle \eqqcolon |n_i\rangle\mbox{.}$$
\item The characters of an arbitrary representation of a group $S$ can be decomposed into the characters of the irreducible representations $i$ by
	$$\chi(R) = \sum_i d_i\,\chi^{(i)}(R) $$
	$$\mbox{ with}\; d_i = \frac{1}{h}\sum_R \cc{\chi}^{(i)}(R) \chi(R)\mbox{.}$$
 	where $\cc{z}$ denotes the complex conjugation of a complex number $z$ and $h=\sum_R$ the number of symmetry operations in the point group.
\item An arbitrary function $|f\rangle$ can be expressed by the complete set of basis functions $|i\alpha\rangle$ of the irreducible representations $i$:
	$$|f\rangle = \sum_{i,\alpha}f_\alpha^{(i)}|i\alpha\rangle = \sum_{i,\alpha} \hat{P}_\alpha^{(i)}|f\rangle$$
	with the operator $\hat{P}_\alpha^{(i)}$ that projects onto $|i\alpha\rangle$ given by
	$$\hat{P}_\alpha^{(i)} = \frac{l_i}{h} \sum_R \cc{D}_{\alpha\alpha}^{(i)}(R)\hat{R}$$
	where $l_i=\sum_\alpha$ is the dimension of the irreducible representation $i$.
\item The basis functions are orthogonal and can be normalized so that we assume for all representations $i$ and $j$ and partners $\alpha$ and $\beta$:
	$$\langle \alpha i|\beta j\rangle = \delta_{ij}\delta_{\alpha\beta}$$
	
\end{enumerate}

Representation theorem (i) is used to classify the band structure by the symmetry behaviour. We first show that the operator of the magnetic wave equation \footnote{The procedure for the electric field equation is analogous. We note however, that the character of any improper rotation of the $E$-field is the negative of the $H$-field character which can be seen for example by using Faraday's equation. While any $I432$ PC does not exhibit any improper rotation, any dielectric PC has time inversion symmetry that is {\it improper} and hence induces the same phase shift between $H$ and $E$-field.}
$\hat{\vartheta}_{\vec{k}}$ of a PC that has a given point symmetry $S$ commutes with any symmetry element $\hat{R}$ that transforms the wave vector $\vec{k}$ into an equivalent wave vector $\vec{k}+\vec{G}$ that is translated by a reciprocal lattice vector $\vec{G}$ only. The set of all those operations $\hat{R}$ form a subgroup $S_{\vec{k}}\le S$ and is called the group of the wave vector.

\begin{figure}[t]
	\begin{overpic}[width=0.7\columnwidth]{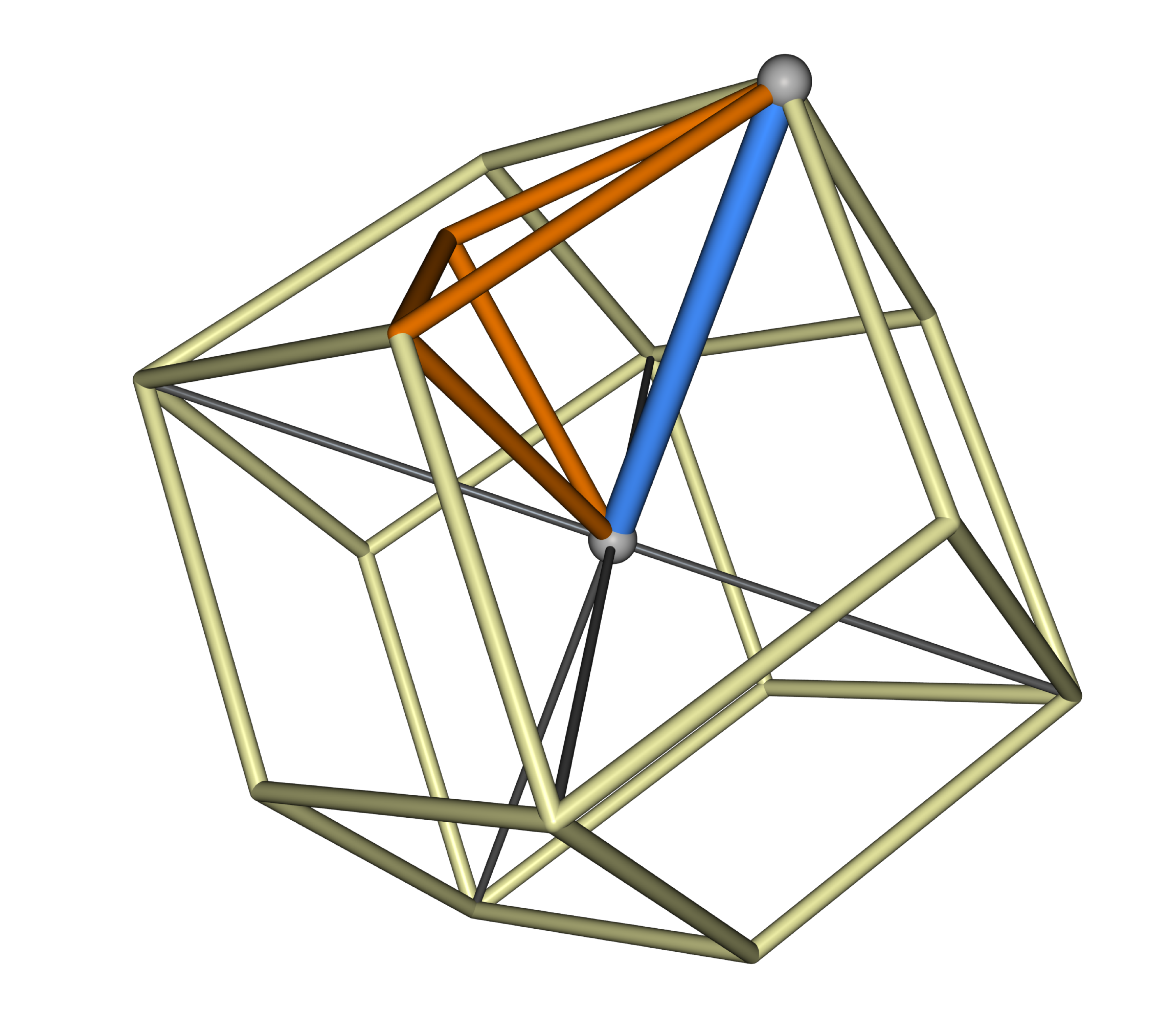}
		\put(70,80){\large $\mathbf{H}$}
		\put(54,33){\large $\mathbf{\Gamma}$}
		\put(60,50){\large $\mathbf{\Delta}$}
	\end{overpic}
	\caption{(Color Online) Brillouin zone of the BCC lattice: A rhombic dodecahedron whose edges are illustrated by yellow bars. The irreducible BZ ($1/48$ of the BZ due to $24$ rotations in $O$ (Tab.~\ref{tab:O-C_4}) and another $24$ time-inverted rotations) is the pyramid framed in orange. The $\Delta$ line marked in blue is one of its edges and connects the $\Gamma$ point in the origin with the $H$ point that is a $4$-connected vertex of the BZ at $2\pi/a\cdot(1,0,0)^T$}
	\label{fig:BZ}
\end{figure}

In our work we consider all modes of a bulk $I432$ PC that can couple to a normally incident plane wave at a $(100)$ interface \footnote{This defines a set of all propagating modes in a semi-infinite structure. However, for any real scattering problem a complete set of PC modes includes evanescent modes. The impact of evanescent waves onto the scattering process becomes dominant if there is no propagating mode present in the transmission channel that has a finite amount of energy in the $00$ Bragg order to which the incident beam couples only. This makes it for example impossible to get any useful information about transmission through a finite slab in the frequency range between $\Omega\coloneqq\omega a/(2\pi c)=0.66$ and $0.71$ from the band structure in figure \ref{fig:transmission-bandstructure} alone.}.
We choose a symmetric set of basis vectors $\vec{a}_1=(-d,d,d);\,\vec{a}_2=(d,-d,d);\,\vec{a}_3=(d,d,-d)$, with $d=a/2$, for the body centered cubic translational symmetry. The Brillouin zone of the BCC lattice illustrated in figure \ref{fig:BZ} and the reciprocal lattice vectors are given by $\vec{b}_1=(0,b,b);\,\vec{b}_2=(b,0,b);\,\vec{b}_3=(b,b,0)$, with $b=2\pi/a$. All coupling modes \footnote{Conservation of the spatio-temporal modulation at the surface implies conservation of frequency and parallel part of the wave vector where the PC acts as a grating and back scattering can occur into several Bragg orders, i.e.~back scattered waves are given by a plane wave Floquet basis.}
lie on a straight line within the Brillouin zone parameterized by $\vec{k}_s=s*(\vec{b}_2+\vec{b}_3-\vec{b}_1),\;s\in(-0.5,0.5]$. The points at $s=0$ and $s=0.5$ are denoted by $\Gamma$ and $H$, respectively. At both points, the group of the wave vector is equal to the full point group that together with the BCC translation gives the full $I432$ space group: That is the $O$ (Schoenfliess) or $432$ (Hermann-Maugin) point group that includes the point symmetries of a cube, i.e.~$6$ $4$-fold ($C_4$) and $3$ $2$-fold ($C_2$) rotations around the $[100]$ axes, $8$ $3$-fold ($C_3$) rotations around the $[111]$ axes and $6$ $2$-fold ($C_{2}^{'}$) rotations around the $[110]$ axes (cf.~Tab.~\ref{tab:O-C_4}). On the $\Delta$ line that connects $\Gamma$ and $H$ ($s\in(0,0.5)$), the (Abelian) group of the wave vector is $C_4$ that only contains the $3$ rotations around the $[100]$ axis where positive ($C_{4_1}$) and negative ($C_{4_3}$) $4$-fold rotations fall into distinct classes \footnote{Technically, a class is a subset $S_n$ of symmetry elements that is obtained by one selected member and all elements that are conjugate. As a result, for most purposes all elements of a class can be treated equally, i.e.~they have same character etc.}.
\begin{table}[t]
	\begin{minipage}{0.52\columnwidth}
        \begin{ruledtabular}
		\begin{tabular}{c|rrrrr}
			$O$	&	$\mathbb{1}$	&	$6C_4$	&	$3C_2$	&	$8C_3$	&	$6C_2^{'}$	\\
			\hline
			$A_1$	&	$1$		&	$1$	&	$1$	&	$1$	&	$1$		\\
			$A_2$	&	$1$		&	$-1$	&	$1$	&	$1$	&	$-1$		\\
			$E$	&	$2$		&	$0$	&	$2$	&	$-1$	&	$0$		\\
			$T_1$	&	$3$		&	$1$	&	$-1$	&	$0$	&	$-1$		\\
			$T_2$	&	$3$		&	$-1$	&	$-1$	&	$0$	&	$1$		\\
		\end{tabular}
        \end{ruledtabular}
	\end{minipage}\hfill
	\begin{minipage}{0.46\columnwidth}
        \begin{ruledtabular}
		\begin{tabular}{c|rrrrr}
			$C_4$	&	$\mathbb{1}$	&	$C_{4_1}$	&	$C_2$	&	$C_{4_3}$	&	TR	\\
			\hline
			$A$	&	$1$		&	$1$		&	$1$	&	$1$		&	(a)	\\
			$B$	&	$1$		&	$-1$		&	$1$	&	$-1$		&	(a)	\\
			$E_+$	&	$1$		&	$i$		&	$-1$	&	$-i$		&	(b)	\\
			$E_-$	&	$1$		&	$-i$		&	$-1$	&	$i$		&	(b)	\\
		\end{tabular}
        \end{ruledtabular}
        \vspace{0.45cm}
	\end{minipage}
	\caption{Character tables for the $O$ and $C_4$ point groups relevant for the $H$ ($\Gamma$) point and $\Delta$ line (Fig.~\ref{fig:BZ}), respectively. The time-reversal symmetry type TR of the irreducible representations of $C_4$ are added in the last column.}
	\label{tab:O-C_4}
\end{table}

The magnetic wave equation is given by
\begin{align*}
	\vartheta_{\vec{k}}\vec{u}_{\vec{k}n}(\vec{r}) &\coloneqq (\imath\vec{k}+\vec{\nabla})\times\left[\frac{1}{\varepsilon(\vec{r})}(\imath\vec{k}+\vec{\nabla})\times\vec{u}_{\vec{k}n}(\vec{r})\right]\\
	&= \frac{\omega_{\vec{k}n}^2}{c^2}\vec{u}_{\vec{k}n}(\vec{r})
\end{align*}
with the periodic part of the Bloch field $\vec{u}_{\vec{k}n}(\vec{r})$, the periodic dielectric function $\varepsilon(\vec{r})$, the eigenfrequency $\omega_{\vec{k}n}$, the imaginary number $\imath\coloneqq\sqrt{-1}$ and the velocity of light $c$.\\
We define the action of the operator $\hat{R}$ on a vector field by $\hat{R}\vec{F}(\vec{r})=\mat{R}\,\vec{F}(\mat{R}^{-1}\vec{r})$. The matrices $\mat{R}$ are here the common $3\times 3$ matrices that transform three-dimensional vectors according to a point symmetry $R$, for example given by Rodriguez formula in case of a rotation etc. In the $432$ group these matrices are a suitable choice for the irreducible representation $T_1$ (Tab.~\ref{tab:O-C_4}). The skew can be understood in terms of an active transformation acting on the vector field itself whereas the change of its position is achieved by a passive transformation that changes space itself. With that definition it is trivial that $\hat{R}\frac{1}{\varepsilon(\vec{r})}\vec{F}(\vec{r})=\frac{1}{\varepsilon(\vec{r})}\hat{R}\vec{F}(\vec{r})$ for any dielectric function that is invariant under $\hat{R}$ so that we are left to show that $\hat{R}$ commutes with the cross products in the wave equation. The proof is done in a Cartesian basis with Einstein convention and the abbreviation $\vec{r}^{'}=\mat{R}^{-1}\vec{r}$ \footnote{An alternative proof is provided in \cite{Sakoda:2005}.}:
	\begin{align*}
		&\left[(\imath\vec{k}+\vec{\nabla})\times\hat{R}\vec{u}(\vec{r})\right]_i
			= \varepsilon_{ijk} (\imath k_j + \frac{\partial}{\partial x_j})R_{kl}u_l(\vec{r}')\\
		&= \varepsilon_{ijk} R_{kl} \left(\imath k_j + 
			\underbrace{\frac{\partial x_m^{'}}{\partial x_j}}_{\overset{(a)}{=}R_{jm}} \frac{\partial }{\partial x_m^{'}}\right) u_l(\vec{r}')\\
		&\overset{(a)}{=} \varepsilon_{ojk}R_{op}R_{ip}R_{kl}R_{jm}\left(\imath R_{qm} k_q + \frac{\partial }{\partial x_m^{'}}\right)u_l(\vec{r}')\\
		&\overset{(b)}{=}\pm R_{ip}\varepsilon_{pml}\left(\imath (k_m-G_m) + \frac{\partial }{\partial x_m^{'}}\right)u_l(\vec{r}')\\
		&=\pm \left[\,\mat{R}\{ (\imath (\vec{k}-\vec{G}) + \vec{\nabla})\times\vec{u} \}\right]_i (\vec{r}')\\
		&=\pm \left[\hat{R} \{(\imath (\vec{k}-\vec{G}) + \vec{\nabla})\times\vec{u}(\vec{r})\}\right]_i
	\end{align*} 
Here we make use of the facts that (a) point symmetries are length preserving and hence the matrix representation is orthogonal, i.e.~$R_{ij}R_{kj}=\delta_{ik}$ with the Kronecker symbol $\delta_{ij}$, and that (b) the Levi-Civita symbol $\varepsilon_{ijk}$ is invariant under proper ($\det(\mat{R})=1$) rotations $\hat{R}^{(+)}$ and changes its sign under improper ($\det(\mat{R})=-1$) rotations $\hat{R}^{(-)}$, i.e.~$R^{(\pm)}_{il}R^{(\pm)}_{jm}R^{(\pm)}_{kn}\varepsilon_{lmn}=\pm\varepsilon_{ijk}$. The identity in (b) is evident through expansion of the Levi-Civita tensor into a triple product of Cartesian basis vectors (so that $\mat{R}$ takes a basic form) and making use of (a).\\
The change of sign for improper rotations cancels out by the double cross product in the $\hat{\vartheta}_{\vec{k}}$ so that:
	\begin{align*}
		&\hat{\vartheta}_{\vec{k}}\hat{R}\,\vec{u}_{\vec{k},n}(\vec{r})
			=(\imath\vec{k}+\vec{\nabla})\times\left[\frac{1}{\varepsilon(\vec{r})}(\imath\vec{k}+\vec{\nabla})\times
			\hat{R}\,\vec{u}_{\vec{k},n}(\vec{r})\right]\\
		&=\hat{R}\left\{(\imath\vec{k}'+\vec{\nabla})\times\left[\frac{1}{\varepsilon(\vec{r})}(\imath\vec{k}'+\vec{\nabla})
			\times\vec{u}_{\vec{k}',n}(\vec{r})\right]\right\}\\
		&=\hat{R}\,\frac{\omega_{\vec{k}',n}^2}{c^2}\vec{u}_{\vec{k}',n}(\vec{r})=\hat{R}\,\frac{\omega_{\vec{k},n}^2}{c^2}\vec{u}_{\vec{k},n}(\vec{r})
			= \hat{R}\,\hat{\vartheta}_{\vec{k}}\,\vec{u}_{\vec{k},n}(\vec{r})
	\end{align*}
where we substitute $\vec{k}'=\vec{k}-\vec{G}$ and use the invariance of the periodic Bloch function $\vec{u}_{\vec{k}+\vec{G}}=\vec{u}_{\vec{k}}$ and the eigenfrequencies $\omega_{\vec{k},n} = \omega_{(\vec{k}+\vec{G}),n}$.

\subsection{Topology and Symmetry Classification of the Bandstructure}

While the exact shape of the PBS strongly depends on geometry and dielectric contrast, the topology, i.e.~degeneracies and hence also band connections at high symmetry points, is generally induced by symmetry alone. This fact combined with symmetry based selection rules is what makes group theory a powerful tool. Here, we derive the topology of the band structure of an $I432$ PC and show that, for the {\bf 8-srs} PC, these results are in perfect agreement with numerical results.
\subsubsection*{(a) Degeneracies at the lowest $\Gamma$ and $H$ points}

To derive the irreducible representation of the eigenmodes at the high symmetry points in the BZ, it is suitable to group the plane wave components $\vec{u}_{\vec{G}\sigma}(\vec{r}) = \vec{u}_{\vec{G}\sigma}e^{\imath (\vec{k}+\vec{G})\cdot\vec{r}}$ of a Bloch mode $\vec{u}_{\vec{k}n_i}(\vec{r}) = \sum_{\vec{G}\sigma}\vec{u}_{\vec{G}\sigma}(\vec{r})$ with $\sigma\in{u,v}$ denoting a linear polarization basis by the length of $\vec{k}+\vec{G}$, i.e.~its vacuum frequency, in equivalence classes $\left[\vec{k}+\vec{G}\right] \coloneqq \left\{\vec{u}_{\vec{G}\sigma}{\vec{G'}}\,:\,|\vec{k}+\vec{G}|=|\vec{k}+\vec{G}'|\right\}$. We sort the equivalence classes at the high symmetry points by ascending vacuum frequency and denote the corresponding class by $\Gamma^{(n)}$ and $H^{(n)}$ ($n\in\mathbb{N}_0$), respectively. All elements within each equivalence class form a basis for a (generally reducible) representation of the group of the wave vector $\vec{k}$ by its definition. This representation can then be reduced into irreducible representations of the group of the wave vector with representation theorem (ii).
\begin{table}[t]
	\renewcommand{\arraystretch}{1.8}
        \begin{ruledtabular}
	\begin{tabular}{c|ccc}
		$\left[\vec{k}\right]$	&	$\Omega_0$	&	$\vec{k}$	&	$\mathcal{R}=$		\\
		\hline
		$\Gamma^{(0)}$		&	$0$		&	$(0,0,0)^T$	&	$T_1^{\dagger}$		\\
		\hline
		$H^{(0)}$		&	$1$		&	$(b,0,0)^T$	&	$2T_1+2T_2$		\\
		\hline
		$\Gamma^{(1)}$		&	$\sqrt{2}$	&	$(b,b,0)^T$	&	$2A_2+2E+4T_1+2T_2$	\\
		\hline
		$H^{(1)}$		&	$\sqrt{3}$	&	$(b,b,b)^T$	&	$2E+2T_1+2T_2$
	\end{tabular}
        \end{ruledtabular}
	\caption{Summary of the results of section B.{\it (a)}. The vacuum frequency $\Omega_0$, a representative wave vector and the compatibility relation are respectively listed for the $4$ lowest points where the group of the wave vector exhibits $432$ symmetry. We use a dagger ($\dagger$) at the $T_1$ representation for $\Gamma^{(0)}$ to indicate that the $T_1$ representation does not split into $A + E_+ + E_-$ but only into $E_+ + E_-$. The static $A$ mode exists at $\Gamma^{(0)}$ itself to give a full basis for a homogeneous 3D vector field. However, it violates the divergence theorem for $k\gtrapprox0$.}
	\label{tab:Gamma-H}
\end{table}

We first consider the $H^{(0)}$ point with $|\vec{k}+\vec{G}|=b$. There are $6$ points in the group of the wave vector with $\vec{k}=\pm b \vec{e}_i$ ($i\in{x,y,z}$). Hence, there are $12$ plane waves that form a closed set of basis functions for a reducible representation $\mathcal{R}\left(H^{(0)}\right)$ within the $432$ point group. The character of this reducible representation is determined by the corresponding representation matrix of a point symmetry $R$. First, we need to know how many of these points are unchanged under $R$ as we only count diagonal elements to obtain the trace, i.e. equivalent to the number of points on the axis (or plane) of symmetry. Then we multiply by the character of the point symmetry acting on a two-dimensional vector in the plane perpendicular to the axis of symmetry which is the contribution of the polarization mode pair on each point: The character of the identity transformation $\mathbb{1}$ is $2$, of the $C_4$ rotation $0$ and of the $C_2$ rotation $-2$. All other operations change all $\vec{k}$ on the $H_n$ or $\Gamma_n$ points into $\vec{k}+\vec{G}$ with $\vec{G}\ne0$ and hence cannot contribute to the trace. The described procedure yields the following characters:
\begin{center}
\begin{ruledtabular}
\begin{tabular}{c|rrrrr}
	$R$					&	$\mathbb{1}$	&	$6C_4$	&	$3C_2$	&	$8C_3$	&	$6C_2^{'}$	\\
	\hline
	$\chi_{\mathcal{R}(H^{(0)})}(R)$	&	$12$		&	$0$	&	$-12$	&	$0$	&	$0$		\\
\end{tabular}
\end{ruledtabular}
\end{center}
Note, that the number $-12$ for the $C_2$ rotation already includes the number of elements in that class, i.e.~$3$. Application of representation theorem (ii) using the characters in Tab.~\ref{tab:O-C_4} we obtain the {\it compatibility relation} $\mathcal{R}=2 T_1 + 2 T_2$ that tells us that the $12$ dimensional representation of the plane waves at the high symmetry points is reduced into respectively $2$ three-dimensional representations $T_1$ and $T_2$ that are irreducible within $432$. This result is also numerically obtained and illustrated in figure \ref{fig:transmission-bandstructure}.
\begin{table}[t]

\begin{ruledtabular}
\begin{tabular}{c|ccccc}
$\mathcal{R}(O)$   & $A_1$ & $A_2$ & $E$   & $T_1$        & $T_2$\\ \hline
$\mathcal{R}(C_4)$ & $A$   & $B$   & $A+B$ & $A+E_++E_-$  & $B+E_++E_-$\\
\end{tabular}
\end{ruledtabular}

%	\begin{tabular}{c|c}
%		$\mathcal{R}(O)$	&	$\mathcal{R}(C_4)$	\\	
%		\hline
%		$A_1$			&	$A$			\\
%		$A_2$			&	$B$			\\
%		$E$			&	$A+B$			\\
%		$T_1$			&	$A+E_++E_-$		\\
%		$T_2$			&	$B+E_++E_-$		\\\hline
%	\end{tabular}
	\caption{Compatibility relations obtained by reduction of the irreducible representations of $O$ in $C_4$.}
	\label{tab:Delta}
\end{table}

Analogously, the irreducible representations for all higher order $\Gamma^{(n)}$ and $H^{(n)}$ points can be obtained. For example, $\Gamma^{(1)}=\left[(b,b,0)^T]\right]$ has $24$ elements yielding the reducible representation with the following characters
\begin{center}
\begin{ruledtabular}
\begin{tabular}{c|rrrrr}
	$R$					&	$\mathbb{1}$	&	$6C_4$	&	$3C_2$	&	$8C_3$	&	$6C_2^{'}$	\\
	\hline
	$\chi_{\mathcal{R}(\Gamma^{(1)})}(R)$	&	$24$		&	$0$	&	$0$	&	$0$	&	$-24$		\\
\end{tabular}
\end{ruledtabular}
\end{center}
and the compatibility relation $\mathcal{R}(\Gamma^{(1)})=2A_2+2E+4T_1+2T_2$. $H^{(1)}=\left[(b,b,b)^T\right]$ has $16$ elements on a $C_3$ axis. The compatibility relation is $\mathcal{R}(H^{(1)})=2E+2T_1+2T_2$. These results are summarized in table \ref{tab:Gamma-H}.
\begin{figure}[t]
	\begin{minipage}{0.49\columnwidth}
		\begin{overpic}[width=\textwidth]{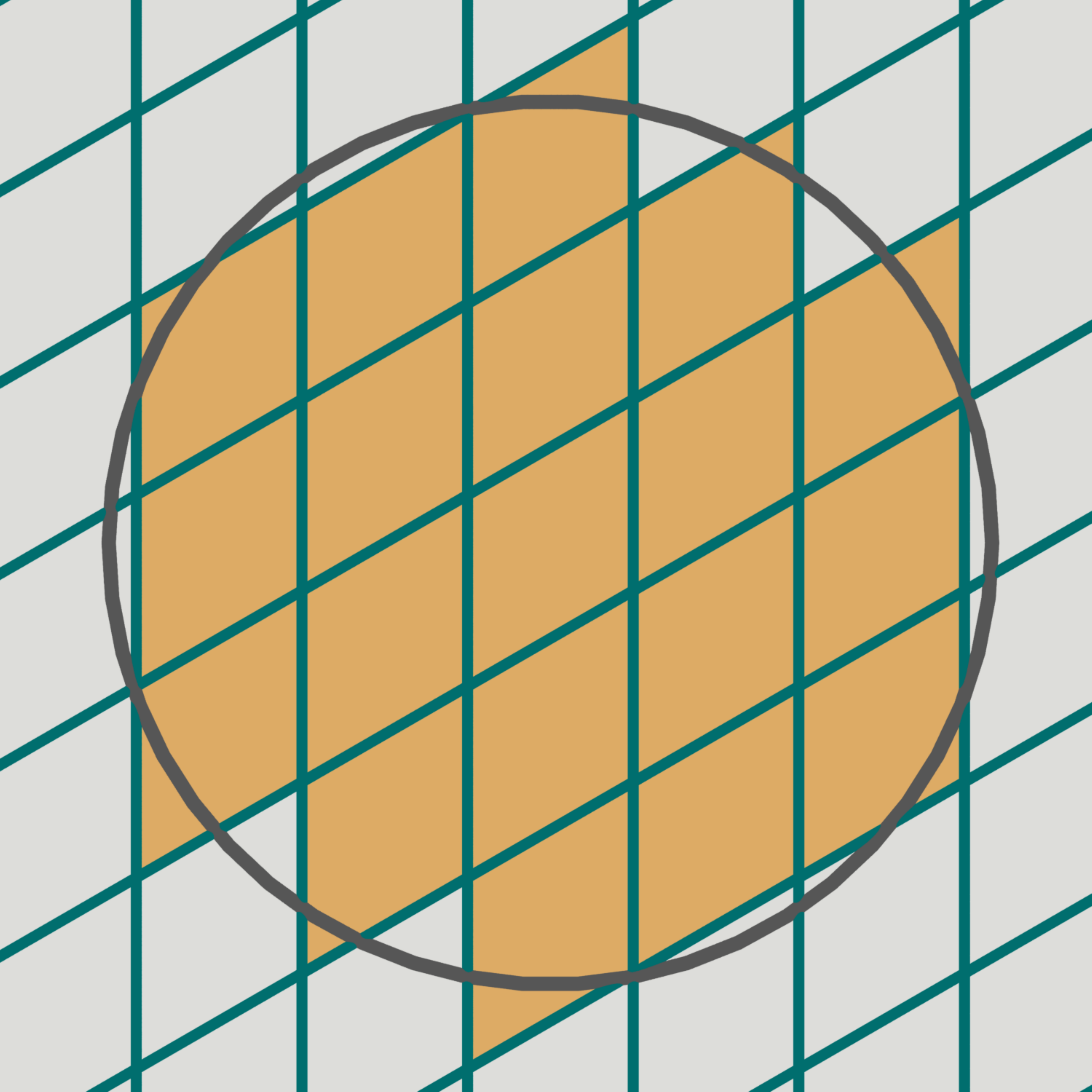}
			\put(0,86){\includegraphics[width=0.16\textwidth,height=0.2\textwidth]{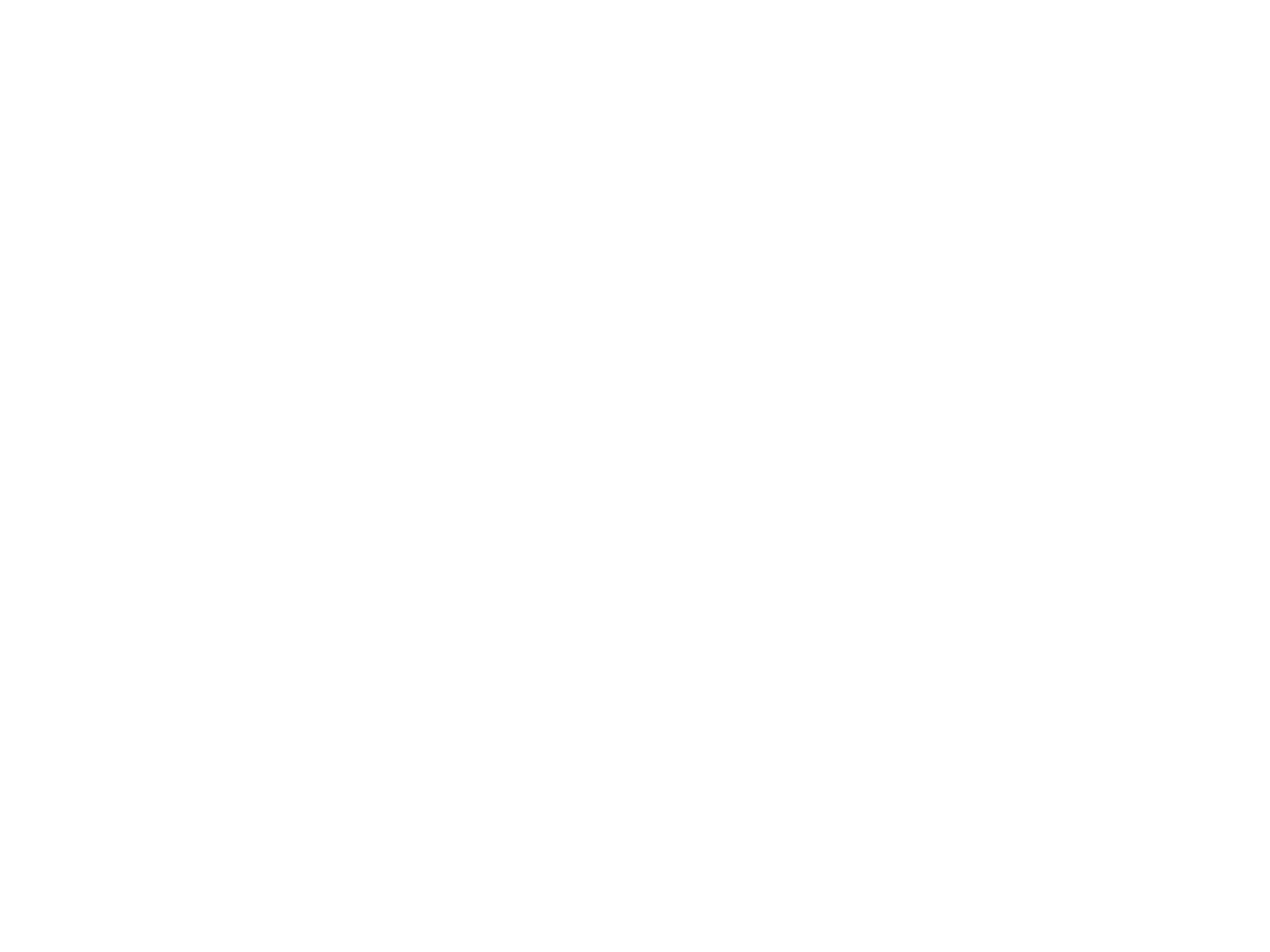}}
			\put(0,90){\large \bf (a)}
		\end{overpic}
	\end{minipage}\hfill
	\begin{minipage}{0.49\columnwidth}
		\begin{overpic}[width=\textwidth]{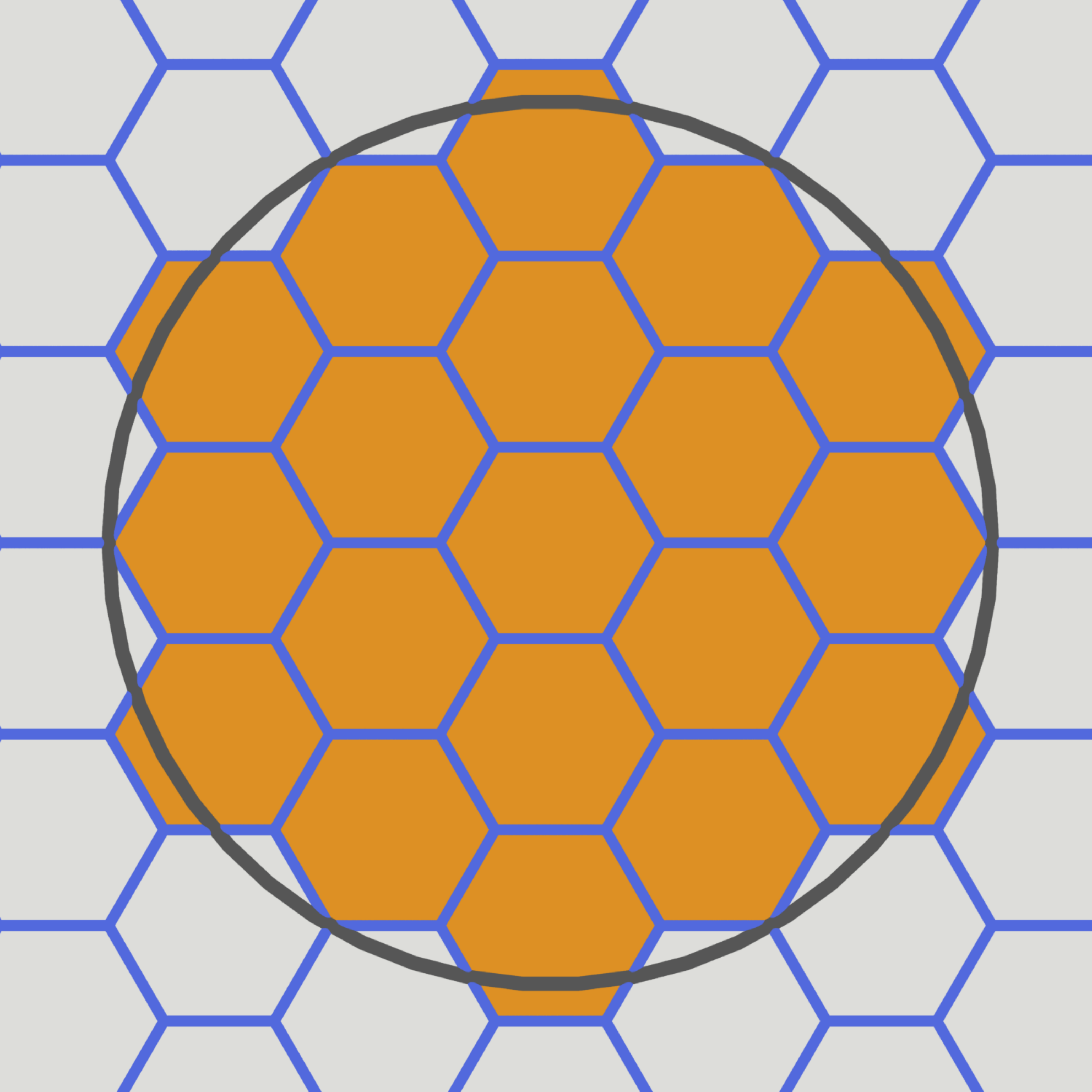}
			\put(0,86){\includegraphics[width=0.17\textwidth,height=0.2\textwidth]{white_bg.pdf}}
			\put(0,90){\large \bf (b)}
		\end{overpic}
	\end{minipage}
	\begin{overpic}[width=\columnwidth]{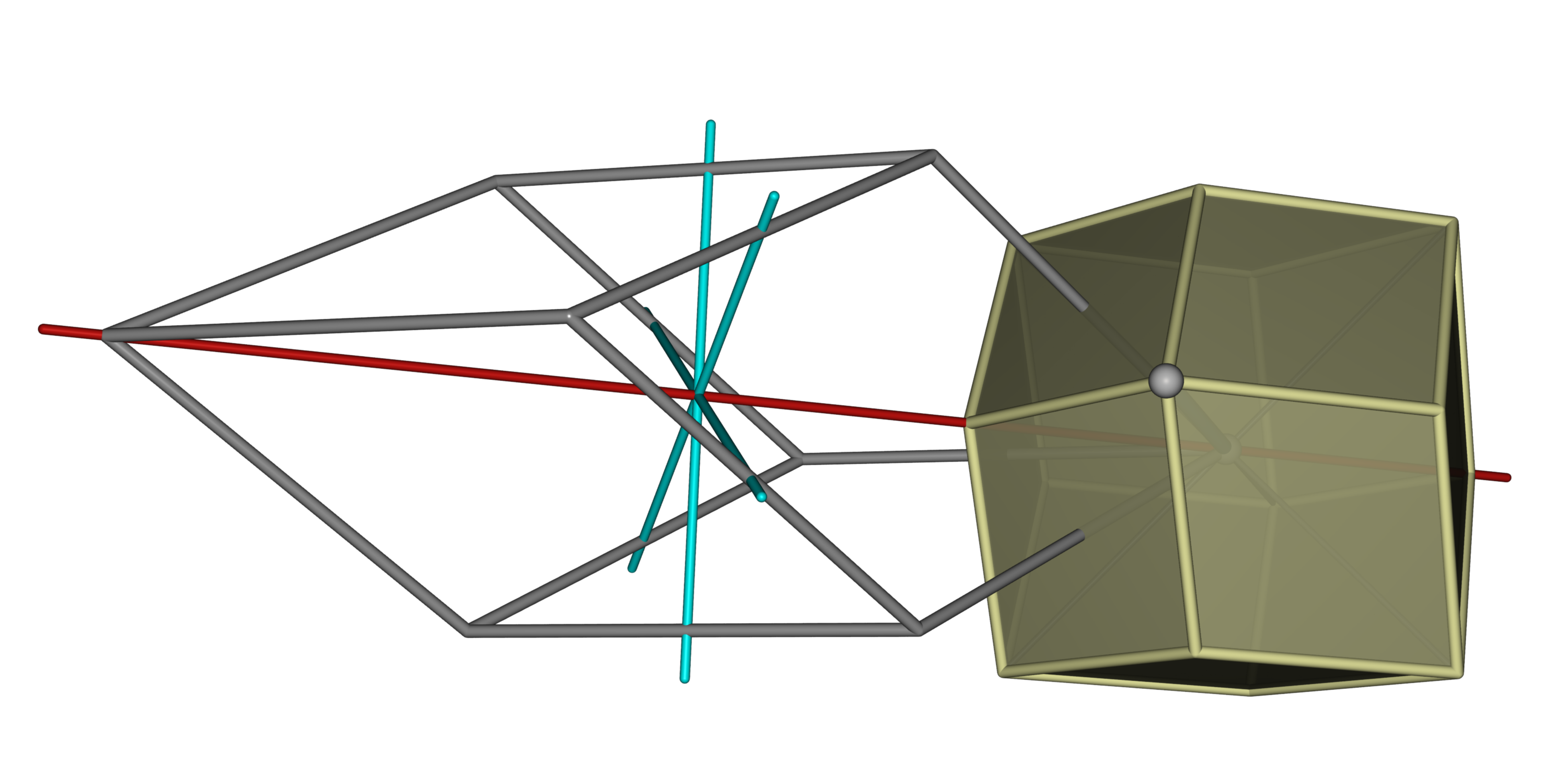}
		\put(3,39){\large \bf (c)}
	\end{overpic}
	\caption{(Color Online) Incompatibility of planar hexagonal and spatial BCC symmetry with the discrete Fourier grid used in PBS plane-wave based frequency domain eigensolvers. For any planar object with hexagonal symmetry, a discretized representation using a Fourier grid with two linearly-independent basis vectors cannot maintain the full hexagonal symmetry, as (a) illustrates for a circle; a discretization by smaller Wigner-Seitz cells, as shown in (b) would alleviate this problem but cannot be implemented for the Fourier analysis. (c) The BCC case represents a 3D analogon, with the same problem. The Brillouin zone (solid cell) is a rhombic dodecahedron with full $432$ symmetry, yet the parallelepiped formed by the basis vectors of the Fourier grid representation (hollow cell) only maintains $D_3$ symmetry with a single $3$-fold axis (red) and three $2$-fold axes (cyan).}
	\label{fig:symm-mismatch}
\end{figure}

\subsubsection*{(b) Degeneracy fully lifted on the $\Delta$ line}
\label{sssec:Delta}
Instead of also reducing an equivalent representation to $\mathcal{R}$ on the respective $\Delta$ line we reduce the irreducible representations of $432$ in the $C_4$ point group to learn how the degeneracies are lifted when going away from the high symmetry points onto $\Delta$. The characters are:
\begin{center}
\begin{ruledtabular}
\begin{tabular}{c|rrrrr}
	$R$				&	$\mathbb{1}$	&	$C_{4_1}$	&	$C_2$	&	$C_{4_3}$	&	\\
	\hline
	$\chi_{A_1}(R)$			&	$1$		&	$1$		&	$1$	&	$1$		&	\\
	$\chi_{A_2}(R)$			&	$1$		&	$-1$		&	$1$	&	$-1$		&	\\
	$\chi_{E}(R)$			&	$2$		&	$0$		&	$2$	&	$0$		&	\\
	$\chi_{T_1}(R)$			&	$3$		&	$1$		&	$-1$	&	$1$		&	\\
	$\chi_{T_2}(R)$			&	$3$		&	$-1$		&	$-1$	&	$-1$		&	\\
\end{tabular}
\end{ruledtabular}
\end{center}
The same reduction procedure as above yields the compatibility relations listed in table \ref{tab:Delta}. The degeneracy is in each case completely lifted. 

For the {\bf 8-srs} PC, the modes perfectly match in the predicted manner in numerical calculations if the Fourier lattice maintains the full $432$ symmetry (figures \ref{fig:SCvsBCC} and \ref{fig:symm-mismatch}).

The eigenmodes of the band structure in Fig.~\ref{fig:transmission-bandstructure} that are coloured by their respective $C_4$ irreducible representation show the predicted behaviour. Each mode representation is obtained numerically by projecting the corresponding normalized magnetic field $|\vec{H}\rangle$ onto the respective basis function with representation theorem (iii). The representation is determined with the norm $N_i = \sum_{\alpha}\left\langle\hat{P}_\alpha^{(i)} \vec{H}|\hat{P}_\alpha^{(i)} \vec{H}\right\rangle$\footnote{The sum over $\alpha$ is superstitious in this special case as all $C_4$ representations are one-dimensional.} of each projection onto one of the $4$ one-dimensional irreducible representations. Up to numerical accuracy, the projection onto the true irreducible representation yields $N_i=1$ while all other projections vanish if a simple cubic lattice with full $O$ symmetry is used for the MPB calculations of the PBS, see Fig.~\ref{fig:SCvsBCC}. The representation algorithm for the band structure in Fig.~\ref{fig:transmission-bandstructure} on the other hand only yields good results if the modes are separated in frequency, i.e.~away from band crossings and high symmetry points. This error is caused by the fact that a BCC Fourier grid by definition is incompatible with the full $432$ symmetry (illustrated in Fig.~\ref{fig:symm-mismatch}) and only has the $D_3$ point symmetry of the parallelepiped spanned by the three (reciprocal) lattice vectors.

The Maxwell version of the Helmann-Feynman theorem yields directly that the band function $\omega_{ni}(\vec{k})$ is continuously differentiable and has a slope given by the average energy flow of the corresponding eigenmode which can be shown to be always less than the velocity of light $c$\footnote{For any dielectric PC \cite{JoannopoulosJohnsonWinnMeade:2008}}. Orthogonality of the irreducible representations further implies that the function $\vec{H}_{ni\vec{k}}(\vec{r})$ does not change its representation $i$ if the $\vec{k}$ vector does not change its point symmetry behaviour under the transformation $\vec{k}\rightarrow\vec{k}+\delta\vec{k}$. The argument also holds at accidental degeneracies where the different representations are still orthogonal and usually even at symmetry triggered degeneracies \footnote{This claim can be proved with degenerate perturbation theory. The irreducible representation of the point group of the wave vector is reduced in the lower symmetry group of $\delta k$ and the matrix elements of the perturbation operator $\hat{\vartheta}_{\delta k}$ in this basis are diagonalized (similar to the $k\cdot p$ analog in quantum mechanics). As all matrix elements satisfy the symmetry relation $<\hat{\vartheta}_{-\delta k}>=-<\hat{\vartheta}_{\delta k}>$, there is always a band with continuous slope passing through the point without changing its representation. The $\Gamma_0$ point where $E$ and $H$ are decoupled and hence the group velocity is zero whereas finite in the vicinity of $\Gamma$ is an exception.}. Bands can therefore either cross ($A$ and $B$ band in the lower half of Fig.~\ref{fig:SCvsBCC} "ii, SC") or {\em anti-cross} (two red bands in Fig.~\ref{fig:SCvsBCC} "i, SC") \footnote{The term {\em anti-crossing} is taken from reference \onlinecite{Dresselhaus:2008} and is used to characterize two bands that come close to each other and seem to interchange group velocities from left to right without actually touching one another.}.

A zoom into the band structure in Fig.~\ref{fig:transmission-bandstructure} reveals (a) that bands do not cross on $\Delta$ but are anti-crossing correlated with an interchange of characters which is visible in the inset in Fig.~\ref{fig:transmission-bandstructure} where the band structure around the point 0.65,0.68 is magnified\footnote{The points are all equal sized in the inset and colored by the color of $i$ with dominating $N_i$ with intensity maximum at $N_i=1$ and continuously decreasing to light grey at $N_i=0.25$.} and (b) that the $3$-fold degeneracies at the high symmetry points are slightly lifted (Fig.~\ref{fig:SCvsBCC}). Both phenomena can be explained with a degenerate perturbation theory model that in case (a) includes only two modes of different representation. The diagonal matrix is perturbed by numerical breaking of the $4$-fold symmetry. Case (a) can be treated analytically with the help of direct product selection rules. The perturbation matrix is orthogonal to the $A$ representation and hence the perturbation matrix does not have diagonal elements. It is however still hermitian so that diagonalization results in a repulsive {\it force} between two closely spaced bands. Any crossing becomes an anti-crossing. The modes mix in equal proportions (and with a phase factor given by the argument of the secondary diagonal entries of the perturbation matrix) at the point of degenerate frequency without perturbation. The modes regain their original character when leaving the degeneracy which can be observed in the inset of Fig.~\ref{fig:transmission-bandstructure} at coordinates $(0.65,0.68)$ where the color is grey at the crossing points and changes to the original $E_-$ ($B$) color when leaving the degeneracy and also in the fundamental bands at $(0.95,0.62)$\footnote{The color code of the $E_\pm$ modes in the main plot is in the common hexadecimal rgb scheme defined by $RGB=C\floor{255(N_{E_+}-N_{E_-})}$ where $C=256^2$ if $N_{E_+}>N_{E_-}$ and $C=1$ else.}.

With representation theorem (iii) we can further explain, why in case (b) the $3$-fold degenerate modes $T_1$ and $T_2$ split into a $2$-fold and a non-degenerate state as illustrated in Fig.~\ref{fig:SCvsBCC} iiz and why the $E$ mode does not split up under the symmetry break. The character table for the $D_3$ point group (cf.~Fig.~\ref{fig:symm-mismatch}) is
\begin{center}
\begin{ruledtabular}
\begin{tabular}{c|rrrrr}
	$D_3$				&	$\mathbb{1}$	&	$2C_3$		&	$3C_2'$	\\
	\hline
	$\chi_{A_1}(R)$			&	$1$		&	$1$		&	$1$	\\
	$\chi_{A_2}(R)$			&	$1$		&	$1$		&	$-1$	\\
	$\chi_{E}(R)$			&	$2$		&	$-1$		&	$0$	\\
\end{tabular}
\end{ruledtabular}
\end{center}
so that we derive:
\begin{center}
\begin{ruledtabular}
\begin{tabular}{c|ccccc}
$\mathcal{R}(O)$   & $A_1$	& $A_2$ 	& $E$	& $T_1$		& $T_2$		\\
\hline
$\mathcal{R}(D_3)$ & $A_1$	& $A_2$		& $E$	& $A_2+E$	& $A_1+E$	\\
\end{tabular}
\end{ruledtabular}
\end{center}
Note that the reduction of the $T_1$ representation in $C_4$ includes a trivial behaviour $A$ whereas the reduction of $T_2$ includes a trivial behaviour $A_1$ in $D_3$. The black and blue colors of the bands in Fig.~\ref{fig:SCvsBCC} "iiz, BCC" are not unique, indicating that the irreducible representation of $C_4$ is not uniquely determined due to the fact that no symmetry except the identity transformation of $C_4$ is a symmetry of the discretization and hence the cannot be perfectly met by the numerics.

All modes have the same irreducible representation in the trivial group of the wave vector $S_{\vec{k}}=\mathbb{1}$ on $\Delta$. Modes of same irreducible representation cannot cross each other, they are exposed to a "fermionic" repulsion\footnote{Electromagnetic eigenmodes (unlike photons in a particle picture) behave like Fermions: There can only be a single mode in a state classified by all symmetries (or quantum numbers) because the two modes are uniquely determined except for a scalar coefficient and hence interchangeable. The symmetries in Maxwell theory are translation invariance in time and space ($\omega$ and $k$), time inversion ($k\leftrightarrow -k$) and point symmetries (character).}. That is the reason why the upper two bands in Fig.~\ref{fig:SCvsBCC} "ii, BCC" cannot cross each other. Fig.~\ref{fig:SCvsBCC} "i, SC" provides another palpable example for a symmetry induced "fermionic" anti-crossing.
\begin{figure}[t]
  \centering
	\includegraphics[width=0.75\columnwidth]{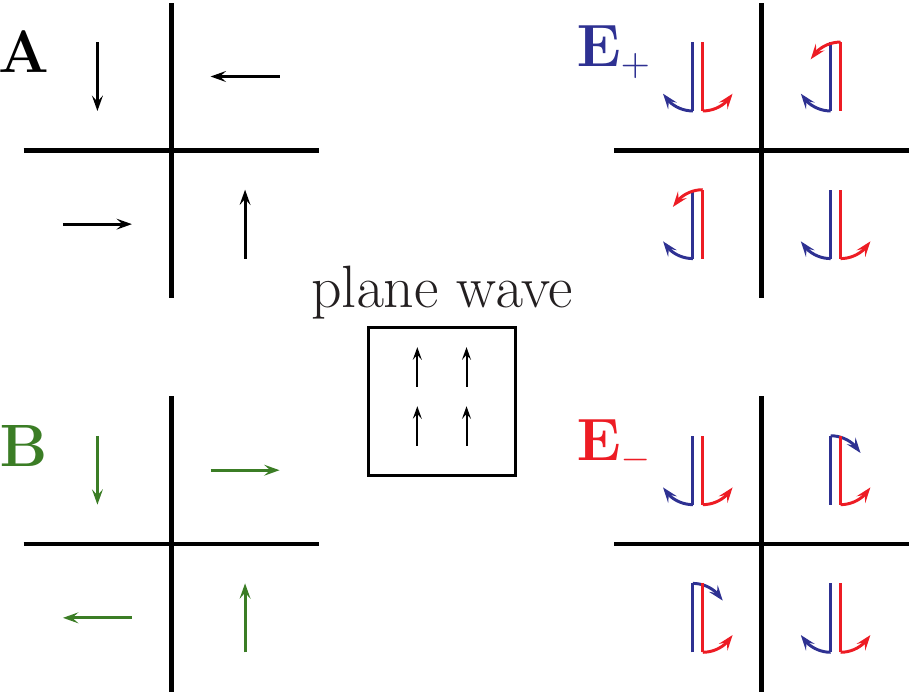}
\caption{(Color Online) Illustration of the four mode structures of a monochromatic vector field that spatially transform according to the respective irreducible representation of $C_4$ (cf.~Tab.~\ref{tab:O-C_4}). The field is shown at an arbitrary point in time at four symmetry equivalent points. For the $E_\pm$ modes for which the characters are not all real, the transformation depends on the polarization state and we split each mode into a red RCP part with temporally right rotation as seen from the receiver and a blue LCP part rotating opposite. The $A$ and $B$ profiles do not couple to a plane wave (center) as opposite contributions with a phase shift of $\pi$ cancel out. The RCP part of an $E_+$ mode and the LCP part of an $E_-$ mode cannot couple to a plane wave for the same reason.}
	\label{fig:mode-coupling}
\end{figure}
\subsubsection*{(c) Time inversion symmetry and slope at $\Gamma$ and $H$}

Due to time-reversal symmetry it is sufficient to show the band structure along the $\Delta$ line (including $\Gamma$ and $H$). To obtain the mode representations of the other half we examine the action of the time-reversal operator $\hat{T}$ on the modes and eigenfrequencies. $\hat{T}$ is defined by $\hat{T}f(t)=f(-t)$. The action on the (complex) spatial part $f(\vec{r})$ of a monochromatic field $f(\vec{r},t)=\mbox{Re}\{f(\vec{r})\exp(-\imath\omega t)\}$ is then given by the antiunitary complex conjugation operator, i.e.~$\hat{T}f(\vec{r})=\cc{f}(\vec{r})$, that obeys $\hat{T}^2=\hat{E}$ and transforms the Bloch-Maxwell operator into one with opposite wave vector:
	$$\hat{\vartheta}_{\vec{k}}'=\hat{T}\hat{\vartheta}_{\vec{k}}\hat{T}^{-1}=\hat{T}\hat{\vartheta}_{\vec{k}}\hat{T}=\hat{\vartheta}_{-\vec{k}}\mbox{.}$$ 
The action on an eigenfunction $|\vec{k}i\rangle$ characterized by its wave vector $\vec{k}$ and representation $i$ is generally given by
	$$\hat{T}|\vec{k}i\rangle = |-\vec{k},i'\rangle\mbox{.}$$
\begin{figure}[t]
	\includegraphics[width=\columnwidth]{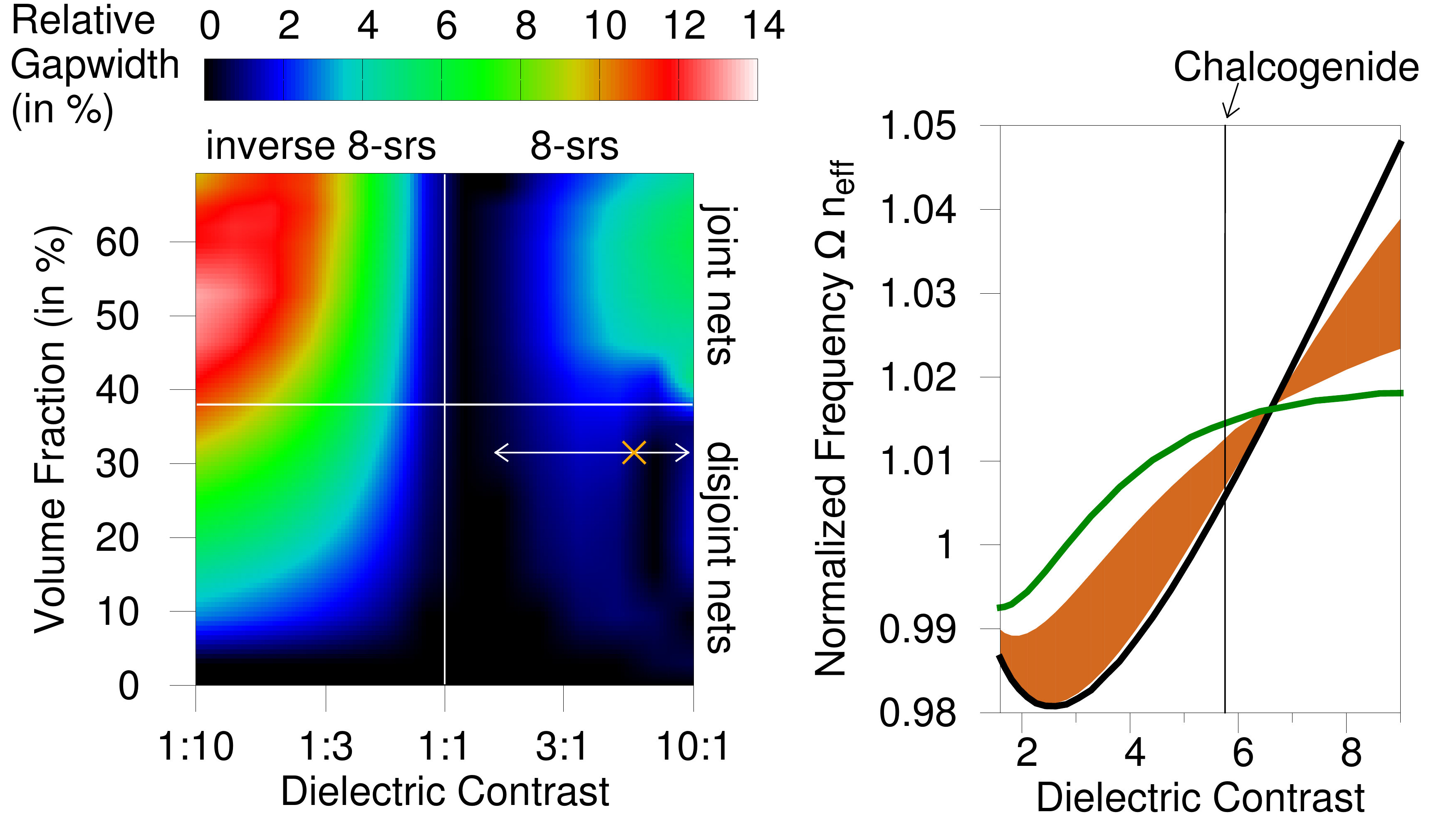}
	\caption{(Color Online) Relative gap width $\delta_G=2(\omega_2-\omega_1)/(\omega_2+\omega_1)$ of the symmetry induced band gap of mid gap frequency $\Omega_G$; $\omega_1$ is the maximum frequency of the two fundamental bands of character $E_+$ and $E_-$, respectively, and $\omega_2$ the minimum of the two air bands of same character. (Left) Color coded map of the $\delta_G$ as a function of volume fraction $\Phi$ and dielectric contrast $\psi=\varepsilon_{\text{8-srs}}:\varepsilon_{\text{background}}$ on a logarithmic scale. The join frequency $\Phi_J\approx38\%$ indicates a topological change: For $\Phi < \Phi_J$, the individual {\bf srs} nets are disconnected, for $\Phi > \Phi_J$ they overlap to form a single connected component. The orange point 'x' marks the choice of parameters for Fig.~\ref{fig:transmission-bandstructure}. (Right) Gap map along the line indicated by the white arrow on the left including the x-point. The frequency is scaled by the effective index $n_{\text{eff}}\coloneqq\sqrt{\sum_{m} \Phi_m\varepsilon_m}$ with $m\in\{\text{8-srs,background}\}$ on the ordinate. The position of the band gap hence does not depend on the choice of the absolute value of dielectric constants. It turns out that the x-point is close to a band structure topology change at $\Phi\approx 31\%$ and $\psi\approx 6.5$ where the $T_1$ (black curve) and $T_2$ (green curve) points at $H$ are accidentally degenerate.}
	\label{fig:gapwidth}
\end{figure}
\begin{figure*}[ht]
	\includegraphics[width=\textwidth]{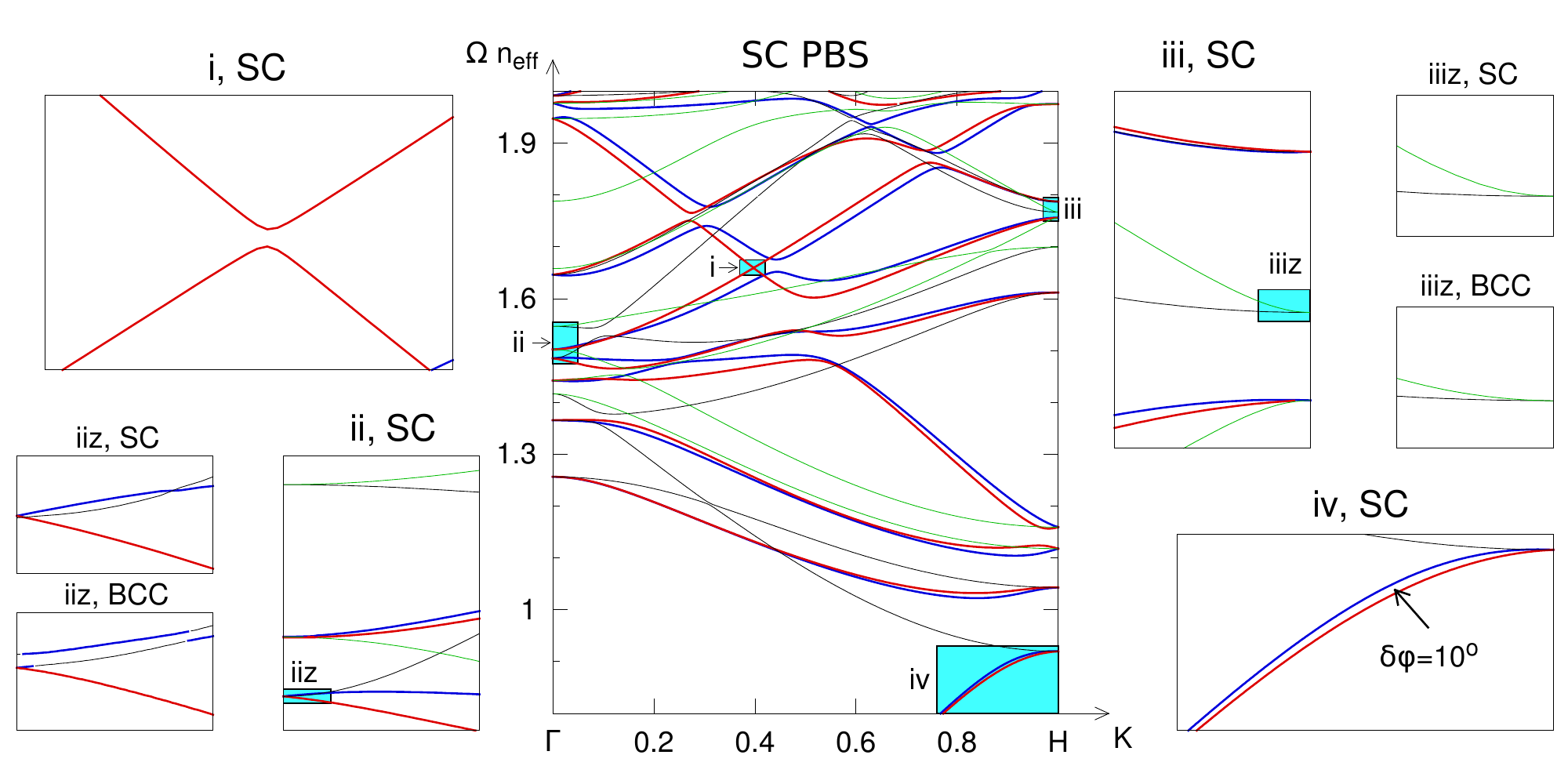}
	\caption{(Color Online) Bandstructure of the inverse {\bf 8-srs} PC for $\varepsilon=1$ within the networks and $\varepsilon=5.76$ outside and volume fraction $\Phi\approx 41\%$ (upper left quadrant in Fig.~\ref{fig:gapwidth}). The PBS is calculated using BCC and simple cubic (SC) symmetry. In the plots labelled 'SC' the band structure of the simple cubic lattice is calculated for a grid discretization of $M^3=32^3$ and unfolded to the BCC $\Delta$ path \cite{details_of_unfolding_in_prb}. The bands are colored by their respective symmetry behaviour corresponding to $i=\{A,B,E_+,E_-\}$. Some of the interesting regions around $H$ and $\Gamma$ are magnified to demonstrate the excellent agreement between theory and numerics in the SC calculations. Plots labelled 'BCC' use a BCC Fourier grid that does not maintain all symmetries of the BCC symmetry, see Fig.~\ref{fig:symm-mismatch}. The BCC calculations illustrate that the violation of correct symmetry lifts the degeneracy  of the bands at $3$-fold degenerate points, even for a fine grid with $M=64$.}
	\label{fig:SCvsBCC}
\end{figure*}
The positive definite\cite{JoannopoulosJohnsonWinnMeade:2008} nature of $\hat{\vartheta}_{\vec{k}}$ yields the frequency "degeneracy" $\omega_i(\vec{k})=\omega_{i'}(-\vec{k})$ \footnote{We have adopted the term degeneracy although the states have opposite Bloch wave vector and generally different representation.}. We distinguish two cases \footnote{Generally, a third case can occur where neither statement (a) nor (b) holds. However, this case only occurs in the presence of a $4$-fold improper rotation or a two-fold screw axis that are both not present and generally cannot be in the group of the wave vector within the Brillouin zone \cite{Herring:1937}.}(cf.~references \onlinecite{Dresselhaus:2008} and \onlinecite{LudwigFalter:1988}) :
\begin{align*}
	\text{(a)}\; \exists R'\in G:\;\forall R\in G \phantom{\hspace{5cm}}\\
		\cc{\mat{D}(R')\,\mat{D}(R)\,\mat{D}(R')^{-1}} = \mat{D}(R')\,\mat{D}(R)\,\mat{D}(R')^{-1},\\
	\text{(b)}\; \nexists R'\in G: \;\forall R\in G \phantom{\hspace{5cm}}\\
		\mat{\cc{D}}_i(R)=\mat{D}(R')\,\mat{D}_i(R)\,\mat{D}(R')^{-1},\\
\end{align*}
As time inversion $\hat{T}$ commutes with the point symmetry operations $\hat{R}$, $i'$ is equal to $i$ in case (a). As a proof for that statement we choose the representation whose matrix entries are all real so that the time-reversal state transforms as a state of irreducible representation $i$:
\begin{align*}
	\hat{R}|-\vec{k} i'\alpha\rangle	&= \hat{R}\hat{T}|\vec{k}i\alpha\rangle = \sum_\beta \hat{T}\mat{D}^{(i)}_{\alpha\beta}(R)|\vec{k}i\beta\rangle \\
						&\overset{\text{(a)}}{=} \sum_\beta \mat{D}^{(i)}_{\alpha\beta}(R)\hat{T}|\vec{k}i\beta\rangle\\
						&= \sum_\beta \mat{D}^{(i)}_{\alpha\beta}(R)|-\vec{k} i'\beta\rangle
\end{align*}
Contrary, $i$ and $i'$ are different irreducible representations (of same dimension) in case (b) that form a {\em quasi-degenerate} pair so that
\begin{align*}
	\omega_i^2(\vec{k})/c^2 &= \langle\vec{k}i|\hat{\vartheta}_{\vec{k}}\vec{k}i\rangle 
				= \langle\hat{T}^2\vec{k}i|\hat{\vartheta}_{\vec{k}}\hat{T}^2\vec{k}i\rangle\\
				&= \langle(\hat{T}\hat{\vartheta}_{\vec{k}}\hat{T})\hat{T}\vec{k}i|\hat{T}\vec{k}i\rangle
				= \langle\hat{\vartheta}_{-\vec{k}}\,-\vec{k},i'|-\vec{k},i'\rangle\\
				&= \omega_{i'}^2(-\vec{k})/c^2\text{.}
\end{align*}
This quasi-degenerate pair naturally meets at both the $\Gamma$ and the $H$ point. The respective case for each irreducible representation can be determined by the Herring rules \cite{Herring:1937}. The pair of two fold rotations in $O$ perpendicular to the $[100]$ direction is responsible for a non-trivial behaviour along $\Delta$. The different cases are listed in table \ref{tab:O-C_4}. At the $\Gamma$ and the $H$ point, the irreducible representations of $O$ are all of time inversion type (a).

In accordance with the compatibility relations above, a pair of $E_+$ and $E_-$ modes along $\Delta$ hence always meets at the $H$ ($\Gamma$) point. Their group velocity in the vicinity is further of same magnitude and opposite sign. The $A$ and $B$ representations on the other hand always have vanishing group velocity close to $\Gamma$ and $H$.

\subsubsection*{(d) $\{A,B,E_+,E_-\}$ correspond to non-interacting scattering channels}
The mode structure of the $C_4$ representations is strongly connected to a plane wave propagating along the axis of symmetry. We derive the symmetry behaviour of such a plane wave using a circular polarization basis, i.e.~we make use of representations theorem (iii) to calculate the composition of an RCP/LCP wave. The modulation factor $\exp\{\imath k z\}$ is obviously unchanged by any operation within $C_4$ so that $\hat{R}$ is only acting on the Jones vector for which we obtain:
	\begin{align*}
		\begin{pmatrix}1\\\mp\imath\\0\end{pmatrix} =
		&\sum_i \frac{l_i}{h} \sum_{R\in C_4} \chi_i(R) \hat{R} \begin{pmatrix}1\\\mp\imath\\0\end{pmatrix}\\
		=&\sum_i \frac{1}{4} \left\{ \chi_i(\mathbb{1}) \begin{pmatrix}1\\\mp\imath\\0\end{pmatrix}
			+ \chi_i(C_{4_1}) \begin{pmatrix}\pm\imath\\1\\0\end{pmatrix}\right. \\
			&\left. + \chi_i(C_2) \begin{pmatrix}-1\\\pm\imath\\0\end{pmatrix} 
			+ \chi_i(C_{4_3}) \begin{pmatrix}\mp\imath\\-1\\0\end{pmatrix}
		\right\}\\
		= & \begin{cases} 
			\begin{pmatrix}
				1\\\mp\imath\\0
			\end{pmatrix} & \text{, if } i=E_\mp\\
			0 & \text{, else}
		\end{cases}
	\end{align*}
Therefore, any LCP/RCP mode transforms purely according to the irreducible representation $E_\pm$ respectively, there is strictly no contribution of an $A$ or $B$ representation to any plane wave travelling along a $C_4$ axis. As a result, the two distinct circular polarization states scatter in orthogonal and hence independent channels corresponding to $E_\pm$ because of representation theorem (iv). Polarization conversion vanishes for any scattering process at a $4$-fold symmetric structure. 

While the rigorous proof is provided above, an intuitive understanding is easily obtained by going the reverse logical way starting from the four irreducible representations and calculate the coupling strength with a plane wave. This idea is illustrated in Fig.~\ref{fig:mode-coupling}.

We finally note that time inversion exchanges the representations $E_\pm$ that are defined respective to a static Cartesian coordinate system whereas it leaves the circular polarization state that is defined respective to a right handed coordinate system depending on the propagation direction $\vec{k}$ that changes its sign under time inversion symmetry.
\begin{figure}[t]
	\includegraphics[width=\columnwidth]{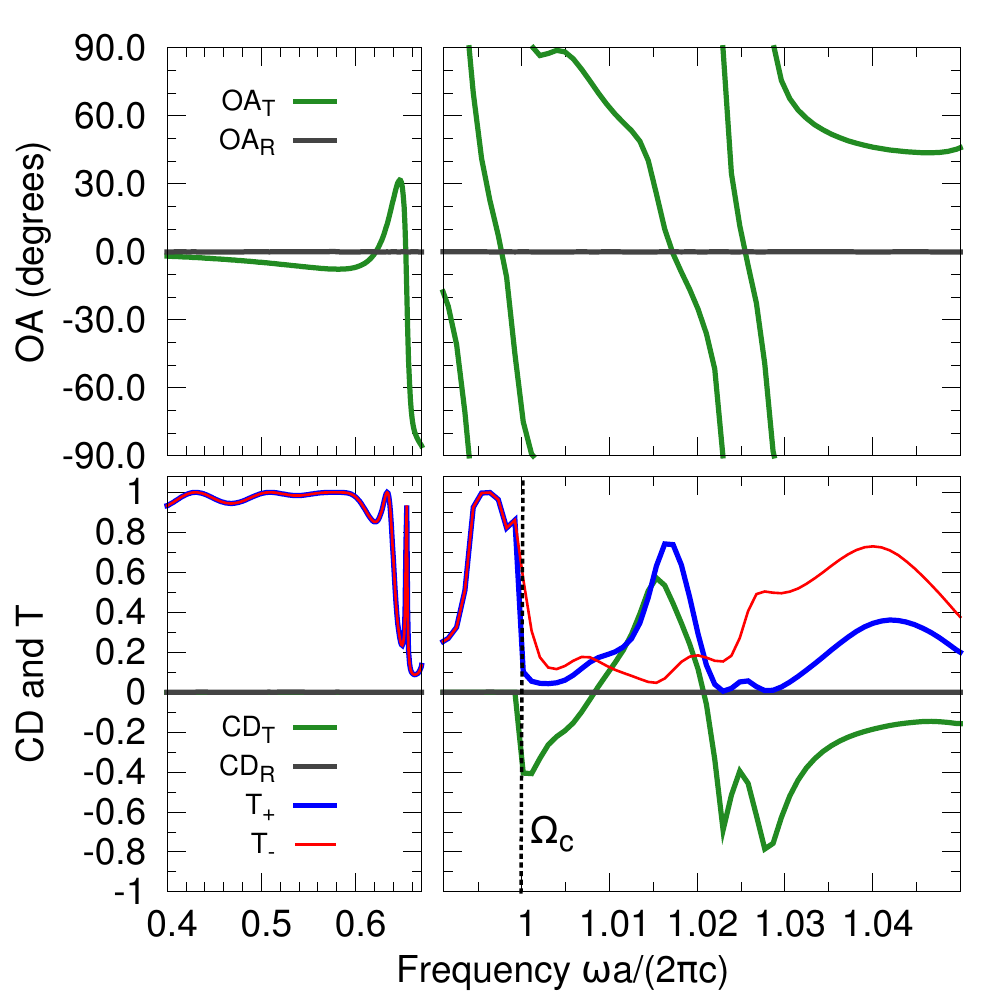}
        \caption{(Color Online) Simulated CD, OA and transmissivity $T_\pm$ in both transmission channels $E_\pm$ for the reflection and transmission of a plane wave at normal incidence: The {\bf 8-srs} PC slab has termination $t=0.25a$ and thickness $N_z=4$. Since optical experiments cannot measure phase differences $> \pm90^\circ$, OA is wrapped onto the interval $[-90^\circ,90^\circ]$ by $\text{OA}\mapsto\arctan\left(\tan(\text{OA})\right)$.}
        \label{fig:optical-activity}
\end{figure}

%%%SECTION SCATTERING PARAMETERS
\section{\label{sec:scattering-parameter-predictions}General Predictions for Scattering Parameters}

General features of the scattering matrix for a finite slab of an $I432$ PC with $(100)$ inclination can be derived by its $4$-fold symmetry alone. In the following, no energy dissipation in any of the PC materials is assumed. Kaschke {\emph et al.}~recently calculated in \cite{KaschkeGanselWegener:2012} that there is no circular dichroism in the reflectance and transmittance for any lossless structure if any higher order scattering is neglected. Here, we show more precisely that the reflectivity matrix of the same system is identical for RCP and LCP light at all frequencies, i.e.~circular dichroism and optical activity are zero. In the transmission spectrum, optical activity is present and particularly strong above the fundamental bands whereas circular dichroism is zero at frequencies below $\Omega_c=1$. Above the critical frequency $\Omega_c$, the $(10)$ Bragg order becomes leaky so that $\text{transmission}+\text{reflection}=1$ is no longer a valid statement if both are just measured within the $(00)$ Bragg order (Fig.~\ref{fig:optical-activity}).

We first show that scattering matrix for any photonic crystal slab with $4$-fold symmetry normal to its clipping planes can be well defined. We use representation theorem (iii) (a) to define a set of symmetry adapted Floquet basis functions of the in-plane vector $\vec{r}=(x,y)^T$ that transform according to the irreducible representations of the $C_4$ point group:
$$
	|(\vec{n}\sigma)i\rangle\equiv f_{\sigma,\vec{n}}^{(i)}(\vec{r})
		= \frac{1}{2}\sum_{R\in C_4}\cc{\chi}_i(R)\hat{R}\,\vec{e}_\sigma e^{\imath\vec{G}_{\vec{n}}\cdot\vec{r}}
$$
Those basis functions are characterized by their irreducible representation $i\in\left\{A,B,E_+,E_-\right\}$ (Tab.~\ref{tab:O-C_4}), polarization $\sigma\in\{s,p\}$ and a corresponding unit vector $e_\sigma$ \footnote{Technically, this is a choice of natural basis states where either the magnetic ($s$) or the electric ($p$) field lies in the plane of incidence of the respective plane wave component, i.e.~the plane spanned by the wave vector and the surface normal. That definition corresponds to polar coordinate system $(r,\varphi,z)$ with the $4$-fold axis as its center so that the polarization state translates as $s=r$ and $p=\varphi$. In that local coordinate system, the action of any $R\in C_4$ onto $e_\sigma$ is the identity transformation. Note that we only use the polar system for the field vectors and not for the position vector so that all spatial derivatives etc.~are still Cartesian.} and Bragg order $\vec{n}=(n_1,n_2)$ (with $n_1\in\mathbb{N}_1$ and $n_2\in\mathbb{N}_0$) that defines the reciprocal grating vector $\vec{G}_{\vec{n}}=(2\pi/a)\vec{n}$. To obtain a complete set for the normal incidence scattering problem, only two additional $(00)$ Bragg order basis functions $f_{x,(0,0)^T}^{E_\pm}(\vec{r})$ are added to the above \footnote{$R(0,0)^T=(0,0)^T\forall \hat{R}\in C_4$ yields $f_{\sigma,(0,0)^T}^{A,B}=0$ and $f_{x,(0,0)^T}^{(i)}=f_{y,(0,0)^T}^{(i)}$.}. The symmetry adapted basis functions are easily shown to be plane orthogonal using the orthogonality of plane waves, the Cartesian basis vectors and the wonderful orthogonality theorem of representation theory:
	$$\frac{1}{a^2}\int_{-a/2}^{a/2}dx\,\int_{-a/2}^{a/2}dy\;\cc{f}_{\sigma,\vec{n}}^{(i)} \, f_{\sigma',\vec{n}'}^{(i')}
		= \delta_{i,i'}\delta_{\sigma,\sigma'}\delta_{\vec{n},\vec{n}'}$$
We use the compact Dirac notation so that the electro-magnetic fields $|F_{(\vec{n}\sigma)i}\rangle$ in vacuum that are involved in the scattering process are expressed by
	$$|F_{(\vec{n}\sigma)i}\rangle \coloneqq \begin{pmatrix} |E_{(\vec{n}\sigma)i}\rangle \\ |H_{(\vec{n}\sigma)i}\rangle \end{pmatrix} 
		= \begin{pmatrix} 1 \\ H_{(\vec{n}\sigma)i,q_d} \end{pmatrix} |(\vec{n}\sigma)i\rangle\otimes|q_d\rangle$$
where $d=\pm$ denotes the sign ambiguity for $|q_d\rangle\equiv e^{\imath q_d z}$ with $q_d=d\,\sqrt{\omega^2/c^2-|\vec{G}_{\vec{n}}|^2}$ defined by the vacuum dispersion relation. The Bloch fields within a thin layer\footnote{So that the structure is assumed to be homogeneous along $z$ within that layer (i.e.~$\varepsilon(z)=\varepsilon(z+z_0)$ if $|z_0|<\delta/2$).} of thickness $\delta$ also transform as one irreducible representation $i$ only and are hence given by
	$$|\mathcal{F}_{\alpha i}\rangle = \begin{pmatrix} |\mathcal{E}_{\alpha i}\rangle \\ |\mathcal{H}_{\alpha i}\rangle\end{pmatrix} =
		\sum_{\vec{n}\sigma} \begin{pmatrix} \mathcal{E}_{(\vec{n}\sigma)i,q_d}^{(\alpha)} \\ \mathcal{H}_{(\vec{n}\sigma)i,q_d}^{(\alpha)} \end{pmatrix}
		|(\vec{n}\sigma)i\rangle\otimes|q_d\rangle$$
where within the layer $|q_d\rangle\equiv e^{\imath q_d z}$ is again a plane wave whose wave vector $q_d$ however has no analytical expression \footnote{It can be numerically computed by conversion of the planar Bloch-Maxwell equation into an eigenvalue equation for $q_d$ \cite{WhittakerCulshaw:1999}.} and $d$ is hence chosen by $\text{sgn}(\text{Im}\{q\})$ if $\text{Im}\{q\}\ne0$ and $\text{sgn}(\text{Re}\{q\})$ else \footnote{$z$-inversion symmetry around the homogeneous layer center and time inversion symmetry guarantee that in each case states come in pairs $q_\pm$}. Because of the orthogonality $\langle ai|bj \rangle = \delta_{ab}\delta_{ij}$ a scattering matrix treatment within each channel $i$ analogous to the calculation described by Whittaker and Culshaw \cite{WhittakerCulshaw:1999} can be performed and (with $\delta\rightarrow0$) yields a well defined scattering matrix for the whole structure that relates the (generally countably infinite number of) amplitudes $c_{O\,\nicefrac{1}{2}}^{(a)}$ of the outgoing symmetry adapted Floquet waves to the amplitudes $c_{I\,\nicefrac{1}{2}}^{(a)}$ the corresponding incoming waves; $\nicefrac{1}{2}$ indicates the $\nicefrac{z_{\text{min}}}{z_{\text{max}}}$ surface of the slab respectively so that outgoing waves are those with the sign $d=\pm$ and incoming waves with $d=\mp$ at the $\nicefrac{1}{2}$ surface. With a vector $\vec{c}$ going over all indices $a$, the scattering matrix $\mat{S}$ is defined within each $4$-fold channel $i\in\{A,B,E_+,E_-\}$ by:
	$$ \begin{pmatrix}\vec{c}_{O1} \\ \vec{c}_{O2}\end{pmatrix}= \begin{pmatrix} \matns{S}_{11} & \matns{S}_{12} \\ \matns{S}_{21} & \matns{S}_{22}\end{pmatrix}
		\begin{pmatrix}\vec{c}_{I1} \\ \vec{c}_{I2}\end{pmatrix}\text{.}$$
The definition of $q_d$ implies that non-zero entries in $\vec{c}_{I\nicefrac{1}{2}}$ are only physical for $\Omega\ge|\vec{n}|$, such that the corresponding fields are propagating. The same condition applies for all entries in $\vec{c}_{O\nicefrac{1}{2}}$ that contribute to the far-field. Further, if the incoming field is at normal incidence only the $4$-fold channels that transform as $E_\pm$ are relevant as shown above with the LCP/RCP Jones vectors. We therefore restrict further arguments to the square submatrices $\mat{S}_{\pm}$ that relate the propagating Floquet modes in the respective scattering channel corresponding to $E_{\pm}$.

\begin{figure*}[t]
	\includegraphics[width=\textwidth]{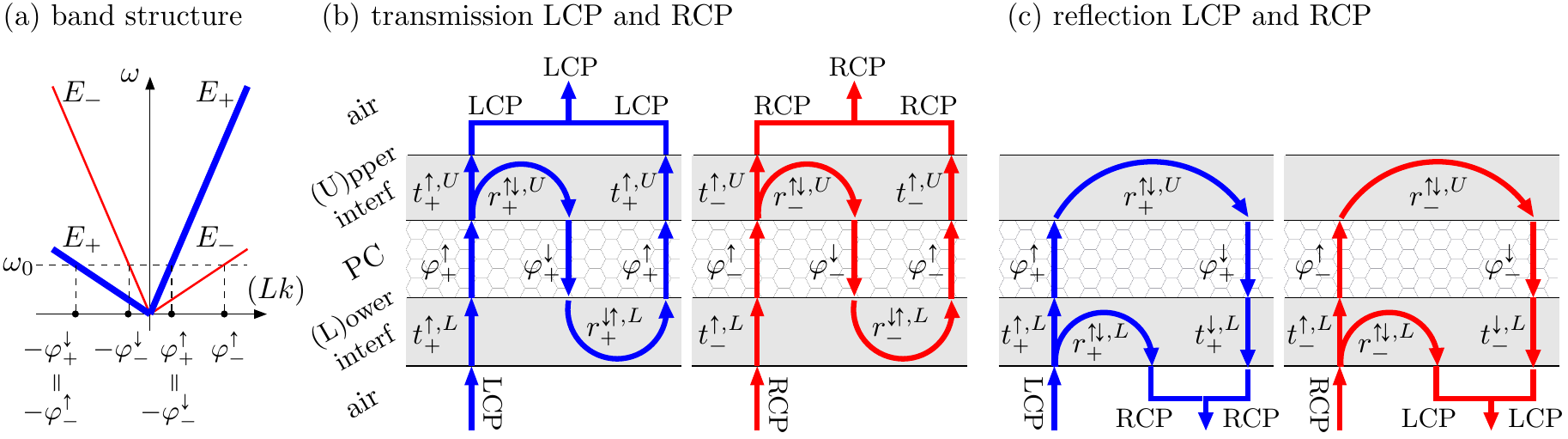}
	\caption{(Color Online) Diagrammatic representation of the processes that lead to different circular-polarisation properties of reflected and transmitted light. Assuming that there is a single propagating mode in each channel and direction (a) at a given frequency $\omega_0$, transmission (b) and reflection (c) can be understood by considering positive, upwards pointing $k$ ($\uparrow$) and their counterparts for negative $k$ ($\downarrow$). If the modes further have a leading Bragg order in propagation direction (i.e.~interference may be neglected) and the thickness of the slab $L$ is an integer multiple of the lattice parameter $a$, the difference in $k$ between two modes $\delta k$ correspond to an optical phase shift $\delta \varphi=\delta k L$. The regions labelled upper (U) and lower (L) interface represent the interfacial two-dimensional planes between free space and the photonic crystal that are inflated to finite thickness for visualization. The sign and coloration represents the $E_\pm$ channels in the same manner as in figures \ref{fig:transmission-bandstructure}, \ref{fig:mode-coupling}, \ref{fig:SCvsBCC} and \ref{fig:optical-activity}.
}
	\label{fig:refl-trans}
\end{figure*}

We now prove by energy conservation and time inversion symmetry that the reflectance spectrum (in the far field) is exactly the same for RCP and LCP incident light whereas the transmittance spectrum has different phases for all frequencies and different magnitudes if $\Omega\ge 1$. These predictions are precisely reproduced for the {\bf 8-srs} by a numerical calculation in Fig.~\ref{fig:optical-activity}.

In scattering theory, the scattering matrix is often assumed to be unitary. We prove that this is also true for the particular problem. Averaging Poynting's theorem over time for the monochromatic fields yields the spatially local equation $\vec{\nabla}\cdot\text{Re}\left\{\cc{\vec{E}}\times\vec{H}\right\} = 0$. We integrate this equation over a box of size $a\times a\times l$ that is centered around the photonic crystal slab with $l\rightarrow\infty$. Because of the mode structure of the $4$-fold modes the parallel components of the Poynting vector vanish at each point so that the divergence theorem yields the equation $\text{Re}\{\int_{A} \cc{\vec{E}}_\parallel\times\vec{H}_\parallel\}=0$ where only the propagating fields parallel to the surface are integrated over the top and bottom $a\times a$ surfaces.

The orthogonality of the basis functions when averaged yields that all mixed contributions in different Bragg orders vanish. We show that for interface $2$ and $s$ polarization ($p$ polarization is analogously done) so that the electric field only has the following $\varphi$ component:
	$$E_{\varphi} = \frac{1}{2}\left[c_O e^{\imath qz} + c_I e^{-\imath qz}\right]\sum_R \cc{\chi}_i(R) e^{\imath \vec{G}_{\mat{R}\vec{n}}\cdot\vec{r}}$$
Since the $z$ component of the magnetic field is irrelevant here we only calculate the in-plane field that is purely radial and obtained by Faraday's law:
	$$H_r = \frac{\imath c}{2\omega} \partial_z E_{\varphi} = \frac{-cq}{2\omega}
		\left[c_O e^{\imath qz} - c_I e^{-\imath qz}\right]\sum_R \cc{\chi}_i(R) e^{\imath \vec{G}_{\mat{R}\vec{n}}\cdot\vec{r}}$$
The left side of the integrated Poynting's theorem for $s$ polarization is therefore given by:
	\begin{align*}
		-\text{Re}\left\{\int_A\cc{E}_{\varphi}\, H_r\right\}
			&= \text{Re}\bigg\{\frac{cqa^2}{4\omega} \left[\cc{c}_O e^{-\imath qz}+\cc{c}_I e^{\imath qz}\right]\\
		&\phantom{=} \times\,\left[c_O e^{\imath qz}-c_I e^{-\imath qz}\right]\sum_R\left|\chi_i(R)\right|^2\bigg\}\\
		& = \frac{cqa^2}{\omega} \left[ |c_O|^2 - |c_I|^2 \right]
	\end{align*}
The purely imaginary cross correlated contributions vanish (i.e.~they average out over time as intuitively expected). The same result is obtained for surface $1$ so that for $q>0$ Poynting's theorem yields for each Bragg order in each $4$-fold channel:
	$$ |\vec{c}_{O1}|^2 + |\vec{c}_{02}^{(a)}|^2 = |\vec{c}_{I1}^{(a)}|^2 + |\vec{c}_{I2}^{(a)}|^2$$
The scattering matrix is hence norm-preserving and therefore unitary by Wigner's theorem.\\

A correlation between the scattering matrices in different channels can be derived by time inversion invariance. We act with $\hat{T}$ upon the equation that defines the scattering matrix and note that the vectors satisfy $\cc{\vec{c}}_{I\pm}(E_+)=\vec{c}_{O\pm}(E_-)$ (and also with $I$ and $O$ exchanged) because the basis functions $|ai\rangle$ stay unchanged by the complex conjugation except for a change in representation $\chi_{E_\pm}(R)\rightarrow\cc{\chi}_{E_\pm}(R)=\chi_{E_\mp}(R)$\footnote{The change of sign in the exponential is irrelevant as the basis functions have a symmetry $f_{\sigma,n}=f_{\sigma,-n}$ that is due the vectorial nature of $f$ and $\chi_{E_\pm}(C_2)=-1$ and can be seen acting with $\hat{C}_2$ twice upon $f$ using the explicit form and the representation form each once.} whereas the function of $z$ changes its sign $\hat{T}|q_+\rangle = |q_-\rangle$. The complex conjugated scattering matrix in one channel is hence the inverse of the matrix in the other channel, i.e.~$ \cc{\matns{S}}_\pm = \matns{S}_{\mp}^{-1} $.

\subsubsection*{(e) No CD and OA in reflectance}
We combine both results and identify the diagonal entries $\matns{S}_{ii}$ with the reflection matrices on either side to obtain $\matns{R}^{(+)}=\matns{R}^{(-)}$, i.e. the reflection matrices are identical and hence OA and CD vanish to any numerical precision for all frequencies (cf.~Fig.~\ref{fig:optical-activity}). We present an illustrative interpretation of this result in figure \ref{fig:refl-trans} based on the assumption that there is only a single mode per channel and propagation direction. If the assumptions made in \ref{fig:refl-trans} are met as for example in the two fundamental bands, the Airy formula in terms of the interfacial scattering parameters defined in \ref{fig:refl-trans} yields for the transmission and reflection amplitudes with $p_\pm^{\uda}\coloneqq\exp\{\varphi_\pm^{\uda}\}$:
\begin{align*}
	t &= t_\pm^{\ua ,L} p_\pm^{\ua} t_\pm^{\ua ,U} \sum_{n=0}^\infty \left(r_\pm^{\uda ,U} p_\pm^{\da} r_\pm^{\dua ,L} p_\pm^{\ua} \right)^n\\
	r &= r_\pm^{\uda ,L} + t_\pm^{\ua ,L} p_\pm^{\ua} r_\pm^{\uda ,U} p_\pm^{\da} t_\pm^{\da ,L} \sum_{n=0}^\infty \left(r_\pm^{\uda ,U} p_\pm^{\da} r_\pm^{\dua ,L} p_\pm^{\ua} \right)^n
\end{align*}
The first two contributions ($n\le1$) of the infinite geometric series are shown for transmission and the first ($n=0$) for reflection. As the surface acts as a planar grating, we can use the previous results of \cite{BaiSvirkoTurunenVallius:2007} that report a special case of our more general treatment: $r_+^{\uda}=r_-^{\uda}$, $r_+^{\dua}=r_-^{\dua}$ and $t_\pm^{\da}=t_\mp^{\ua}$ can be used. Hence $\text{OA}_R=0$, but there is no general restriction on $\text{OA}_T$, because the number of passes through the structure is even for reflectivity and each contribution to the sum and hence $t$ on the lower interface and $p$ come only in pairs so that $t_+^{\da ,L}t_+^{\ua ,L}=t_-^{\da ,L}t_-^{\ua ,L}$ and $p_+^{\ua}p_+^{\da}=p_-^{\ua}p_-^{\da}$ because of time inversion symmetry (see Fig.~\ref{fig:refl-trans}).

The phase shift at the surfaces caused by the respective scattering amplitudes is usually negligible for frequencies in the fundamental bands and therefore optical activity is essentially due to the difference in the wave vectors of both eigenmodes (cf.~Fig.~\ref{fig:refl-trans}). This straightforward interpretation yields easy and fast numerical analysis of OA that is comparable in magnitude to that of planar, metallic meta-materials \cite{DeckerRutherKrieglerZhouSoukoulisLindenWegener:2009} and therefore makes the frequency region below the fundamental band edges ($0.5\lessapprox\Omega\lessapprox0.6$ in Fig.~\ref{fig:optical-activity}) particularly interesting for engineering of optical devices such as beam splitters based on the {\bf 8-srs} or similar structures with $432$ point symmetry.

\subsubsection*{(f) No CD in transmission below a critical frequency}
For the transmission matrices we derive analogously $\matns{S}_{21}^{(+)}=\matns{S}_{12}^{(-)}$. This statement is however of less practical relevance because it identifies transmission in the $E_+$ channel where the source is on one side of the structure with transmission in the $E_-$ channel with the source on the other side and hence corresponds to two distinct experiments. Comparing transmission matrices for the same experiment we derive from the diagonal entries of the unitarity condition in both channels and the identity of the reflection matrices that the norm of transmission $\matns{T}^\dagger\matns{T}$ is the same in both channels. Therefore, 
$\text{CD}_T$ is zero below the critical frequency $\Omega_c=1$ above which energy is able to leak into the $(10)$ Bragg order and hence energy conservation within the $(00)$ order with $\vec{k}$ perpendicular to the plane of interface is not valid. There is no restriction on $\text{OA}_T$ that can be finite for all frequencies (see Fig.~\ref{fig:optical-activity}).

Particularly interesting is the magnified region labelled $iv$ in Fig.~\ref{fig:SCvsBCC}. Transmission is governed by the respective calculated mode of the fundamental band in each channel so that the situation well fits the one illustrated in Fig.~\ref{fig:refl-trans}. If we further assume that the interfacial reflections are sufficiently small (which is expected as transmission is almost $100\%$ in the fundamental bands, see Fig.~\ref{fig:optical-activity}), optical activity is given by $\delta\varphi=\delta k L$ which is estimated to be around $10\%$ at $\Omega n_{\text{eff}}=1.1$.

%%%SECTION CONCLUSIONS
\section{Conclusions and Outlook}

We have provided consistent analytical and numerical results that demonstrate the potential of the {\bf 8-srs} geometry as a dielectric photonic material that should be realisable by current nano-fabrication technology. Analytic results, based on versatile group theory and scattering matrix treatment and applicable to any photonic crystal with $I432$ symmetry, are obtained for transmission and reflection amplitudes, and for the topology of the band structure. For the {\bf 8-srs} PC, these results are in perfect agreement with numerical simulations that further demonstrate that the {\bf 8-srs} exhibits strong optical activity in transmission, comparable to metallic meta-materials, yet without losses and without any ellipticity.\\  

The potential of the {\bf 8-srs} for applications as an optically active nanofabricated material, say for the relevant communication wavelength of $\lambda  \approx 1.5 \mu m$, can be gauged from the following considerations. 

First, if robustness of the optical activity with respect to slight changes in $\lambda$ is desirable, the data in Fig.~\ref{fig:optical-activity} suggests a realisation of the {\bf 8-srs} with a lattice parameter $a=0.55 \lambda \approx 0.83 \mu m$ such that $\omega a/(2\pi c) = a / \lambda \approx = 0.55$; the rotation of the polarisation plane through a layer of four unit cells (of total thickness $3.5\mu m$) would then correspond to approximately $-8^o$ and close to 100\% transmission for both RCP and LCP light and with zero ellipticity, assuming the same values for $\phi$ and $\epsilon$ as in Fig.~\ref{fig:optical-activity}. Second, if a strong dependence of OA on the value of the frequency is acceptable, or even desired, a realisation of the {\bf 8-srs} with $a=1.44\mu m$, corresponding to $\omega a/(2\pi c) \approx 0.95$, gives a very strong optical rotation of $\approx 50^o$, in the optical communication window, again with close to 100\% transmission for both RCP and LCP light and with zero ellipticity.

Importantly, Fig.~\ref{fig:gapwidth} suggests that the parameters chosen here (corresponding to the cross on the white line in Fig.~\ref{fig:gapwidth}) are not optimized to give the strongest photonic response. Specifically, the inverse structure, consisting of hollow air channels along the edges of the {\bf 8-srs} structure embedded in a dielectric matrix, has a wider band gap. While the $E_\pm$ frequency split that is desired for many applications is generally stronger for the original {\bf 8-srs} structure (cf.~higher order bands in figures \ref{fig:transmission-bandstructure} and \ref{fig:SCvsBCC}) due to the light-guiding topology of the networks, it is found relatively weak in the fundamental bands here as those cross close to the band edge (Fig.~\ref{fig:transmission-bandstructure}).

These values for the degree of optical activity should be compared to those in other systems. For quartz, the rotary power varies from $3.2^o/\mathrm{mm}$ at wavelength $1.42\mu\mathrm{m}$ to $776^o/\mathrm{mm}$ at $152\mathrm{nm}$ \cite{Radhakrishnan:1947}; for the liquid crystalline $BP_{SmA}1$ blue phase between $0$ and $\approx 250^o/\mathrm{mm}$ over the visible spectral range \cite{GreletCollingsLiNguyen:2001}; for a smectic phase $SmC_AP_A$ a value of $100-1000 ^o/\mathrm{mm}$ has been reported \cite{HoughZhuNakataChatthamDantlgraberTschierskeClark:2007}. For a quasi-planar ``twisted-cross'' gold meta-material of thickness $87\mathrm{nm}$, Decker {\em et al.} have reported 'strong' rotations of the polarisation plane of up to $4^o$, yet with significant losses \cite{DeckerRutherKrieglerZhouSoukoulisLindenWegener:2009}; for the split-ring-resonator meta-material of thickness $205\mathrm{nm}$, 'huge' rotations of up to $30^o$ are observed with different transmittances for RCP and LCP light of less than 50\% (see Fig.~3 in \cite{DeckerZhaoSoukoulisLindenWegener:2010}). Song {\em et al.} have reported even 'more gigantic' polarisation rotation in a microwave experiment using a chiral composite material, yet also with the losses typical for metallic meta-materials. Note that the dependence on sample thickness (that can be varied in the {\bf 8-srs} by varying the number $N_z$ of unit cells) is non-linear, making a comparison of polarisation rotation normalised to the sample thickness less meaningful. 

Even for the likely non-optimal parameters chosen here, the lossless dielectric {\bf 8-srs} photonic crystal hence has a significant degree of optical activity, that is combined with the complete absence of ellipticity, absence of losses and transmission rates of close to 100\%. Further, for applications such as beam-splitters where optical properties in various transmission directions are important, the cubic symmetry of the {\bf 8-srs} structure is a further benefit to the uni-axial designs of the quasi-planar meta-materials. Given that a single {\bf srs} net with $a_0=2a=1.2\mu m$ has been realised \cite{TurnerSabaZhangCummingSchroederTurkGu:2013}, it appears likely that future advances in direct laser writing technology make a realisation of the {\bf 8-srs} or its inverse structure with a lattice size $a=1\mu m$ a genuine possibility. Taking all of these considerations into a account, we believe that the {\bf 8-srs} is a design for a chiral-optical material that is worth further investigation.\\

Evidently, the validity of all theoretic predictions of this article is not limited to the {\bf 8-srs} geometry, but applies more generally to all geometries with symmetry $I432$. It is therefore worth exploring structural data bases for other designs with that symmetry; this may include network structures such as the {\bf fcd} net \cite{DelgadoFriedrichsFosterOKeeffeProserpioTreacyYaghi:2005}, sphere packings such as Fischer's {\bf 4/4/c14} or {\bf 3/4/c3} packings \cite{Fischer:2004}, minimal surface geometries such as Koch's $NO3^2-c2$ structure \cite{Koch:1999}, rod packings such as {\bf utz-b} \cite{RosiKimEddaoudiChenOKeeffeYaghi:2005} or woven filaments such as Evans' $P_{129RL}(\cosh^{-1}(3/2))$ structure (see Fig.~8 of \cite{EvansRobinsHyde:2013b}).

Beyond the specific symmetry group $I432$, our group theoretic arguments can be adapted to apply more widely to other crystalline chiral materials. For example, the suppression of polarisation conversion is expected to be valid for any chiral structure that has $m$-fold rotations (as opposed to screw rotations) with $m\ge3$ along the $[100]$ propagation direction. These conditions are for example met for all cubic structures with symmetry groups $I432$ (no.~211), $F432$ (no.~209) and $P432$ (no.~207); therefore other network structures with these symmetries may be alternative photonic designs with similar properties. From a perspective of photonic materials, particularly structures with simple cubic $P$ symmetries may be attractive, because of their robustness with respect to the incident wave vector angle; for transmission along the $[100]$ direction of a simple cubic crystal, the $X$ point of the Brillouin zone represents the center of a facet, in contrast to the $H$ point which represents a vertex of the BCC Brillouin zone. Therefore, a small variation in the incident angle will, to first order, not affect the distance between $\Gamma$ and $X$ in the SC case, again in contrast to the distance between $\Gamma$ and $H$ in the BCC case. As is the case for the $I432$ symmetry, advanced geometry and structural data bases can provide suitable structure candidates, such as {\bf dgn} or {\bf fce} \cite{OKeeffePeskovRamsdenYaghi:2008} or $P_{118RL}(\cosh^{-1}(\sqrt{6}))$ (see Fig.~14 of \cite{EvansRobinsHyde:2013b}).

As a by-product of our exact group theoretic results for the topology of the band structure, we have identified a general problem with numeric Fourier-based methods for the band structure of structures with BCC (and hexagonal and face centered cubic etc.) symmetry. The Fourier grid for BCC structures can never maintain the full point symmetry (Fig.~\ref{fig:symm-mismatch}), and hence discretization artefacts are unavoidable and, as Fig.~\ref{fig:SCvsBCC} shows, significant even for relatively large sizes of the Fourier grid. This problem, that is not specific to our application, needs to be considered for all numerical Fourier methods.

In a broader context, our study of the {\bf 8-srs} geometry demonstrates the benefit of using advanced unconventional concepts of modern real-space geometry, such as the maximally symmetric intergrowth of multiple minimal nets, for the informed design process of functional photonic materials. While numerical photonic methods for transmission parameters and band structures clearly are indispensable tools to translate geometric structure into photonic properties, we have here demonstrated the ability of group theory to provide a firm theoretical understanding of the relationship between photonic functionality and geometric properties for complex three-dimensional chiral photonic crystals.

\begin{acknowledgments}
We are grateful to Stephen Hyde for the inspiration to investigate multiple entangled {\bf srs} nets. We thank Michael Fischer for comments on the manuscript. MS, KM and GST gratefully acknowledge financial support by the German Science Foundation through the excellence cluster "Engineering of Advanced Materials" at the Friedrich-Alexander-University Erlangen-N\"urnberg. MS is grateful to Nadav Gutman for inspiration and suggestion group theory as an approach to analyse symmetry behaviour. MT and MG are grateful for financial support by the Australian Research Council through the "Centre for Ultrahigh-Bandwidth Devices for Optical Systems".

\end{acknowledgments}

\bibliography{8srs-literature}

%merlin.mbs apsrev4-1.bst 2010-07-25 4.21a (PWD, AO, DPC) hacked
%Control: key (0)
%Control: author (8) initials jnrlst
%Control: editor formatted (1) identically to author
%Control: production of article title (-1) disabled
%Control: page (0) single
%Control: year (1) truncated
%Control: production of eprint (0) enabled
\begin{thebibliography}{104}%
\makeatletter
\providecommand \@ifxundefined [1]{%
 \@ifx{#1\undefined}
}%
\providecommand \@ifnum [1]{%
 \ifnum #1\expandafter \@firstoftwo
 \else \expandafter \@secondoftwo
 \fi
}%
\providecommand \@ifx [1]{%
 \ifx #1\expandafter \@firstoftwo
 \else \expandafter \@secondoftwo
 \fi
}%
\providecommand \natexlab [1]{#1}%
\providecommand \enquote  [1]{``#1''}%
\providecommand \bibnamefont  [1]{#1}%
\providecommand \bibfnamefont [1]{#1}%
\providecommand \citenamefont [1]{#1}%
\providecommand \href@noop [0]{\@secondoftwo}%
\providecommand \href [0]{\begingroup \@sanitize@url \@href}%
\providecommand \@href[1]{\@@startlink{#1}\@@href}%
\providecommand \@@href[1]{\endgroup#1\@@endlink}%
\providecommand \@sanitize@url [0]{\catcode `\\12\catcode `\$12\catcode
  `\&12\catcode `\#12\catcode `\^12\catcode `\_12\catcode `\%12\relax}%
\providecommand \@@startlink[1]{}%
\providecommand \@@endlink[0]{}%
\providecommand \url  [0]{\begingroup\@sanitize@url \@url }%
\providecommand \@url [1]{\endgroup\@href {#1}{\urlprefix }}%
\providecommand \urlprefix  [0]{URL }%
\providecommand \Eprint [0]{\href }%
\providecommand \doibase [0]{http://dx.doi.org/}%
\providecommand \selectlanguage [0]{\@gobble}%
\providecommand \bibinfo  [0]{\@secondoftwo}%
\providecommand \bibfield  [0]{\@secondoftwo}%
\providecommand \translation [1]{[#1]}%
\providecommand \BibitemOpen [0]{}%
\providecommand \bibitemStop [0]{}%
\providecommand \bibitemNoStop [0]{.\EOS\space}%
\providecommand \EOS [0]{\spacefactor3000\relax}%
\providecommand \BibitemShut  [1]{\csname bibitem#1\endcsname}%
\let\auto@bib@innerbib\@empty
%</preamble>
\bibitem [{\citenamefont {Johnson}(1990)}]{Johnson:1990}%
  \BibitemOpen
  \bibfield  {author} {\bibinfo {author} {\bibfnamefont {W.~C.}\ \bibnamefont
  {Johnson}},\ }\href@noop {} {\bibfield  {journal} {\bibinfo  {journal}
  {Proteins: Structure, Function, and Bioinformatics}\ }\textbf {\bibinfo
  {volume} {7}},\ \bibinfo {pages} {205} (\bibinfo {year} {1990})}\BibitemShut
  {NoStop}%
\bibitem [{\citenamefont {Barron}(2009)}]{Barron:2009}%
  \BibitemOpen
  \bibfield  {author} {\bibinfo {author} {\bibfnamefont {L.}~\bibnamefont
  {Barron}},\ }\href@noop {} {\emph {\bibinfo {title} {Molecular Light
  Scattering and Optical Activity}}}\ (\bibinfo  {publisher} {Cambridge
  University Press},\ \bibinfo {year} {2009})\BibitemShut {NoStop}%
\bibitem [{\citenamefont {Schadt}\ and\ \citenamefont
  {Helfrich}(1971)}]{SchadtHelfrich:1971}%
  \BibitemOpen
  \bibfield  {author} {\bibinfo {author} {\bibfnamefont {M.}~\bibnamefont
  {Schadt}}\ and\ \bibinfo {author} {\bibfnamefont {W.}~\bibnamefont
  {Helfrich}},\ }\href@noop {} {\bibfield  {journal} {\bibinfo  {journal}
  {Applied Physics Letters}\ }\textbf {\bibinfo {volume} {18}},\ \bibinfo
  {pages} {127} (\bibinfo {year} {1971})}\BibitemShut {NoStop}%
\bibitem [{\citenamefont {Grelet}\ \emph {et~al.}(2001)\citenamefont {Grelet},
  \citenamefont {{J. Collings}}, \citenamefont {Li},\ and\ \citenamefont
  {Nguyen}}]{GreletCollingsLiNguyen:2001}%
  \BibitemOpen
  \bibfield  {author} {\bibinfo {author} {\bibfnamefont {E.}~\bibnamefont
  {Grelet}}, \bibinfo {author} {\bibfnamefont {P.}~\bibnamefont {{J.
  Collings}}}, \bibinfo {author} {\bibfnamefont {M.-H.}\ \bibnamefont {Li}}, \
  and\ \bibinfo {author} {\bibfnamefont {H.-T.}\ \bibnamefont {Nguyen}},\
  }\href@noop {} {\bibfield  {journal} {\bibinfo  {journal} {The European
  Physical Journal E}\ }\textbf {\bibinfo {volume} {6}},\ \bibinfo {pages}
  {157} (\bibinfo {year} {2001})}\BibitemShut {NoStop}%
\bibitem [{\citenamefont {Bensimon}\ \emph {et~al.}(1983)\citenamefont
  {Bensimon}, \citenamefont {Domany},\ and\ \citenamefont
  {Shtrikman}}]{BensimonDomanyShtrikman:1983}%
  \BibitemOpen
  \bibfield  {author} {\bibinfo {author} {\bibfnamefont {D.}~\bibnamefont
  {Bensimon}}, \bibinfo {author} {\bibfnamefont {E.}~\bibnamefont {Domany}}, \
  and\ \bibinfo {author} {\bibfnamefont {S.}~\bibnamefont {Shtrikman}},\
  }\href@noop {} {\bibfield  {journal} {\bibinfo  {journal} {Phys. Rev. A}\
  }\textbf {\bibinfo {volume} {28}},\ \bibinfo {pages} {427} (\bibinfo {year}
  {1983})}\BibitemShut {NoStop}%
\bibitem [{\citenamefont {Michelson}(1911)}]{Michelson:1911}%
  \BibitemOpen
  \bibfield  {author} {\bibinfo {author} {\bibfnamefont {A.}~\bibnamefont
  {Michelson}},\ }\href@noop {} {\bibfield  {journal} {\bibinfo  {journal}
  {Philosophical Magazine Series 6}\ }\textbf {\bibinfo {volume} {21}},\
  \bibinfo {pages} {554} (\bibinfo {year} {1911})}\BibitemShut {NoStop}%
\bibitem [{\citenamefont {Jewell}\ \emph {et~al.}(2007)\citenamefont {Jewell},
  \citenamefont {Vukusic},\ and\ \citenamefont
  {Roberts}}]{JewellVukusicRoberts:2007}%
  \BibitemOpen
  \bibfield  {author} {\bibinfo {author} {\bibfnamefont {S.~A.}\ \bibnamefont
  {Jewell}}, \bibinfo {author} {\bibfnamefont {P.}~\bibnamefont {Vukusic}}, \
  and\ \bibinfo {author} {\bibfnamefont {N.~W.}\ \bibnamefont {Roberts}},\
  }\href@noop {} {\bibfield  {journal} {\bibinfo  {journal} {New Journal of
  Physics}\ }\textbf {\bibinfo {volume} {9}},\ \bibinfo {pages} {99} (\bibinfo
  {year} {2007})}\BibitemShut {NoStop}%
\bibitem [{\citenamefont {Sharma}\ \emph {et~al.}(2009)\citenamefont {Sharma},
  \citenamefont {Crne}, \citenamefont {Park},\ and\ \citenamefont
  {Srinivasarao}}]{SharmaCrneParkSrinivasarao:2009}%
  \BibitemOpen
  \bibfield  {author} {\bibinfo {author} {\bibfnamefont {V.}~\bibnamefont
  {Sharma}}, \bibinfo {author} {\bibfnamefont {M.}~\bibnamefont {Crne}},
  \bibinfo {author} {\bibfnamefont {J.}~\bibnamefont {Park}}, \ and\ \bibinfo
  {author} {\bibfnamefont {M.}~\bibnamefont {Srinivasarao}},\ }\href@noop {}
  {\bibfield  {journal} {\bibinfo  {journal} {Science}\ }\textbf {\bibinfo
  {volume} {325}},\ \bibinfo {pages} {449} (\bibinfo {year}
  {2009})}\BibitemShut {NoStop}%
\bibitem [{\citenamefont {Kleinlogel}\ and\ \citenamefont
  {White}(2008)}]{KleinlogelWhite:2008}%
  \BibitemOpen
  \bibfield  {author} {\bibinfo {author} {\bibfnamefont {S.}~\bibnamefont
  {Kleinlogel}}\ and\ \bibinfo {author} {\bibfnamefont {A.~G.}\ \bibnamefont
  {White}},\ }\href {\doibase 10.1371/journal.pone.0002190} {\bibfield
  {journal} {\bibinfo  {journal} {PLoS ONE}\ }\textbf {\bibinfo {volume} {3}},\
  \bibinfo {pages} {e2190} (\bibinfo {year} {2008})}\BibitemShut {NoStop}%
\bibitem [{\citenamefont {Chiou}\ \emph {et~al.}(2008)\citenamefont {Chiou},
  \citenamefont {Kleinlogel}, \citenamefont {Cronin}, \citenamefont {Caldwell},
  \citenamefont {Loeffler}, \citenamefont {Siddiqi}, \citenamefont {Goldizen},\
  and\ \citenamefont
  {Marshall}}]{ChiouKleinlogelCroninCaldwellLoefflerSiddiqiGoldizenMarshal:2008}%
  \BibitemOpen
  \bibfield  {author} {\bibinfo {author} {\bibfnamefont {T.-H.}\ \bibnamefont
  {Chiou}}, \bibinfo {author} {\bibfnamefont {S.}~\bibnamefont {Kleinlogel}},
  \bibinfo {author} {\bibfnamefont {T.}~\bibnamefont {Cronin}}, \bibinfo
  {author} {\bibfnamefont {R.}~\bibnamefont {Caldwell}}, \bibinfo {author}
  {\bibfnamefont {B.}~\bibnamefont {Loeffler}}, \bibinfo {author}
  {\bibfnamefont {A.}~\bibnamefont {Siddiqi}}, \bibinfo {author} {\bibfnamefont
  {A.}~\bibnamefont {Goldizen}}, \ and\ \bibinfo {author} {\bibfnamefont
  {J.}~\bibnamefont {Marshall}},\ }\href@noop {} {\bibfield  {journal}
  {\bibinfo  {journal} {Current Biology}\ }\textbf {\bibinfo {volume} {18}},\
  \bibinfo {pages} {429 } (\bibinfo {year} {2008})}\BibitemShut {NoStop}%
\bibitem [{\citenamefont {Saba}\ \emph {et~al.}(2011)\citenamefont {Saba},
  \citenamefont {Thiel}, \citenamefont {Turner}, \citenamefont {Hyde},
  \citenamefont {Gu}, \citenamefont {Grosse-Brauckmann}, \citenamefont
  {Neshev}, \citenamefont {Mecke},\ and\ \citenamefont
  {Schr{\"o}der-Turk}}]{SabaThielTurnerHydeGuBrauckmannNeshevMeckeSchroederTurk:2011}%
  \BibitemOpen
  \bibfield  {author} {\bibinfo {author} {\bibfnamefont {M.}~\bibnamefont
  {Saba}}, \bibinfo {author} {\bibfnamefont {M.}~\bibnamefont {Thiel}},
  \bibinfo {author} {\bibfnamefont {M.}~\bibnamefont {Turner}}, \bibinfo
  {author} {\bibfnamefont {S.~T.}\ \bibnamefont {Hyde}}, \bibinfo {author}
  {\bibfnamefont {M.}~\bibnamefont {Gu}}, \bibinfo {author} {\bibfnamefont
  {K.}~\bibnamefont {Grosse-Brauckmann}}, \bibinfo {author} {\bibfnamefont
  {D.}~\bibnamefont {Neshev}}, \bibinfo {author} {\bibfnamefont
  {K.}~\bibnamefont {Mecke}}, \ and\ \bibinfo {author} {\bibfnamefont {G.~E.}\
  \bibnamefont {Schr{\"o}der-Turk}},\ }\href@noop {} {\bibfield  {journal}
  {\bibinfo  {journal} {Phys.~Rev.~Lett.}\ }\textbf {\bibinfo {volume} {106}},\
  \bibinfo {pages} {103902} (\bibinfo {year} {2011})}\BibitemShut {NoStop}%
\bibitem [{\citenamefont {Vignolini}\ \emph {et~al.}(2012)\citenamefont
  {Vignolini}, \citenamefont {Rudall}, \citenamefont {Rowland}, \citenamefont
  {Reed}, \citenamefont {Moyroud}, \citenamefont {Faden}, \citenamefont
  {Baumberg}, \citenamefont {Glover},\ and\ \citenamefont
  {Steiner}}]{VignoliniRudallRowlandReedMoyroudFadenBaumbergGloverSteiner:2012}%
  \BibitemOpen
  \bibfield  {author} {\bibinfo {author} {\bibfnamefont {S.}~\bibnamefont
  {Vignolini}}, \bibinfo {author} {\bibfnamefont {P.~J.}\ \bibnamefont
  {Rudall}}, \bibinfo {author} {\bibfnamefont {A.~V.}\ \bibnamefont {Rowland}},
  \bibinfo {author} {\bibfnamefont {A.}~\bibnamefont {Reed}}, \bibinfo {author}
  {\bibfnamefont {E.}~\bibnamefont {Moyroud}}, \bibinfo {author} {\bibfnamefont
  {R.~B.}\ \bibnamefont {Faden}}, \bibinfo {author} {\bibfnamefont {J.~J.}\
  \bibnamefont {Baumberg}}, \bibinfo {author} {\bibfnamefont {B.~J.}\
  \bibnamefont {Glover}}, \ and\ \bibinfo {author} {\bibfnamefont
  {U.}~\bibnamefont {Steiner}},\ }\href@noop {} {\bibfield  {journal} {\bibinfo
   {journal} {Proceedings of the National Academy of Sciences}\ } (\bibinfo
  {year} {2012})}\BibitemShut {NoStop}%
\bibitem [{\citenamefont {Thiel}\ \emph {et~al.}(2009)\citenamefont {Thiel},
  \citenamefont {Rill}, \citenamefont {{von Freymann}},\ and\ \citenamefont
  {Wegener}}]{ThielRillVonFreymannWegener:2009}%
  \BibitemOpen
  \bibfield  {author} {\bibinfo {author} {\bibfnamefont {M.}~\bibnamefont
  {Thiel}}, \bibinfo {author} {\bibfnamefont {M.}~\bibnamefont {Rill}},
  \bibinfo {author} {\bibfnamefont {G.}~\bibnamefont {{von Freymann}}}, \ and\
  \bibinfo {author} {\bibfnamefont {M.}~\bibnamefont {Wegener}},\ }\href@noop
  {} {\bibfield  {journal} {\bibinfo  {journal} {Advanced Materials}\ }\textbf
  {\bibinfo {volume} {21}},\ \bibinfo {pages} {4680} (\bibinfo {year}
  {2009})}\BibitemShut {NoStop}%
\bibitem [{\citenamefont {Turner}\ \emph {et~al.}(2011)\citenamefont {Turner},
  \citenamefont {Schr{\"o}der-Turk},\ and\ \citenamefont
  {Gu}}]{TurnerSchroederTurkGu:2011}%
  \BibitemOpen
  \bibfield  {author} {\bibinfo {author} {\bibfnamefont {M.~D.}\ \bibnamefont
  {Turner}}, \bibinfo {author} {\bibfnamefont {G.~E.}\ \bibnamefont
  {Schr{\"o}der-Turk}}, \ and\ \bibinfo {author} {\bibfnamefont
  {M.}~\bibnamefont {Gu}},\ }\href {\doibase 10.1364/OE.19.010001} {\bibfield
  {journal} {\bibinfo  {journal} {Optics Express}\ }\textbf {\bibinfo {volume}
  {19}},\ \bibinfo {pages} {10001} (\bibinfo {year} {2011})}\BibitemShut
  {NoStop}%
\bibitem [{\citenamefont {Turner}\ \emph {et~al.}(2013)\citenamefont {Turner},
  \citenamefont {Saba}, \citenamefont {Zhang}, \citenamefont {Cumming},
  \citenamefont {Schr\"oder-Turk},\ and\ \citenamefont
  {Gu}}]{TurnerSabaZhangCummingSchroederTurkGu:2013}%
  \BibitemOpen
  \bibfield  {author} {\bibinfo {author} {\bibfnamefont {M.}~\bibnamefont
  {Turner}}, \bibinfo {author} {\bibfnamefont {M.}~\bibnamefont {Saba}},
  \bibinfo {author} {\bibfnamefont {Q.}~\bibnamefont {Zhang}}, \bibinfo
  {author} {\bibfnamefont {B.}~\bibnamefont {Cumming}}, \bibinfo {author}
  {\bibfnamefont {G.}~\bibnamefont {Schr\"oder-Turk}}, \ and\ \bibinfo {author}
  {\bibfnamefont {M.}~\bibnamefont {Gu}},\ }\href@noop {} {\bibfield  {journal}
  {\bibinfo  {journal} {Nature Photonics}\ }\textbf {\bibinfo {volume} {in
  press}} (\bibinfo {year} {2013})}\BibitemShut {NoStop}%
\bibitem [{\citenamefont {Decker}\ \emph {et~al.}(2009)\citenamefont {Decker},
  \citenamefont {Ruther}, \citenamefont {Kriegler}, \citenamefont {Zhou},
  \citenamefont {Soukoulis}, \citenamefont {Linden},\ and\ \citenamefont
  {Wegener}}]{DeckerRutherKrieglerZhouSoukoulisLindenWegener:2009}%
  \BibitemOpen
  \bibfield  {author} {\bibinfo {author} {\bibfnamefont {M.}~\bibnamefont
  {Decker}}, \bibinfo {author} {\bibfnamefont {M.}~\bibnamefont {Ruther}},
  \bibinfo {author} {\bibfnamefont {C.~E.}\ \bibnamefont {Kriegler}}, \bibinfo
  {author} {\bibfnamefont {J.}~\bibnamefont {Zhou}}, \bibinfo {author}
  {\bibfnamefont {C.~M.}\ \bibnamefont {Soukoulis}}, \bibinfo {author}
  {\bibfnamefont {S.}~\bibnamefont {Linden}}, \ and\ \bibinfo {author}
  {\bibfnamefont {M.}~\bibnamefont {Wegener}},\ }\href@noop {} {\bibfield
  {journal} {\bibinfo  {journal} {Opt. Lett.}\ }\textbf {\bibinfo {volume}
  {34}},\ \bibinfo {pages} {2501} (\bibinfo {year} {2009})}\BibitemShut
  {NoStop}%
\bibitem [{\citenamefont {Gansel}\ \emph {et~al.}(2009)\citenamefont {Gansel},
  \citenamefont {Thiel}, \citenamefont {Rill}, \citenamefont {Decker},
  \citenamefont {Bade}, \citenamefont {Saile}, \citenamefont {{von Freymann}},
  \citenamefont {Linden},\ and\ \citenamefont
  {Wegener}}]{GanselThielRillDeckerBadeSaileVonFreymannLindenWegener:2009}%
  \BibitemOpen
  \bibfield  {author} {\bibinfo {author} {\bibfnamefont {J.}~\bibnamefont
  {Gansel}}, \bibinfo {author} {\bibfnamefont {M.}~\bibnamefont {Thiel}},
  \bibinfo {author} {\bibfnamefont {M.}~\bibnamefont {Rill}}, \bibinfo {author}
  {\bibfnamefont {M.}~\bibnamefont {Decker}}, \bibinfo {author} {\bibfnamefont
  {K.}~\bibnamefont {Bade}}, \bibinfo {author} {\bibfnamefont {V.}~\bibnamefont
  {Saile}}, \bibinfo {author} {\bibfnamefont {G.}~\bibnamefont {{von
  Freymann}}}, \bibinfo {author} {\bibfnamefont {S.}~\bibnamefont {Linden}}, \
  and\ \bibinfo {author} {\bibfnamefont {M.}~\bibnamefont {Wegener}},\
  }\href@noop {} {\bibfield  {journal} {\bibinfo  {journal} {Science}\ }\textbf
  {\bibinfo {volume} {325}},\ \bibinfo {pages} {1513} (\bibinfo {year}
  {2009})}\BibitemShut {NoStop}%
\bibitem [{\citenamefont {Kuwata-Gonokami}\ \emph {et~al.}(2005)\citenamefont
  {Kuwata-Gonokami}, \citenamefont {Saito}, \citenamefont {Ino}, \citenamefont
  {Kauranen}, \citenamefont {Jefimovs}, \citenamefont {Vallius}, \citenamefont
  {Turunen},\ and\ \citenamefont
  {Svirko}}]{KuwataGonokamiSaitoInoKauranenJefimovsValliusTurunenSvirko:2005}%
  \BibitemOpen
  \bibfield  {author} {\bibinfo {author} {\bibfnamefont {M.}~\bibnamefont
  {Kuwata-Gonokami}}, \bibinfo {author} {\bibfnamefont {N.}~\bibnamefont
  {Saito}}, \bibinfo {author} {\bibfnamefont {Y.}~\bibnamefont {Ino}}, \bibinfo
  {author} {\bibfnamefont {M.}~\bibnamefont {Kauranen}}, \bibinfo {author}
  {\bibfnamefont {K.}~\bibnamefont {Jefimovs}}, \bibinfo {author}
  {\bibfnamefont {T.}~\bibnamefont {Vallius}}, \bibinfo {author} {\bibfnamefont
  {J.}~\bibnamefont {Turunen}}, \ and\ \bibinfo {author} {\bibfnamefont
  {Y.}~\bibnamefont {Svirko}},\ }\href@noop {} {\bibfield  {journal} {\bibinfo
  {journal} {Phys. Rev. Lett.}\ }\textbf {\bibinfo {volume} {95}},\ \bibinfo
  {pages} {227401} (\bibinfo {year} {2005})}\BibitemShut {NoStop}%
\bibitem [{\citenamefont {Decker}\ \emph {et~al.}(2007)\citenamefont {Decker},
  \citenamefont {Klein}, \citenamefont {Wegener},\ and\ \citenamefont
  {Linden}}]{DeckerKleinWegenerLinden:2007}%
  \BibitemOpen
  \bibfield  {author} {\bibinfo {author} {\bibfnamefont {M.}~\bibnamefont
  {Decker}}, \bibinfo {author} {\bibfnamefont {M.~W.}\ \bibnamefont {Klein}},
  \bibinfo {author} {\bibfnamefont {M.}~\bibnamefont {Wegener}}, \ and\
  \bibinfo {author} {\bibfnamefont {S.}~\bibnamefont {Linden}},\ }\href@noop {}
  {\bibfield  {journal} {\bibinfo  {journal} {Opt. Lett.}\ }\textbf {\bibinfo
  {volume} {32}},\ \bibinfo {pages} {856} (\bibinfo {year} {2007})}\BibitemShut
  {NoStop}%
\bibitem [{\citenamefont {Zhang}\ \emph {et~al.}(2009)\citenamefont {Zhang},
  \citenamefont {Park}, \citenamefont {Li}, \citenamefont {Lu}, \citenamefont
  {Zhang},\ and\ \citenamefont {Zhang}}]{ZhangParkLiLuZhangZhang:2009}%
  \BibitemOpen
  \bibfield  {author} {\bibinfo {author} {\bibfnamefont {S.}~\bibnamefont
  {Zhang}}, \bibinfo {author} {\bibfnamefont {Y.-S.}\ \bibnamefont {Park}},
  \bibinfo {author} {\bibfnamefont {J.}~\bibnamefont {Li}}, \bibinfo {author}
  {\bibfnamefont {X.}~\bibnamefont {Lu}}, \bibinfo {author} {\bibfnamefont
  {W.}~\bibnamefont {Zhang}}, \ and\ \bibinfo {author} {\bibfnamefont
  {X.}~\bibnamefont {Zhang}},\ }\href@noop {} {\bibfield  {journal} {\bibinfo
  {journal} {Phys. Rev. Lett.}\ }\textbf {\bibinfo {volume} {102}},\ \bibinfo
  {pages} {023901} (\bibinfo {year} {2009})}\BibitemShut {NoStop}%
\bibitem [{\citenamefont {Plum}\ \emph {et~al.}()\citenamefont {Plum},
  \citenamefont {Zhou}, \citenamefont {Dong}, \citenamefont {Fedotov},
  \citenamefont {Koschny}, \citenamefont {Soukoulis},\ and\ \citenamefont
  {Zheludev}}]{PlumZhouDongFedotovKoschnySoukoulisZheludev:2009}%
  \BibitemOpen
  \bibfield  {author} {\bibinfo {author} {\bibfnamefont {E.}~\bibnamefont
  {Plum}}, \bibinfo {author} {\bibfnamefont {J.}~\bibnamefont {Zhou}}, \bibinfo
  {author} {\bibfnamefont {J.}~\bibnamefont {Dong}}, \bibinfo {author}
  {\bibfnamefont {V.~A.}\ \bibnamefont {Fedotov}}, \bibinfo {author}
  {\bibfnamefont {T.}~\bibnamefont {Koschny}}, \bibinfo {author} {\bibfnamefont
  {C.~M.}\ \bibnamefont {Soukoulis}}, \ and\ \bibinfo {author} {\bibfnamefont
  {N.~I.}\ \bibnamefont {Zheludev}},\ }\href@noop {} {\bibfield  {journal}
  {\bibinfo  {journal} {Phys. Rev. B}\ }\textbf {\bibinfo {volume} {79}},\
  \bibinfo {pages} {035407}}\BibitemShut {NoStop}%
\bibitem [{\citenamefont {Pendry}(2004)}]{Pendry:2004}%
  \BibitemOpen
  \bibfield  {author} {\bibinfo {author} {\bibfnamefont {J.}~\bibnamefont
  {Pendry}},\ }\href@noop {} {\bibfield  {journal} {\bibinfo  {journal}
  {Science}\ }\textbf {\bibinfo {volume} {306}},\ \bibinfo {pages} {1353}
  (\bibinfo {year} {2004})}\BibitemShut {NoStop}%
\bibitem [{\citenamefont {Liu}\ \emph {et~al.}(2010)\citenamefont {Liu},
  \citenamefont {Zentgraf}, \citenamefont {Liu}, \citenamefont {Bartal},\ and\
  \citenamefont {Zhang}}]{LiuZentgrafLiuBartalZhang:2010}%
  \BibitemOpen
  \bibfield  {author} {\bibinfo {author} {\bibfnamefont {M.}~\bibnamefont
  {Liu}}, \bibinfo {author} {\bibfnamefont {T.}~\bibnamefont {Zentgraf}},
  \bibinfo {author} {\bibfnamefont {Y.}~\bibnamefont {Liu}}, \bibinfo {author}
  {\bibfnamefont {G.}~\bibnamefont {Bartal}}, \ and\ \bibinfo {author}
  {\bibfnamefont {X.}~\bibnamefont {Zhang}},\ }\href {\doibase
  http://dx.doi.org/10.1038/nnano.2010.128} {\bibfield  {journal} {\bibinfo
  {journal} {Nat Nano}\ }\textbf {\bibinfo {volume} {5}},\ \bibinfo {pages}
  {570} (\bibinfo {year} {2010})},\ \bibinfo {note}
  {10.1038/nnano.2010.128}\BibitemShut {NoStop}%
\bibitem [{\citenamefont {{Zhang Shuang}}\ \emph {et~al.}(2012)\citenamefont
  {{Zhang Shuang}}, \citenamefont {{Zhou Jiangfeng}}, \citenamefont {{Park
  Yong-Shik}}, \citenamefont {{Rho Junsuk}}, \citenamefont {{Singh Ranjan}},
  \citenamefont {{Nam Sunghyun}}, \citenamefont {{Azad Abul K.}}, \citenamefont
  {{Chen Hou-Tong}}, \citenamefont {{Yin Xiaobo}}, \citenamefont {{Taylor
  Antoinette J.}},\ and\ \citenamefont {{Zhang
  Xiang}}}]{ZhangZhouParkRhoSinghNamAzadChenYinTaylorZhang:2012}%
  \BibitemOpen
  \bibfield  {author} {\bibinfo {author} {\bibnamefont {{Zhang Shuang}}},
  \bibinfo {author} {\bibnamefont {{Zhou Jiangfeng}}}, \bibinfo {author}
  {\bibnamefont {{Park Yong-Shik}}}, \bibinfo {author} {\bibnamefont {{Rho
  Junsuk}}}, \bibinfo {author} {\bibnamefont {{Singh Ranjan}}}, \bibinfo
  {author} {\bibnamefont {{Nam Sunghyun}}}, \bibinfo {author} {\bibnamefont
  {{Azad Abul K.}}}, \bibinfo {author} {\bibnamefont {{Chen Hou-Tong}}},
  \bibinfo {author} {\bibnamefont {{Yin Xiaobo}}}, \bibinfo {author}
  {\bibnamefont {{Taylor Antoinette J.}}}, \ and\ \bibinfo {author}
  {\bibnamefont {{Zhang Xiang}}},\ }\href {\doibase
  http://dx.doi.org/10.1038/ncomms1908} {\bibfield  {journal} {\bibinfo
  {journal} {Nat Commun}\ }\textbf {\bibinfo {volume} {3}},\ \bibinfo {pages}
  {942} (\bibinfo {year} {2012})},\ \bibinfo {note}
  {10.1038/ncomms1908}\BibitemShut {NoStop}%
\bibitem [{\citenamefont {Ren}\ \emph {et~al.}(2012)\citenamefont {Ren},
  \citenamefont {Plum}, \citenamefont {Xu},\ and\ \citenamefont
  {Zheludev}}]{RenPlumXuZheludev:2012}%
  \BibitemOpen
  \bibfield  {author} {\bibinfo {author} {\bibfnamefont {M.}~\bibnamefont
  {Ren}}, \bibinfo {author} {\bibfnamefont {E.}~\bibnamefont {Plum}}, \bibinfo
  {author} {\bibfnamefont {J.}~\bibnamefont {Xu}}, \ and\ \bibinfo {author}
  {\bibfnamefont {N.~I.}\ \bibnamefont {Zheludev}},\ }\href {\doibase
  http://dx.doi.org/10.1038/ncomms1805} {\bibfield  {journal} {\bibinfo
  {journal} {Nat Commun}\ }\textbf {\bibinfo {volume} {3}},\ \bibinfo {pages}
  {833} (\bibinfo {year} {2012})},\ \bibinfo {note}
  {10.1038/ncomms1805}\BibitemShut {NoStop}%
\bibitem [{\citenamefont {Sch\"aferling}\ \emph {et~al.}(2012)\citenamefont
  {Sch\"aferling}, \citenamefont {Dregely}, \citenamefont {Hentschel},\ and\
  \citenamefont {Giessen}}]{SchaeferlingDregelyHentschelGiessen:2012}%
  \BibitemOpen
  \bibfield  {author} {\bibinfo {author} {\bibfnamefont {M.}~\bibnamefont
  {Sch\"aferling}}, \bibinfo {author} {\bibfnamefont {D.}~\bibnamefont
  {Dregely}}, \bibinfo {author} {\bibfnamefont {M.}~\bibnamefont {Hentschel}},
  \ and\ \bibinfo {author} {\bibfnamefont {H.}~\bibnamefont {Giessen}},\
  }\href@noop {} {\bibfield  {journal} {\bibinfo  {journal} {Phys. Rev. X}\
  }\textbf {\bibinfo {volume} {2}},\ \bibinfo {pages} {031010} (\bibinfo {year}
  {2012})}\BibitemShut {NoStop}%
\bibitem [{\citenamefont
  {Hahn}(1992)}]{InternationalTablesForCrystallography:1992}%
  \BibitemOpen
  \bibinfo {editor} {\bibfnamefont {T.}~\bibnamefont {Hahn}},\ ed.,\ \href@noop
  {} {\emph {\bibinfo {title} {International Tables For Crystallography}}}\
  (\bibinfo  {publisher} {Kluwer Academic Publishers},\ \bibinfo {address}
  {Dordrecht},\ \bibinfo {year} {1992})\BibitemShut {NoStop}%
\bibitem [{\citenamefont {Joannopoulos}\ \emph {et~al.}(2008)\citenamefont
  {Joannopoulos}, \citenamefont {Johnson}, \citenamefont {Winn},\ and\
  \citenamefont {Meade}}]{JoannopoulosJohnsonWinnMeade:2008}%
  \BibitemOpen
  \bibfield  {author} {\bibinfo {author} {\bibfnamefont {J.}~\bibnamefont
  {Joannopoulos}}, \bibinfo {author} {\bibfnamefont {S.}~\bibnamefont
  {Johnson}}, \bibinfo {author} {\bibfnamefont {J.}~\bibnamefont {Winn}}, \
  and\ \bibinfo {author} {\bibfnamefont {R.}~\bibnamefont {Meade}},\
  }\href@noop {} {\emph {\bibinfo {title} {Photonic Crystals: Molding the flow
  of light}}}\ (\bibinfo  {publisher} {Princeton University Press},\ \bibinfo
  {address} {Princeton},\ \bibinfo {year} {2008})\BibitemShut {NoStop}%
\bibitem [{\citenamefont {Kaschke}\ \emph {et~al.}(2012)\citenamefont
  {Kaschke}, \citenamefont {Gansel},\ and\ \citenamefont
  {Wegener}}]{KaschkeGanselWegener:2012}%
  \BibitemOpen
  \bibfield  {author} {\bibinfo {author} {\bibfnamefont {J.}~\bibnamefont
  {Kaschke}}, \bibinfo {author} {\bibfnamefont {J.~K.}\ \bibnamefont {Gansel}},
  \ and\ \bibinfo {author} {\bibfnamefont {M.}~\bibnamefont {Wegener}},\ }\href
  {\doibase 10.1364/OE.20.026012} {\bibfield  {journal} {\bibinfo  {journal}
  {Opt. Express}\ }\textbf {\bibinfo {volume} {20}},\ \bibinfo {pages} {26012}
  (\bibinfo {year} {2012})}\BibitemShut {NoStop}%
\bibitem [{\citenamefont {Bai}\ \emph {et~al.}(2007)\citenamefont {Bai},
  \citenamefont {Svirko}, \citenamefont {Turunen},\ and\ \citenamefont
  {Vallius}}]{BaiSvirkoTurunenVallius:2007}%
  \BibitemOpen
  \bibfield  {author} {\bibinfo {author} {\bibfnamefont {B.}~\bibnamefont
  {Bai}}, \bibinfo {author} {\bibfnamefont {Y.}~\bibnamefont {Svirko}},
  \bibinfo {author} {\bibfnamefont {J.}~\bibnamefont {Turunen}}, \ and\
  \bibinfo {author} {\bibfnamefont {T.}~\bibnamefont {Vallius}},\ }\href@noop
  {} {\bibfield  {journal} {\bibinfo  {journal} {Phys. Rev. A}\ }\textbf
  {\bibinfo {volume} {76}},\ \bibinfo {pages} {023811} (\bibinfo {year}
  {2007})}\BibitemShut {NoStop}%
\bibitem [{\citenamefont {Bai}\ \emph {et~al.}(2012)\citenamefont {Bai},
  \citenamefont {Ventola}, \citenamefont {Tervo},\ and\ \citenamefont
  {Zhang}}]{BaiVentolaTervoZhang:2012}%
  \BibitemOpen
  \bibfield  {author} {\bibinfo {author} {\bibfnamefont {B.}~\bibnamefont
  {Bai}}, \bibinfo {author} {\bibfnamefont {K.}~\bibnamefont {Ventola}},
  \bibinfo {author} {\bibfnamefont {J.}~\bibnamefont {Tervo}}, \ and\ \bibinfo
  {author} {\bibfnamefont {Y.}~\bibnamefont {Zhang}},\ }\href@noop {}
  {\bibfield  {journal} {\bibinfo  {journal} {Phys. Rev. A}\ }\textbf {\bibinfo
  {volume} {85}},\ \bibinfo {pages} {053808} (\bibinfo {year}
  {2012})}\BibitemShut {NoStop}%
\bibitem [{\citenamefont {Menzel}\ \emph {et~al.}(2010)\citenamefont {Menzel},
  \citenamefont {Rockstuhl},\ and\ \citenamefont
  {Lederer}}]{MenzelRockstuhlLederer:2010}%
  \BibitemOpen
  \bibfield  {author} {\bibinfo {author} {\bibfnamefont {C.}~\bibnamefont
  {Menzel}}, \bibinfo {author} {\bibfnamefont {C.}~\bibnamefont {Rockstuhl}}, \
  and\ \bibinfo {author} {\bibfnamefont {F.}~\bibnamefont {Lederer}},\
  }\href@noop {} {\bibfield  {journal} {\bibinfo  {journal} {Phys. Rev. A}\
  }\textbf {\bibinfo {volume} {82}},\ \bibinfo {pages} {053811} (\bibinfo
  {year} {2010})}\BibitemShut {NoStop}%
\bibitem [{\citenamefont {Rockstuhl}\ \emph {et~al.}(2009)\citenamefont
  {Rockstuhl}, \citenamefont {Menzel}, \citenamefont {Paul},\ and\
  \citenamefont {Lederer}}]{RockstuhlMenzelPaulLederer:2009}%
  \BibitemOpen
  \bibfield  {author} {\bibinfo {author} {\bibfnamefont {C.}~\bibnamefont
  {Rockstuhl}}, \bibinfo {author} {\bibfnamefont {C.}~\bibnamefont {Menzel}},
  \bibinfo {author} {\bibfnamefont {T.}~\bibnamefont {Paul}}, \ and\ \bibinfo
  {author} {\bibfnamefont {F.}~\bibnamefont {Lederer}},\ }\href@noop {}
  {\bibfield  {journal} {\bibinfo  {journal} {Phys. Rev. B}\ }\textbf {\bibinfo
  {volume} {79}},\ \bibinfo {pages} {035321} (\bibinfo {year}
  {2009})}\BibitemShut {NoStop}%
\bibitem [{\citenamefont {Maslovski}\ \emph {et~al.}(2009)\citenamefont
  {Maslovski}, \citenamefont {Morits},\ and\ \citenamefont
  {Tretyakov}}]{MaslovskiMoritsTretyakov:2009}%
  \BibitemOpen
  \bibfield  {author} {\bibinfo {author} {\bibfnamefont {S.~I.}\ \bibnamefont
  {Maslovski}}, \bibinfo {author} {\bibfnamefont {D.~K.}\ \bibnamefont
  {Morits}}, \ and\ \bibinfo {author} {\bibfnamefont {S.~A.}\ \bibnamefont
  {Tretyakov}},\ }\href@noop {} {\bibfield  {journal} {\bibinfo  {journal}
  {Journal of Optics A: Pure and Applied Optics}\ }\textbf {\bibinfo {volume}
  {11}},\ \bibinfo {pages} {074004} (\bibinfo {year} {2009})}\BibitemShut
  {NoStop}%
\bibitem [{\citenamefont {Dmitriev}(2011)}]{Dmitriev:2011}%
  \BibitemOpen
  \bibfield  {author} {\bibinfo {author} {\bibfnamefont {V.}~\bibnamefont
  {Dmitriev}},\ }\href@noop {} {\bibfield  {journal} {\bibinfo  {journal}
  {Metamaterials}\ }\textbf {\bibinfo {volume} {5}},\ \bibinfo {pages} {141 }
  (\bibinfo {year} {2011})}\BibitemShut {NoStop}%
\bibitem [{\citenamefont {Dmitriev}(2013)}]{Dmitriev:2013}%
  \BibitemOpen
  \bibfield  {author} {\bibinfo {author} {\bibfnamefont {V.}~\bibnamefont
  {Dmitriev}},\ }\href@noop {} {\bibfield  {journal} {\bibinfo  {journal}
  {Antennas and Propagation, IEEE Transactions on}\ }\textbf {\bibinfo {volume}
  {61}},\ \bibinfo {pages} {185} (\bibinfo {year} {2013})}\BibitemShut
  {NoStop}%
\bibitem [{\citenamefont {Kwon}\ \emph {et~al.}(2008)\citenamefont {Kwon},
  \citenamefont {Werner},\ and\ \citenamefont
  {Werner}}]{KwonWernerWerner:2008}%
  \BibitemOpen
  \bibfield  {author} {\bibinfo {author} {\bibfnamefont {D.-H.}\ \bibnamefont
  {Kwon}}, \bibinfo {author} {\bibfnamefont {P.~L.}\ \bibnamefont {Werner}}, \
  and\ \bibinfo {author} {\bibfnamefont {D.~H.}\ \bibnamefont {Werner}},\
  }\href@noop {} {\bibfield  {journal} {\bibinfo  {journal} {Optics Express}\
  }\textbf {\bibinfo {volume} {16}},\ \bibinfo {pages} {11802} (\bibinfo {year}
  {2008})}\BibitemShut {NoStop}%
\bibitem [{\citenamefont {Dresselhaus}(2008)}]{Dresselhaus:2008}%
  \BibitemOpen
  \bibfield  {author} {\bibinfo {author} {\bibfnamefont {M.~S.}\ \bibnamefont
  {Dresselhaus}},\ }\href@noop {} {\emph {\bibinfo {title} {Group Theory:
  Application to the Physics of Condensed Matter}}}\ (\bibinfo  {publisher}
  {Springer-Verlag},\ \bibinfo {year} {2008})\BibitemShut {NoStop}%
\bibitem [{\citenamefont {Sakoda}(2005)}]{Sakoda:2005}%
  \BibitemOpen
  \bibfield  {author} {\bibinfo {author} {\bibfnamefont {K.}~\bibnamefont
  {Sakoda}},\ }\href@noop {} {\emph {\bibinfo {title} {Optical Properties of
  Photonic Crystals}}}\ (\bibinfo  {publisher} {Springer-Verlag},\ \bibinfo
  {year} {2005})\BibitemShut {NoStop}%
\bibitem [{\citenamefont {Hyde}\ \emph {et~al.}(2008)\citenamefont {Hyde},
  \citenamefont {{O'{K}eeffe}},\ and\ \citenamefont
  {Proserpio}}]{HydeOKeeffeProserpio:2008}%
  \BibitemOpen
  \bibfield  {author} {\bibinfo {author} {\bibfnamefont {S.~T.}\ \bibnamefont
  {Hyde}}, \bibinfo {author} {\bibfnamefont {M.}~\bibnamefont {{O'{K}eeffe}}},
  \ and\ \bibinfo {author} {\bibfnamefont {D.~M.}\ \bibnamefont {Proserpio}},\
  }\href@noop {} {\bibfield  {journal} {\bibinfo  {journal}
  {Angew.~Chem.~Int.~Ed.}\ }\textbf {\bibinfo {volume} {47}},\ \bibinfo {pages}
  {7996} (\bibinfo {year} {2008})}\BibitemShut {NoStop}%
\bibitem [{Note1()}]{Note1}%
  \BibitemOpen
  \bibinfo {note} {The network’s single type of 3-coordinated vertices are at
  Wyckhoff sites 8a (.32) in $I4_132$ (no. 214 in \cite
  {InternationalTablesForCrystallography:1992})}\BibitemShut {NoStop}%
\bibitem [{\citenamefont {Lu}\ \emph {et~al.}(2013)\citenamefont {Lu},
  \citenamefont {Fu}, \citenamefont {Joannopoulos},\ and\ \citenamefont
  {Solja\v{c}i\'c}}]{LuFuJoannopoulosSoljacic:2013}%
  \BibitemOpen
  \bibfield  {author} {\bibinfo {author} {\bibfnamefont {L.}~\bibnamefont
  {Lu}}, \bibinfo {author} {\bibfnamefont {L.}~\bibnamefont {Fu}}, \bibinfo
  {author} {\bibfnamefont {J.~D.}\ \bibnamefont {Joannopoulos}}, \ and\
  \bibinfo {author} {\bibfnamefont {M.}~\bibnamefont {Solja\v{c}i\'c}},\ }\href
  {\doibase 10.1038/NPHOTON.2013.42} {\bibfield  {journal} {\bibinfo  {journal}
  {Nature Photonics}\ }\textbf {\bibinfo {volume} {7}},\ \bibinfo {pages} {294}
  (\bibinfo {year} {2013})}\BibitemShut {NoStop}%
\bibitem [{\citenamefont {Oh}\ \emph {et~al.}(2013)\citenamefont {Oh},
  \citenamefont {Demetriadou}, \citenamefont {Wuestner},\ and\ \citenamefont
  {Hess}}]{OhDemetriadouWuestnerHess:2013}%
  \BibitemOpen
  \bibfield  {author} {\bibinfo {author} {\bibfnamefont {S.~S.}\ \bibnamefont
  {Oh}}, \bibinfo {author} {\bibfnamefont {A.}~\bibnamefont {Demetriadou}},
  \bibinfo {author} {\bibfnamefont {S.}~\bibnamefont {Wuestner}}, \ and\
  \bibinfo {author} {\bibfnamefont {O.}~\bibnamefont {Hess}},\ }\href@noop {}
  {\bibfield  {journal} {\bibinfo  {journal} {Advanced Materials}\ }\textbf
  {\bibinfo {volume} {25}},\ \bibinfo {pages} {612} (\bibinfo {year}
  {2013})}\BibitemShut {NoStop}%
\bibitem [{\citenamefont {Hur}\ \emph {et~al.}(2011)\citenamefont {Hur},
  \citenamefont {Francescato}, \citenamefont {Giannini}, \citenamefont {Maier},
  \citenamefont {Hennig},\ and\ \citenamefont
  {Wiesner}}]{HurFrancescatoGianniniMaierHennigWiesner:2011}%
  \BibitemOpen
  \bibfield  {author} {\bibinfo {author} {\bibfnamefont {K.}~\bibnamefont
  {Hur}}, \bibinfo {author} {\bibfnamefont {Y.}~\bibnamefont {Francescato}},
  \bibinfo {author} {\bibfnamefont {V.}~\bibnamefont {Giannini}}, \bibinfo
  {author} {\bibfnamefont {S.~A.}\ \bibnamefont {Maier}}, \bibinfo {author}
  {\bibfnamefont {R.~G.}\ \bibnamefont {Hennig}}, \ and\ \bibinfo {author}
  {\bibfnamefont {U.}~\bibnamefont {Wiesner}},\ }\href {\doibase
  10.1002/anie.201104888} {\bibfield  {journal} {\bibinfo  {journal}
  {Angew.~Chem.~Int.~Edit.}\ ,\ \bibinfo {pages} {n/a}} (\bibinfo {year}
  {2011})}\BibitemShut {NoStop}%
\bibitem [{\citenamefont {Michielsen}\ and\ \citenamefont
  {Stavenga}(2008)}]{MichielsenStavenga:2008}%
  \BibitemOpen
  \bibfield  {author} {\bibinfo {author} {\bibfnamefont {K.}~\bibnamefont
  {Michielsen}}\ and\ \bibinfo {author} {\bibfnamefont {D.}~\bibnamefont
  {Stavenga}},\ }\href@noop {} {\bibfield  {journal} {\bibinfo  {journal} {J.
  R. Soc. Interface}\ }\textbf {\bibinfo {volume} {5}},\ \bibinfo {pages} {85}
  (\bibinfo {year} {2008})}\BibitemShut {NoStop}%
\bibitem [{\citenamefont {Saranathan}\ \emph {et~al.}(2010)\citenamefont
  {Saranathan}, \citenamefont {Osuji}, \citenamefont {Mochrie}, \citenamefont
  {Noh}, \citenamefont {Narayanan}, \citenamefont {Sandy}, \citenamefont
  {Dufresne},\ and\ \citenamefont
  {Prum}}]{SaranathanOsujiMochrieNohNarayananSandyDufresnePrum:2010}%
  \BibitemOpen
  \bibfield  {author} {\bibinfo {author} {\bibfnamefont {V.}~\bibnamefont
  {Saranathan}}, \bibinfo {author} {\bibfnamefont {C.}~\bibnamefont {Osuji}},
  \bibinfo {author} {\bibfnamefont {S.}~\bibnamefont {Mochrie}}, \bibinfo
  {author} {\bibfnamefont {H.}~\bibnamefont {Noh}}, \bibinfo {author}
  {\bibfnamefont {S.}~\bibnamefont {Narayanan}}, \bibinfo {author}
  {\bibfnamefont {A.}~\bibnamefont {Sandy}}, \bibinfo {author} {\bibfnamefont
  {E.}~\bibnamefont {Dufresne}}, \ and\ \bibinfo {author} {\bibfnamefont
  {R.}~\bibnamefont {Prum}},\ }\href {\doibase 10.1073/pnas.0909616107}
  {\bibfield  {journal} {\bibinfo  {journal} {PNAS}\ }\textbf {\bibinfo
  {volume} {107}},\ \bibinfo {pages} {11676} (\bibinfo {year}
  {2010})}\BibitemShut {NoStop}%
\bibitem [{\citenamefont {Schr{\"o}der-Turk}\ \emph {et~al.}(2011)\citenamefont
  {Schr{\"o}der-Turk}, \citenamefont {Wickham}, \citenamefont {Averdunk},
  \citenamefont {Large}, \citenamefont {Poladian}, \citenamefont {Brink},
  \citenamefont {{Fitz Gerald}},\ and\ \citenamefont
  {Hyde}}]{SchroederTurkWickhamAverdunkBrinkFitzGeraldPoladianLargeHyde:2011}%
  \BibitemOpen
  \bibfield  {author} {\bibinfo {author} {\bibfnamefont {G.}~\bibnamefont
  {Schr{\"o}der-Turk}}, \bibinfo {author} {\bibfnamefont {S.}~\bibnamefont
  {Wickham}}, \bibinfo {author} {\bibfnamefont {H.}~\bibnamefont {Averdunk}},
  \bibinfo {author} {\bibfnamefont {M.}~\bibnamefont {Large}}, \bibinfo
  {author} {\bibfnamefont {L.}~\bibnamefont {Poladian}}, \bibinfo {author}
  {\bibfnamefont {F.}~\bibnamefont {Brink}}, \bibinfo {author} {\bibfnamefont
  {J.}~\bibnamefont {{Fitz Gerald}}}, \ and\ \bibinfo {author} {\bibfnamefont
  {S.~T.}\ \bibnamefont {Hyde}},\ }\href {\doibase
  doi:10.1016/j.jsb.2011.01.004} {\bibfield  {journal} {\bibinfo  {journal}
  {J.~Struct.~Biol.}\ }\textbf {\bibinfo {volume} {174}},\ \bibinfo {pages}
  {290} (\bibinfo {year} {2011})}\BibitemShut {NoStop}%
\bibitem [{\citenamefont {Vignolini}\ \emph {et~al.}(2011)\citenamefont
  {Vignolini}, \citenamefont {Yufa}, \citenamefont {Cunha}, \citenamefont
  {Guldin}, \citenamefont {Rushkin}, \citenamefont {Stefik}, \citenamefont
  {Hur}, \citenamefont {Wiesner}, \citenamefont {Baumberg},\ and\ \citenamefont
  {Steiner}}]{VignoliniYufaCunhaGuldinRushkinStefikHurWiesnerBaumbergSteiner:2011}%
  \BibitemOpen
  \bibfield  {author} {\bibinfo {author} {\bibfnamefont {S.}~\bibnamefont
  {Vignolini}}, \bibinfo {author} {\bibfnamefont {N.~A.}\ \bibnamefont {Yufa}},
  \bibinfo {author} {\bibfnamefont {P.~S.}\ \bibnamefont {Cunha}}, \bibinfo
  {author} {\bibfnamefont {S.}~\bibnamefont {Guldin}}, \bibinfo {author}
  {\bibfnamefont {I.}~\bibnamefont {Rushkin}}, \bibinfo {author} {\bibfnamefont
  {M.}~\bibnamefont {Stefik}}, \bibinfo {author} {\bibfnamefont
  {K.}~\bibnamefont {Hur}}, \bibinfo {author} {\bibfnamefont {U.}~\bibnamefont
  {Wiesner}}, \bibinfo {author} {\bibfnamefont {J.~J.}\ \bibnamefont
  {Baumberg}}, \ and\ \bibinfo {author} {\bibfnamefont {U.}~\bibnamefont
  {Steiner}},\ }\href@noop {} {\bibfield  {journal} {\bibinfo  {journal}
  {Advanced Materials}\ ,\ \bibinfo {pages} {n/a}} (\bibinfo {year}
  {2011})}\BibitemShut {NoStop}%
\bibitem [{\citenamefont {Mille}\ \emph {et~al.}(2013)\citenamefont {Mille},
  \citenamefont {Tyrode},\ and\ \citenamefont
  {Corkery}}]{MilleTyrodeCorkery:2013}%
  \BibitemOpen
  \bibfield  {author} {\bibinfo {author} {\bibfnamefont {C.}~\bibnamefont
  {Mille}}, \bibinfo {author} {\bibfnamefont {E.~C.}\ \bibnamefont {Tyrode}}, \
  and\ \bibinfo {author} {\bibfnamefont {R.~W.}\ \bibnamefont {Corkery}},\
  }\href@noop {} {\bibfield  {journal} {\bibinfo  {journal} {RSC Adv.}\
  }\textbf {\bibinfo {volume} {3}},\ \bibinfo {pages} {3109} (\bibinfo {year}
  {2013})}\BibitemShut {NoStop}%
\bibitem [{\citenamefont {Mille}\ \emph {et~al.}(2011)\citenamefont {Mille},
  \citenamefont {Tyrode},\ and\ \citenamefont
  {Corkery}}]{MilleTyrodeCorkery:2011}%
  \BibitemOpen
  \bibfield  {author} {\bibinfo {author} {\bibfnamefont {C.}~\bibnamefont
  {Mille}}, \bibinfo {author} {\bibfnamefont {E.~C.}\ \bibnamefont {Tyrode}}, \
  and\ \bibinfo {author} {\bibfnamefont {R.~W.}\ \bibnamefont {Corkery}},\
  }\href@noop {} {\bibfield  {journal} {\bibinfo  {journal} {Chem. Commun.}\
  }\textbf {\bibinfo {volume} {47}},\ \bibinfo {pages} {9873} (\bibinfo {year}
  {2011})}\BibitemShut {NoStop}%
\bibitem [{\citenamefont {Pouya}\ and\ \citenamefont
  {Vukusic}(2012)}]{PouyaVukusic:2012}%
  \BibitemOpen
  \bibfield  {author} {\bibinfo {author} {\bibfnamefont {C.}~\bibnamefont
  {Pouya}}\ and\ \bibinfo {author} {\bibfnamefont {P.}~\bibnamefont
  {Vukusic}},\ }\href {\doibase 10.1098/rsfs.2011.0091} {\bibfield  {journal}
  {\bibinfo  {journal} {Interface Focus}\ }\textbf {\bibinfo {volume} {2}},\
  \bibinfo {pages} {645} (\bibinfo {year} {2012})}\BibitemShut {NoStop}%
\bibitem [{\citenamefont {Hyde}\ and\ \citenamefont
  {Ramsden}(2000)}]{HydeRamsden:2000}%
  \BibitemOpen
  \bibfield  {author} {\bibinfo {author} {\bibfnamefont {S.~T.}\ \bibnamefont
  {Hyde}}\ and\ \bibinfo {author} {\bibfnamefont {S.}~\bibnamefont {Ramsden}},\
  }\href@noop {} {\bibfield  {journal} {\bibinfo  {journal} {Europhys.~Lett.}\
  }\textbf {\bibinfo {volume} {50}},\ \bibinfo {pages} {135} (\bibinfo {year}
  {2000})}\BibitemShut {NoStop}%
\bibitem [{\citenamefont {Hyde}\ and\ \citenamefont
  {Oguey}(2000)}]{HydeOguey:2000}%
  \BibitemOpen
  \bibfield  {author} {\bibinfo {author} {\bibfnamefont {S.~T.}\ \bibnamefont
  {Hyde}}\ and\ \bibinfo {author} {\bibfnamefont {C.}~\bibnamefont {Oguey}},\
  }\href@noop {} {\bibfield  {journal} {\bibinfo  {journal} {Eur.~Phys.~J.~B}\
  }\textbf {\bibinfo {volume} {16}},\ \bibinfo {pages} {613} (\bibinfo {year}
  {2000})}\BibitemShut {NoStop}%
\bibitem [{\citenamefont {Evans}\ \emph
  {et~al.}(2013{\natexlab{a}})\citenamefont {Evans}, \citenamefont {Robins},\
  and\ \citenamefont {Hyde}}]{EvansRobinsHyde:2013a}%
  \BibitemOpen
  \bibfield  {author} {\bibinfo {author} {\bibfnamefont {M.~E.}\ \bibnamefont
  {Evans}}, \bibinfo {author} {\bibfnamefont {V.}~\bibnamefont {Robins}}, \
  and\ \bibinfo {author} {\bibfnamefont {S.~T.}\ \bibnamefont {Hyde}},\
  }\href@noop {} {\bibfield  {journal} {\bibinfo  {journal} {Acta
  Crystallographica Section A}\ }\textbf {\bibinfo {volume} {69}},\ \bibinfo
  {pages} {241} (\bibinfo {year} {2013}{\natexlab{a}})}\BibitemShut {NoStop}%
\bibitem [{\citenamefont {Saba}\ \emph {et~al.}(2013)\citenamefont {Saba},
  \citenamefont {Turner}, \citenamefont {Mecke}, \citenamefont {Gu},\ and\
  \citenamefont {Schr{\"o}der-Turk}}]{SabaTurnerGuSchroeder_prl:2013}%
  \BibitemOpen
  \bibfield  {author} {\bibinfo {author} {\bibfnamefont {M.}~\bibnamefont
  {Saba}}, \bibinfo {author} {\bibfnamefont {M.}~\bibnamefont {Turner}},
  \bibinfo {author} {\bibfnamefont {K.}~\bibnamefont {Mecke}}, \bibinfo
  {author} {\bibfnamefont {M.}~\bibnamefont {Gu}}, \ and\ \bibinfo {author}
  {\bibfnamefont {G.~E.}\ \bibnamefont {Schr{\"o}der-Turk}},\ }\href@noop {}
  {\bibfield  {journal} {\bibinfo  {journal} {submitted for review to
  Phys.~Rev.~Lett}\ } (\bibinfo {year} {2013})}\BibitemShut {NoStop}%
\bibitem [{Note2()}]{Note2}%
  \BibitemOpen
  \bibinfo {note} {The BCC Wigner-Seitz cell forms a Kelvin body that is a cube
  truncated along its [100] directions to give a polytope with 14 facets, eight
  of which are regular hexagons and six of which are squares, also known as
  truncated octahedron. The use of the term {\protect \em Kelvin body} for this
  cell is motivated by it being Kelvin's proposition for the cell with least
  surface area (given fixed volume) that can tessellate space \cite
  {Thomson:1888}.}\BibitemShut {Stop}%
\bibitem [{Note3()}]{Note3}%
  \BibitemOpen
  \bibinfo {note} {See also the reticular chemistry structure resource
  {\protect \tt www.rscr.anu.edu.au} \cite {OKeeffePeskovRamsdenYaghi:2008} for
  details, where the {\protect \bf 2-srs}, the {\protect \bf 4-srs} and the
  {\protect \bf 8-srs} are denoted srs-c2*, srs-c4 and srs-c8,
  respectively.}\BibitemShut {Stop}%
\bibitem [{Note4()}]{Note4}%
  \BibitemOpen
  \bibinfo {note} {Transmission simulations are performed with the open source
  finite difference time domain package MEEP \cite
  {OskooiRoundyIbanescuBermelJoannopoulosJohnson:2010}. Simulations are for a
  slab of $53$ unit cells of {\protect \bf 8-srs} structure. Periodic boundary
  conditions are assumed in all three directions of the simulation box with a
  Yee grid of $64$ points per unit cell and a size of $1x1x61$ unit cells. A
  combination of respectively $2$ unit cells vacuo and perfectly matched layers
  on each side are used with a Gaussian source in vacuum on one side and energy
  flux measured on the other. Band structure frequencies and eigenmodes are
  calculated with the open source plane wave frequency domain eigensolver MPB
  \cite {JohnsonJoannopoulos:2001} with a $128^3$ structural resolution and a
  $32^3$ Fourier grid in the primitive body centered cubic basis.}\BibitemShut
  {Stop}%
\bibitem [{Note5()}]{Note5}%
  \BibitemOpen
  \bibinfo {note} {Simulations are performed with the commercial software
  package CST Microwave Studio. We used the finite element frequency domain
  solver with periodic boundary conditions in the transverse plane and a $3$
  [$4$] unit cells wide slab of the {\protect \bf 8-srs} structure. Perfectly
  matched layers are used for the $z$ boundaries of the simulation
  box.}\BibitemShut {Stop}%
\bibitem [{\citenamefont {Nicoletti}\ \emph {et~al.}(2008)\citenamefont
  {Nicoletti}, \citenamefont {Zhou}, \citenamefont {Jia}, \citenamefont
  {Venture}, \citenamefont {Bulla}, \citenamefont {Luther-Davies},\ and\
  \citenamefont {Gu}}]{NicolettiZhouJiaVenturaDouglasLutherDaviesGu:2008}%
  \BibitemOpen
  \bibfield  {author} {\bibinfo {author} {\bibfnamefont {E.}~\bibnamefont
  {Nicoletti}}, \bibinfo {author} {\bibfnamefont {G.}~\bibnamefont {Zhou}},
  \bibinfo {author} {\bibfnamefont {B.}~\bibnamefont {Jia}}, \bibinfo {author}
  {\bibfnamefont {M.~J.}\ \bibnamefont {Venture}}, \bibinfo {author}
  {\bibfnamefont {D.}~\bibnamefont {Bulla}}, \bibinfo {author} {\bibfnamefont
  {B.}~\bibnamefont {Luther-Davies}}, \ and\ \bibinfo {author} {\bibfnamefont
  {M.}~\bibnamefont {Gu}},\ }\href {\doibase 10.1364/OL.33.002311} {\bibfield
  {journal} {\bibinfo  {journal} {Opt. Lett.}\ }\textbf {\bibinfo {volume}
  {33}},\ \bibinfo {pages} {2311} (\bibinfo {year} {2008})}\BibitemShut
  {NoStop}%
\bibitem [{Note6()}]{Note6}%
  \BibitemOpen
  \bibinfo {note} {The infinite size in $x$ and $y$ direction is achieved by
  use of periodic boundary conditions with a single unit cell of the {\protect
  \bf 8-srs} structure}\BibitemShut {NoStop}%
\bibitem [{Note7()}]{Note7}%
  \BibitemOpen
  \bibinfo {note} {The termination determines the planes at which the infinite
  periodic crystal is clipped to give the slab of height $n_z\protect \tmspace
  +\thinmuskip {.1667em}a$. The termination is indicated by the vertical
  clipping position $t$ in crystallographic coordinates, with $t=0$
  corresponding to a clipping plane through the point with symmetry $432$ (the
  origin in the notation of ref.~\cite
  {InternationalTablesForCrystallography:1992}).}\BibitemShut {Stop}%
\bibitem [{\citenamefont {Han}\ \emph {et~al.}(2009)\citenamefont {Han},
  \citenamefont {Zhang}, \citenamefont {Chng}, \citenamefont {Sun},
  \citenamefont {Zhao}, \citenamefont {Zou},\ and\ \citenamefont
  {Ying}}]{HanZhangChngSunZhaoZouYing:2009}%
  \BibitemOpen
  \bibfield  {author} {\bibinfo {author} {\bibfnamefont {Y.}~\bibnamefont
  {Han}}, \bibinfo {author} {\bibfnamefont {D.}~\bibnamefont {Zhang}}, \bibinfo
  {author} {\bibfnamefont {L.}~\bibnamefont {Chng}}, \bibinfo {author}
  {\bibfnamefont {J.}~\bibnamefont {Sun}}, \bibinfo {author} {\bibfnamefont
  {L.}~\bibnamefont {Zhao}}, \bibinfo {author} {\bibfnamefont {X.}~\bibnamefont
  {Zou}}, \ and\ \bibinfo {author} {\bibfnamefont {J.}~\bibnamefont {Ying}},\
  }\href@noop {} {\bibfield  {journal} {\bibinfo  {journal} {Nat.~Chem.}\
  }\textbf {\bibinfo {volume} {1}},\ \bibinfo {pages} {123} (\bibinfo {year}
  {2009})}\BibitemShut {NoStop}%
\bibitem [{\citenamefont {Hyde}\ \emph {et~al.}(2009)\citenamefont {Hyde},
  \citenamefont {{Di Campo}},\ and\ \citenamefont
  {Oguey}}]{HydeDiCampOguey:2009}%
  \BibitemOpen
  \bibfield  {author} {\bibinfo {author} {\bibfnamefont {S.~T.}\ \bibnamefont
  {Hyde}}, \bibinfo {author} {\bibfnamefont {L.}~\bibnamefont {{Di Campo}}}, \
  and\ \bibinfo {author} {\bibfnamefont {C.}~\bibnamefont {Oguey}},\
  }\href@noop {} {\bibfield  {journal} {\bibinfo  {journal} {Soft Matter}\
  }\textbf {\bibinfo {volume} {5}},\ \bibinfo {pages} {2782} (\bibinfo {year}
  {2009})}\BibitemShut {NoStop}%
\bibitem [{\citenamefont {Kirkensgaard}(2012)}]{Kirkensgaard:2012}%
  \BibitemOpen
  \bibfield  {author} {\bibinfo {author} {\bibfnamefont {J.~J.~K.}\
  \bibnamefont {Kirkensgaard}},\ }\href {\doibase 10.1103/PhysRevE.85.031802}
  {\bibfield  {journal} {\bibinfo  {journal} {Phys. Rev. E}\ }\textbf {\bibinfo
  {volume} {85}},\ \bibinfo {pages} {031802} (\bibinfo {year}
  {2012})}\BibitemShut {NoStop}%
\bibitem [{\citenamefont {Schroder-Turk}\ \emph {et~al.}(2013)\citenamefont
  {Schroder-Turk}, \citenamefont {de~Campo}, \citenamefont {Evans},
  \citenamefont {Saba}, \citenamefont {Kapfer}, \citenamefont {Varslot},
  \citenamefont {Grosse-Brauckmann}, \citenamefont {Ramsden},\ and\
  \citenamefont {Hyde}}]{SchroederTurk:2013}%
  \BibitemOpen
  \bibfield  {author} {\bibinfo {author} {\bibfnamefont {G.~E.}\ \bibnamefont
  {Schroder-Turk}}, \bibinfo {author} {\bibfnamefont {L.}~\bibnamefont
  {de~Campo}}, \bibinfo {author} {\bibfnamefont {M.~E.}\ \bibnamefont {Evans}},
  \bibinfo {author} {\bibfnamefont {M.}~\bibnamefont {Saba}}, \bibinfo {author}
  {\bibfnamefont {S.~C.}\ \bibnamefont {Kapfer}}, \bibinfo {author}
  {\bibfnamefont {T.}~\bibnamefont {Varslot}}, \bibinfo {author} {\bibfnamefont
  {K.}~\bibnamefont {Grosse-Brauckmann}}, \bibinfo {author} {\bibfnamefont
  {S.}~\bibnamefont {Ramsden}}, \ and\ \bibinfo {author} {\bibfnamefont
  {S.~T.}\ \bibnamefont {Hyde}},\ }\href {\doibase 10.1039/C2FD20112G}
  {\bibfield  {journal} {\bibinfo  {journal} {Faraday Discuss.}\ }\textbf
  {\bibinfo {volume} {161}},\ \bibinfo {pages} {215} (\bibinfo {year}
  {2013})}\BibitemShut {NoStop}%
\bibitem [{\citenamefont {Cumming}\ \emph {et~al.}(2012)\citenamefont
  {Cumming}, \citenamefont {Turner}, \citenamefont {Debbarma}, \citenamefont
  {Luther-Davies}, \citenamefont {Schr\"{o}der-Turk},\ and\ \citenamefont
  {Gu}}]{Cumming:2012}%
  \BibitemOpen
  \bibfield  {author} {\bibinfo {author} {\bibfnamefont {B.~P.}\ \bibnamefont
  {Cumming}}, \bibinfo {author} {\bibfnamefont {M.~D.}\ \bibnamefont {Turner}},
  \bibinfo {author} {\bibfnamefont {S.}~\bibnamefont {Debbarma}}, \bibinfo
  {author} {\bibfnamefont {B.}~\bibnamefont {Luther-Davies}}, \bibinfo {author}
  {\bibfnamefont {G.}~\bibnamefont {Schr\"{o}der-Turk}}, \ and\ \bibinfo
  {author} {\bibfnamefont {M.}~\bibnamefont {Gu}},\ }in\ \href@noop {} {\emph
  {\bibinfo {booktitle} {Frontiers in Optics Conference}}}\ (\bibinfo
  {publisher} {Optical Society of America},\ \bibinfo {year} {2012})\ p.\
  \bibinfo {pages} {FW1F.7}\BibitemShut {NoStop}%
\bibitem [{Note8()}]{Note8}%
  \BibitemOpen
  \bibinfo {note} {Note, that for compactness of notation we suppress that
  $\alpha $ itself depends on $i$, i.e.~$\alpha =\alpha _i$.}\BibitemShut
  {Stop}%
\bibitem [{Note9()}]{Note9}%
  \BibitemOpen
  \bibinfo {note} {The procedure for the electric field equation is analogous.
  We note however, that the character of any improper rotation of the $E$-field
  is the negative of the $H$-field character which can be seen for example by
  using Faraday's equation. While any $I432$ PC does not exhibit any improper
  rotation, any dielectric PC has time inversion symmetry that is {\protect \it
  improper} and hence induces the same phase shift between $H$ and
  $E$-field.}\BibitemShut {Stop}%
\bibitem [{Note10()}]{Note10}%
  \BibitemOpen
  \bibinfo {note} {This defines a set of all propagating modes in a
  semi-infinite structure. However, for any real scattering problem a complete
  set of PC modes includes evanescent modes. The impact of evanescent waves
  onto the scattering process becomes dominant if there is no propagating mode
  present in the transmission channel that has a finite amount of energy in the
  $00$ Bragg order to which the incident beam couples only. This makes it for
  example impossible to get any useful information about transmission through a
  finite slab in the frequency range between $\Omega \mathrel {\mathrel
  {\mathop :}\mkern -1.2mu=}\omega a/(2\pi c)=0.66$ and $0.71$ from the band
  structure in figure \ref {fig:transmission-bandstructure} alone.}\BibitemShut
  {Stop}%
\bibitem [{Note11()}]{Note11}%
  \BibitemOpen
  \bibinfo {note} {Conservation of the spatio-temporal modulation at the
  surface implies conservation of frequency and parallel part of the wave
  vector where the PC acts as a grating and back scattering can occur into
  several Bragg orders, i.e.~back scattered waves are given by a plane wave
  Floquet basis.}\BibitemShut {Stop}%
\bibitem [{Note12()}]{Note12}%
  \BibitemOpen
  \bibinfo {note} {Technically, a class is a subset $S_n$ of symmetry elements
  that is obtained by one selected member and all elements that are conjugate.
  As a result, for most purposes all elements of a class can be treated
  equally, i.e.~they have same character etc.}\BibitemShut {Stop}%
\bibitem [{Note13()}]{Note13}%
  \BibitemOpen
  \bibinfo {note} {An alternative proof is provided in \cite
  {Sakoda:2005}.}\BibitemShut {Stop}%
\bibitem [{Note14()}]{Note14}%
  \BibitemOpen
  \bibinfo {note} {The sum over $\alpha $ is superstitious in this special case
  as all $C_4$ representations are one-dimensional.}\BibitemShut {Stop}%
\bibitem [{Note15()}]{Note15}%
  \BibitemOpen
  \bibinfo {note} {For any dielectric PC \cite
  {JoannopoulosJohnsonWinnMeade:2008}}\BibitemShut {NoStop}%
\bibitem [{Note16()}]{Note16}%
  \BibitemOpen
  \bibinfo {note} {This claim can be proved with degenerate perturbation
  theory. The irreducible representation of the point group of the wave vector
  is reduced in the lower symmetry group of $\delta k$ and the matrix elements
  of the perturbation operator $\protect \mathaccentV {hat}05E{\vartheta
  }_{\delta k}$ in this basis are diagonalized (similar to the $k\cdot p$
  analog in quantum mechanics). As all matrix elements satisfy the symmetry
  relation $<\protect \mathaccentV {hat}05E{\vartheta }_{-\delta k}>=-<\protect
  \mathaccentV {hat}05E{\vartheta }_{\delta k}>$, there is always a band with
  continuous slope passing through the point without changing its
  representation. The $\Gamma _0$ point where $E$ and $H$ are decoupled and
  hence the group velocity is zero whereas finite in the vicinity of $\Gamma $
  is an exception.}\BibitemShut {Stop}%
\bibitem [{Note17()}]{Note17}%
  \BibitemOpen
  \bibinfo {note} {The term {\protect \em anti-crossing} is taken from
  reference \protect \rev@citealpnum {Dresselhaus:2008} and is used to
  characterize two bands that come close to each other and seem to interchange
  group velocities from left to right without actually touching one
  another.}\BibitemShut {Stop}%
\bibitem [{Note18()}]{Note18}%
  \BibitemOpen
  \bibinfo {note} {The points are all equal sized in the inset and colored by
  the color of $i$ with dominating $N_i$ with intensity maximum at $N_i=1$ and
  continuously decreasing to light grey at $N_i=0.25$.}\BibitemShut {Stop}%
\bibitem [{Note19()}]{Note19}%
  \BibitemOpen
  \bibinfo {note} {The color code of the $E_\pm $ modes in the main plot is in
  the common hexadecimal rgb scheme defined by $RGB=C\protect \floor
  {255(N_{E_+}-N_{E_-})}$ where $C=256^2$ if $N_{E_+}>N_{E_-}$ and $C=1$
  else.}\BibitemShut {Stop}%
\bibitem [{Note20()}]{Note20}%
  \BibitemOpen
  \bibinfo {note} {Electromagnetic eigenmodes (unlike photons in a particle
  picture) behave like Fermions: There can only be a single mode in a state
  classified by all symmetries (or quantum numbers) because the two modes are
  uniquely determined except for a scalar coefficient and hence
  interchangeable. The symmetries in Maxwell theory are translation invariance
  in time and space ($\omega $ and $k$), time inversion ($k\leftrightarrow -k$)
  and point symmetries (character).}\BibitemShut {Stop}%
\bibitem [{det()}]{details_of_unfolding_in_prb}%
  \BibitemOpen
  \href@noop {} {}\bibinfo {note} {The main Bragg contribution $\vec{G}$ to the
  SC Bloch mode is determined and the band structure unfolded plotting $1-k$
  and exchanging $E_+\leftrightarrow E_-$ if
  $2\pi\vec{G}\cdot\vec{e}_{\vec{k}}/a$ yields an odd number.}\BibitemShut
  {Stop}%
\bibitem [{Note21()}]{Note21}%
  \BibitemOpen
  \bibinfo {note} {We have adopted the term degeneracy although the states have
  opposite Bloch wave vector and generally different
  representation.}\BibitemShut {Stop}%
\bibitem [{Note22()}]{Note22}%
  \BibitemOpen
  \bibinfo {note} {Generally, a third case can occur where neither statement
  (a) nor (b) holds. However, this case only occurs in the presence of a
  $4$-fold improper rotation or a two-fold screw axis that are both not present
  and generally cannot be in the group of the wave vector within the Brillouin
  zone \cite {Herring:1937}.}\BibitemShut {Stop}%
\bibitem [{\citenamefont {Ludwig}\ and\ \citenamefont
  {Falter}(1988)}]{LudwigFalter:1988}%
  \BibitemOpen
  \bibfield  {author} {\bibinfo {author} {\bibfnamefont {W.}~\bibnamefont
  {Ludwig}}\ and\ \bibinfo {author} {\bibfnamefont {C.}~\bibnamefont
  {Falter}},\ }\href@noop {} {\emph {\bibinfo {title} {Symmetries in Physics:
  Group Theory applied to Physical Problems}}}\ (\bibinfo  {publisher}
  {Springer-Verlag},\ \bibinfo {year} {1988})\BibitemShut {NoStop}%
\bibitem [{\citenamefont {Herring}(1937)}]{Herring:1937}%
  \BibitemOpen
  \bibfield  {author} {\bibinfo {author} {\bibfnamefont {C.}~\bibnamefont
  {Herring}},\ }\href@noop {} {\bibfield  {journal} {\bibinfo  {journal}
  {Phys.~Rev.}\ } (\bibinfo {year} {1937})}\BibitemShut {NoStop}%
\bibitem [{Note23()}]{Note23}%
  \BibitemOpen
  \bibinfo {note} {Technically, this is a choice of natural basis states where
  either the magnetic ($s$) or the electric ($p$) field lies in the plane of
  incidence of the respective plane wave component, i.e.~the plane spanned by
  the wave vector and the surface normal. That definition corresponds to polar
  coordinate system $(r,\varphi ,z)$ with the $4$-fold axis as its center so
  that the polarization state translates as $s=r$ and $p=\varphi $. In that
  local coordinate system, the action of any $R\in C_4$ onto $e_\sigma $ is the
  identity transformation. Note that we only use the polar system for the field
  vectors and not for the position vector so that all spatial derivatives
  etc.~are still Cartesian.}\BibitemShut {Stop}%
\bibitem [{Note24()}]{Note24}%
  \BibitemOpen
  \bibinfo {note} {$R(0,0)^T=(0,0)^T\forall \protect \mathaccentV
  {hat}05E{R}\in C_4$ yields $f_{\sigma ,(0,0)^T}^{A,B}=0$ and
  $f_{x,(0,0)^T}^{(i)}=f_{y,(0,0)^T}^{(i)}$.}\BibitemShut {Stop}%
\bibitem [{Note25()}]{Note25}%
  \BibitemOpen
  \bibinfo {note} {So that the structure is assumed to be homogeneous along $z$
  within that layer (i.e.~$\varepsilon (z)=\varepsilon (z+z_0)$ if
  $|z_0|<\delta /2$).}\BibitemShut {Stop}%
\bibitem [{Note26()}]{Note26}%
  \BibitemOpen
  \bibinfo {note} {It can be numerically computed by conversion of the planar
  Bloch-Maxwell equation into an eigenvalue equation for $q_d$ \cite
  {WhittakerCulshaw:1999}.}\BibitemShut {Stop}%
\bibitem [{Note27()}]{Note27}%
  \BibitemOpen
  \bibinfo {note} {$z$-inversion symmetry around the homogeneous layer center
  and time inversion symmetry guarantee that in each case states come in pairs
  $q_\pm $}\BibitemShut {NoStop}%
\bibitem [{\citenamefont {Whittaker}\ and\ \citenamefont
  {Culshaw}(1999)}]{WhittakerCulshaw:1999}%
  \BibitemOpen
  \bibfield  {author} {\bibinfo {author} {\bibfnamefont {D.~M.}\ \bibnamefont
  {Whittaker}}\ and\ \bibinfo {author} {\bibfnamefont {I.~S.}\ \bibnamefont
  {Culshaw}},\ }\href {\doibase 10.1103/PhysRevB.60.2610} {\bibfield  {journal}
  {\bibinfo  {journal} {Phys. Rev. B}\ }\textbf {\bibinfo {volume} {60}},\
  \bibinfo {pages} {2610} (\bibinfo {year} {1999})}\BibitemShut {NoStop}%
\bibitem [{Note28()}]{Note28}%
  \BibitemOpen
  \bibinfo {note} {The change of sign in the exponential is irrelevant as the
  basis functions have a symmetry $f_{\sigma ,n}=f_{\sigma ,-n}$ that is due
  the vectorial nature of $f$ and $\chi _{E_\pm }(C_2)=-1$ and can be seen
  acting with $\protect \mathaccentV {hat}05E{C}_2$ twice upon $f$ using the
  explicit form and the representation form each once.}\BibitemShut {Stop}%
\bibitem [{\citenamefont {Radhakrishnan}(1947)}]{Radhakrishnan:1947}%
  \BibitemOpen
  \bibfield  {author} {\bibinfo {author} {\bibfnamefont {T.}~\bibnamefont
  {Radhakrishnan}},\ }\href@noop {} {\bibfield  {journal} {\bibinfo  {journal}
  {Proceedings of the Indian Academy of Sciences - Section A}\ }\textbf
  {\bibinfo {volume} {25}},\ \bibinfo {pages} {260} (\bibinfo {year}
  {1947})}\BibitemShut {NoStop}%
\bibitem [{\citenamefont {Hough}\ \emph {et~al.}(2007)\citenamefont {Hough},
  \citenamefont {Zhu}, \citenamefont {Nakata}, \citenamefont {Chattham},
  \citenamefont {Dantlgraber}, \citenamefont {Tschierske},\ and\ \citenamefont
  {Clark}}]{HoughZhuNakataChatthamDantlgraberTschierskeClark:2007}%
  \BibitemOpen
  \bibfield  {author} {\bibinfo {author} {\bibfnamefont {L.~E.}\ \bibnamefont
  {Hough}}, \bibinfo {author} {\bibfnamefont {C.}~\bibnamefont {Zhu}}, \bibinfo
  {author} {\bibfnamefont {M.}~\bibnamefont {Nakata}}, \bibinfo {author}
  {\bibfnamefont {N.}~\bibnamefont {Chattham}}, \bibinfo {author}
  {\bibfnamefont {G.}~\bibnamefont {Dantlgraber}}, \bibinfo {author}
  {\bibfnamefont {C.}~\bibnamefont {Tschierske}}, \ and\ \bibinfo {author}
  {\bibfnamefont {N.~A.}\ \bibnamefont {Clark}},\ }\href@noop {} {\bibfield
  {journal} {\bibinfo  {journal} {Phys. Rev. Lett.}\ }\textbf {\bibinfo
  {volume} {98}},\ \bibinfo {pages} {037802} (\bibinfo {year}
  {2007})}\BibitemShut {NoStop}%
\bibitem [{\citenamefont {Decker}\ \emph {et~al.}(2010)\citenamefont {Decker},
  \citenamefont {Zhao}, \citenamefont {Soukoulis}, \citenamefont {Linden},\
  and\ \citenamefont {Wegener}}]{DeckerZhaoSoukoulisLindenWegener:2010}%
  \BibitemOpen
  \bibfield  {author} {\bibinfo {author} {\bibfnamefont {M.}~\bibnamefont
  {Decker}}, \bibinfo {author} {\bibfnamefont {R.}~\bibnamefont {Zhao}},
  \bibinfo {author} {\bibfnamefont {C.~M.}\ \bibnamefont {Soukoulis}}, \bibinfo
  {author} {\bibfnamefont {S.}~\bibnamefont {Linden}}, \ and\ \bibinfo {author}
  {\bibfnamefont {M.}~\bibnamefont {Wegener}},\ }\href@noop {} {\bibfield
  {journal} {\bibinfo  {journal} {Opt. Lett.}\ }\textbf {\bibinfo {volume}
  {35}},\ \bibinfo {pages} {1593} (\bibinfo {year} {2010})}\BibitemShut
  {NoStop}%
\bibitem [{\citenamefont {Delgado-Friedrichs}\ \emph
  {et~al.}(2005)\citenamefont {Delgado-Friedrichs}, \citenamefont {Foster},
  \citenamefont {O’Keeffe}, \citenamefont {Proserpio}, \citenamefont
  {Treacy},\ and\ \citenamefont
  {Yaghi}}]{DelgadoFriedrichsFosterOKeeffeProserpioTreacyYaghi:2005}%
  \BibitemOpen
  \bibfield  {author} {\bibinfo {author} {\bibfnamefont {O.}~\bibnamefont
  {Delgado-Friedrichs}}, \bibinfo {author} {\bibfnamefont {M.~D.}\ \bibnamefont
  {Foster}}, \bibinfo {author} {\bibfnamefont {M.}~\bibnamefont {O’Keeffe}},
  \bibinfo {author} {\bibfnamefont {D.~M.}\ \bibnamefont {Proserpio}}, \bibinfo
  {author} {\bibfnamefont {M.~M.}\ \bibnamefont {Treacy}}, \ and\ \bibinfo
  {author} {\bibfnamefont {O.~M.}\ \bibnamefont {Yaghi}},\ }\href {\doibase
  http://dx.doi.org/10.1016/j.jssc.2005.06.037} {\bibfield  {journal} {\bibinfo
   {journal} {Journal of Solid State Chemistry}\ }\textbf {\bibinfo {volume}
  {178}},\ \bibinfo {pages} {2533 } (\bibinfo {year} {2005})},\ \bibinfo {note}
  {<ce:title>Reticular Chemistry: Design, Synthesis, Properties and
  Applications of Metal-Organic Polyhedra and
  Frameworks</ce:title>}\BibitemShut {NoStop}%
\bibitem [{\citenamefont {Fischer}(2004)}]{Fischer:2004}%
  \BibitemOpen
  \bibfield  {author} {\bibinfo {author} {\bibfnamefont {W.}~\bibnamefont
  {Fischer}},\ }\href@noop {} {\bibfield  {journal} {\bibinfo  {journal} {Acta
  Crystallographica Section A}\ }\textbf {\bibinfo {volume} {60}},\ \bibinfo
  {pages} {246} (\bibinfo {year} {2004})}\BibitemShut {NoStop}%
\bibitem [{\citenamefont {Koch}(2000)}]{Koch:1999}%
  \BibitemOpen
  \bibfield  {author} {\bibinfo {author} {\bibfnamefont {E.}~\bibnamefont
  {Koch}},\ }\href@noop {} {\bibfield  {journal} {\bibinfo  {journal} {Acta
  Crystallographica Section A}\ }\textbf {\bibinfo {volume} {56}},\ \bibinfo
  {pages} {15} (\bibinfo {year} {2000})}\BibitemShut {NoStop}%
\bibitem [{\citenamefont {Rosi}\ \emph {et~al.}(2005)\citenamefont {Rosi},
  \citenamefont {Kim}, \citenamefont {Eddaoudi}, \citenamefont {Chen},
  \citenamefont {O'Keeffe},\ and\ \citenamefont
  {Yaghi}}]{RosiKimEddaoudiChenOKeeffeYaghi:2005}%
  \BibitemOpen
  \bibfield  {author} {\bibinfo {author} {\bibfnamefont {N.~L.}\ \bibnamefont
  {Rosi}}, \bibinfo {author} {\bibfnamefont {J.}~\bibnamefont {Kim}}, \bibinfo
  {author} {\bibfnamefont {M.}~\bibnamefont {Eddaoudi}}, \bibinfo {author}
  {\bibfnamefont {B.}~\bibnamefont {Chen}}, \bibinfo {author} {\bibfnamefont
  {M.}~\bibnamefont {O'Keeffe}}, \ and\ \bibinfo {author} {\bibfnamefont
  {O.~M.}\ \bibnamefont {Yaghi}},\ }\href@noop {} {\bibfield  {journal}
  {\bibinfo  {journal} {Journal of the American Chemical Society}\ }\textbf
  {\bibinfo {volume} {127}},\ \bibinfo {pages} {1504} (\bibinfo {year}
  {2005})}\BibitemShut {NoStop}%
\bibitem [{\citenamefont {Evans}\ \emph
  {et~al.}(2013{\natexlab{b}})\citenamefont {Evans}, \citenamefont {Robins},\
  and\ \citenamefont {Hyde}}]{EvansRobinsHyde:2013b}%
  \BibitemOpen
  \bibfield  {author} {\bibinfo {author} {\bibfnamefont {M.~E.}\ \bibnamefont
  {Evans}}, \bibinfo {author} {\bibfnamefont {V.}~\bibnamefont {Robins}}, \
  and\ \bibinfo {author} {\bibfnamefont {S.~T.}\ \bibnamefont {Hyde}},\
  }\href@noop {} {\bibfield  {journal} {\bibinfo  {journal} {Acta
  Crystallographica Section A}\ }\textbf {\bibinfo {volume} {69}} (\bibinfo
  {year} {2013}{\natexlab{b}})}\BibitemShut {NoStop}%
\bibitem [{\citenamefont {O’Keeffe}\ \emph {et~al.}(2008)\citenamefont
  {O’Keeffe}, \citenamefont {Peskov}, \citenamefont {Ramsden},\ and\
  \citenamefont {Yaghi}}]{OKeeffePeskovRamsdenYaghi:2008}%
  \BibitemOpen
  \bibfield  {author} {\bibinfo {author} {\bibfnamefont {M.}~\bibnamefont
  {O’Keeffe}}, \bibinfo {author} {\bibfnamefont {M.~A.}\ \bibnamefont
  {Peskov}}, \bibinfo {author} {\bibfnamefont {S.~J.}\ \bibnamefont {Ramsden}},
  \ and\ \bibinfo {author} {\bibfnamefont {O.~M.}\ \bibnamefont {Yaghi}},\
  }\href@noop {} {\bibfield  {journal} {\bibinfo  {journal} {Accounts of
  Chemical Research}\ }\textbf {\bibinfo {volume} {41}},\ \bibinfo {pages}
  {1782} (\bibinfo {year} {2008})}\BibitemShut {NoStop}%
\bibitem [{\citenamefont {Thomson}(1888)}]{Thomson:1888}%
  \BibitemOpen
  \bibfield  {author} {\bibinfo {author} {\bibfnamefont {W.}~\bibnamefont
  {Thomson}},\ }\href@noop {} {\bibfield  {journal} {\bibinfo  {journal} {Acta
  mathematica}\ }\textbf {\bibinfo {volume} {1}},\ \bibinfo {pages} {121}
  (\bibinfo {year} {1888})}\BibitemShut {NoStop}%
\bibitem [{\citenamefont {Oskooi}\ \emph {et~al.}(2010)\citenamefont {Oskooi},
  \citenamefont {Roundy}, \citenamefont {Ibanescu}, \citenamefont {Bermel},
  \citenamefont {Joannopoulos},\ and\ \citenamefont
  {Johnson}}]{OskooiRoundyIbanescuBermelJoannopoulosJohnson:2010}%
  \BibitemOpen
  \bibfield  {author} {\bibinfo {author} {\bibfnamefont {A.~F.}\ \bibnamefont
  {Oskooi}}, \bibinfo {author} {\bibfnamefont {D.}~\bibnamefont {Roundy}},
  \bibinfo {author} {\bibfnamefont {M.}~\bibnamefont {Ibanescu}}, \bibinfo
  {author} {\bibfnamefont {P.}~\bibnamefont {Bermel}}, \bibinfo {author}
  {\bibfnamefont {J.~D.}\ \bibnamefont {Joannopoulos}}, \ and\ \bibinfo
  {author} {\bibfnamefont {S.~G.}\ \bibnamefont {Johnson}},\ }\href@noop {}
  {\bibfield  {journal} {\bibinfo  {journal} {Computer Physics Communications}\
  }\textbf {\bibinfo {volume} {181}},\ \bibinfo {pages} {687} (\bibinfo {year}
  {2010})}\BibitemShut {NoStop}%
\bibitem [{\citenamefont {Johnson}\ and\ \citenamefont
  {Joannopoulos}(2001)}]{JohnsonJoannopoulos:2001}%
  \BibitemOpen
  \bibfield  {author} {\bibinfo {author} {\bibfnamefont {S.}~\bibnamefont
  {Johnson}}\ and\ \bibinfo {author} {\bibfnamefont {J.}~\bibnamefont
  {Joannopoulos}},\ }\href@noop {} {\bibfield  {journal} {\bibinfo  {journal}
  {Opt. Express}\ }\textbf {\bibinfo {volume} {8}},\ \bibinfo {pages} {173}
  (\bibinfo {year} {2001})}\BibitemShut {NoStop}%
\end{thebibliography}%

\end{document}